\documentclass[twocolumn]{aastex7}

\usepackage{amsmath}
\usepackage{natbib}

\usepackage{booktabs}

\usepackage{xcolor}
\usepackage[inline,ignoremode]{trackchanges}

\newcommand{\pL}{p_{\mathrm{L}}}
\newcommand{\pR}{p_{\mathrm{R}}}

\newcommand{\Ct}{$^{12}${\rm{C} }}
\newcommand{\Nf}{$^{14}${\rm{N}}}
\newcommand{\Os}{$^{16}${\rm{O}}}
\newcommand{\Lis}{$^{7}${\rm{Li}}} 
 
\newcommand{\Bt}{$^{10}${\rm{B}}} 
\newcommand{\Ben}{$^{9}${\rm{Be} }}
\newcommand{\Bet}{$^{10}${\rm{Be} }}
\newcommand{\Bel}{$^{11}${\rm{B} }}

\newcommand{\BtoC}{{\rm{B/C} }}
\newcommand{\BBtoC}{{(\Bet+\Bel)/\Ct}}
\newcommand{\BettoBen}{\Bet/\Ben }

\newcommand{\kpc}{{\mathrm{kpc}}}
\newcommand{\Myr}{{\mathrm{Myr}}} 

\newcommand{\eV}{{\mathrm{eV}}} % alias
\newcommand{\GeV}{{{\mathrm{G}}\eV}} % giga electronvolt

\begin{document}

\title{Implementation of CR Energy SPectrum (CRESP) algorithm in PIERNIK MHD code.
II. Propagation of Primary and Secondary nuclei in a magneto-hydrodynamical environment}

\author[0000-0002-5376-3322]{Antoine Baldacchino-Jordan}
\affiliation{Institute of Astronomy, Faculty of Physics, Astronomy and Informatics, Nicolaus Copernicus University in Toruń,
ul. Grudziadzka 5/7, PL-87-100 Toruń, Poland}
\email{503332@doktorant.umk.pl}
\author[0000-0002-2370-5631]{Michal Hanasz}
\affiliation{Institute of Astronomy, Faculty of Physics, Astronomy and Informatics, Nicolaus Copernicus University in Toruń,
ul. Grudziadzka 5/7, PL-87-100 Toruń, Poland}
\email{mhanasz@umk.pl}
\author[0000-0003-4810-2244]{Mateusz Ogrodnik}
\affiliation{Institute of Astronomy, Faculty of Physics, Astronomy and Informatics, Nicolaus Copernicus University in Toruń,
ul. Grudziadzka 5/7, PL-87-100 Toruń, Poland}
\email{mogrodnik@abs.umk.pl}
\author[0000-0003-3054-6461]{Dominik Wóltański}
\affiliation{Institute of Astronomy, Faculty of Physics, Astronomy and Informatics, Nicolaus Copernicus University in Toruń,
ul. Grudziadzka 5/7, PL-87-100 Toruń, Poland}
\email{wolt@umk.pl}
\author[0000-0002-4435-4594]{Artur Gawryszczak}
\affiliation{Nicolaus Copernicus Astronomical Center, Bartycka 18, PL-00-716 Warsaw, Poland}
\email{gawrysz@gmail.com}
\author[0000-0003-3799-5489]{Andrew W. Strong}
\affiliation{Max-Planck-Institut für extraterrestrische Physik, 85748 Garching, Germany}
\email{aws@mpe.mpg.de}
\author[0000-0002-9300-9914]{Philipp Girichidis}
\affiliation{Universität Heidelberg, Zentrum für Astronomie, Institut für Theoretische Astrophysik, Albert-Ueberle-Str. 2, 69120 Heidelberg,
Germany}
\email{philipp@girichidis.com}
\begin{abstract}

We developed a new model for the production and propagation of spectrally resolved primary and secondary Cosmic Ray (CR) nuclei elements within the framework of the Cosmic Ray Energy Spectrum (CRESP) module of the PIERNIK MHD code.  We extend the algorithm to several CR nuclei and demonstrate our code's capability to model primary and secondary CR species simultaneously. Primary C, N, and O are accelerated in supernova (SN) remnants. The spallation collisions of the primary nuclei against the thermal ISM protons lead to secondary Li, Be, and B products. All the CR species evolve according to the momentum-dependent Fokker-Planck equations that are dynamically coupled to the MHD system of equations governing the evolution of the ISM.  We demonstrate the operation of this system in the gravity stratified box reproducing the Milky Way conditions in the Sun's local environment. We perform a parameter study by investigating the impact of the SN rate, the CR parallel diffusion coefficient $D_\parallel$, and the rigidity-dependent diffusion coefficient power index $\delta$. A novel result of our investigation is that the secondary-to-primary flux ratio \BtoC increases with increasing diffusion coefficient, due to the weaker vertical magnetic field resulting from CR buoyancy effects.
Moreover, a higher SN rate leads to lower values of \BtoC because of stronger winds and the shorter residence time of primary CR particles in dense disk regions.

\end{abstract}

\keywords{Cosmic Rays --- Astroparticle physics --- Interstellar Medium --- MHD --- Numerical Method}

\section{Introduction} \label{sec:intro}

Cosmic Rays (CRs), relativistic charged subatomic particles traveling through the (inter)galactic medium, represent an essential component of the ISM \citep{2015ARA&A..53..199G} and take part in high-energy processes like gamma-ray production or synchrotron radiation \citep{2023A&ARv..31....4R}. Understanding their propagation challenges experiments, theory, and numerical simulations.

 Collisions between primary CR nuclei and ISM atoms lead, via hadronic processes, to the production of secondary CR particles. Leaky-box models demonstrated decades ago the dependency of secondary to primary and unstable to stable isotope flux ratios on CR transport parameters such as escape time or diffusion coefficient \citep{2023arXiv230900298E}. Experiments such as the AMS-02 spectrometer in the International Space Station have measured fluxes of secondary \Lis, stable and unstable \Ben,  \Bet, and \Bel \citep{2018PhRvL.120b1101A}, as well as \BtoC \citep{2016PhRvL.117w1102A}, $e^+/e^-$ \citep{2018PhRvL.121e1102A} or $p^-/p^+$ \citep{2016PhRvL.117i1103A} flux ratios, providing experimental data to constrain the transport parameters and CR propagation models \citep{2019PhRvD..99j3023E}. On the other hand, CR transport models are needed to interpret radio data on galactic halos \citep{2023A&A...670A.158S}. A massive amount of  CR data from various experiments became accessible through the Cosmic Ray Data Base tool \citep{2014A&A...569A..32M,2020Univ....6..102M,2023EPJC...83..971M}.

Two distinct numerical approaches have been developed to model the propagation of CRs in galaxies: the phenomenological or GALPROP-type modeling \citep{Strong&Moskalenko1998} based on the Fokker-Planck equation for CR population propagating in a stationary ISM environment, and a class of self-consistent models \citep{2007ARNPS..57..285S}, which combine the propagation of CRs with the time-dependent system of MHD equations that include the dynamical coupling of CRs with thermal gas and the galactic magnetic field \citep[see][for the review]{2021LRCA....7....2H}.

There are several models of different complexity within this class: the most basic two-fluid diffusion-advection model includes an additional CR pressure term in the MHD system, including a momentum-integrated CR propagation equation. This model treats the whole CR population as a spectrally unresolved relativistic gas (referred to as grey approximation) diffusing anisotropically along magnetic field lines, with interactions between CRs and magnetic fields \citep{2003A&A...412..331H}.
The following extension includes spectrally resolved CRs by introducing momentum dependence as the fourth dimension in the physical description of a single-component CR proton or electron population \citep{2020MNRAS.491..993G,2021ApJS..253...18O,2022MNRAS.tmp.1768H}. Another extension includes the streaming process relying on CR scattering on self-excited Alfv\'en waves \citep{sharma_2010,2018apj...854....5j,2019MNRAS.485.2977T}. The work of \cite{2022MNRAS.tmp.1768H} introduced the whole suit of CR species (e$^+$, e$^-$, p$^+$, p$^-$ and nuclei), transport and interaction processes with spectrally resolved modeling in galactic-scaled simulations, including advection, diffusion and streaming.
A few simulation attempts to model CR transport using self-consistent models have demonstrated the influence of CRs on the cosmological evolution of galactic discs, galactic gas outflows, and ISM evolution \citep{2013ApJ...777L..38H,2021MNRAS.508.4269P,2023mnras.522.5529p,2022MNRAS.510.3917G,2024MNRAS.52710897G,2024ApJ...964...99A}.

Here we extend the CRESP algorithm \citep{2021ApJS..253...18O}, added to the PIERNIK MHD multifluid environment \citep{2010EAS....42..275H,2010EAS....42..281H,2012EAS....56..363H,2012EAS....56..367H}, which involves the momentum-dependent transport of spectrally resolved CR electrons on Eulerian grids.
The present project aims to introduce and validate the numerical algorithm for the production and propagation of multiple spectrally resolved hadronic CR species. We aim to compare the computed \BtoC and \BettoBen values to the experimental data from AMS02. While our attention focuses on the method rather than a precise description of the whole CR environment, we simplify the physical layout in many respects. We restrict the computational domain to a local rectangular patch of adiabatic interstellar medium, described using a low-resolution gravity-stratified box that represents a portion of the ISM in the vicinity of the Solar System. We also take into account randomly generated supernovae in the interstellar medium, considering only the dynamic effects of the proton component of CRs, ignoring thermal and kinetic SN feedback, as well as their clustering.
Similarly to \cite{2022MNRAS.tmp.1768H} but within a narrower set of CR physics processes, we include the production and propagation of secondary CR nuclei elements, following the preliminary work presented in \cite{2023ecrs.confE.138B}. We assume that primary CRs are accelerated in astrophysical MHD shocks, and secondary CRs result from hadronic collisions of CR primaries against the kernels of the ISM. We focus on spectrally resolved primary \Ct, \Nf, \Os, and secondary \Lis, \Ben, \Bet, \Bt\, and \Bel nuclei isotopes and investigate their spatial and momentum evolution. We apply the spectral description only to CR nucleons. We describe protons using a gray approximation, taking into account only diffusion-advection mechanisms and the adiabatic process, while we omit the CR streaming effects. In the spectrally resolved description of CR nucleons, we account for the above processes, including spallation against nuclei of the thermal ISM and Coulomb cooling, while omitting hadronic losses and plan to include these elements in a subsequent article. The overall goal of this paper is not to accurately fit the CR propagation parameters from these simplified experiments, but rather to verify the qualitative consistency with observational data and the response of the observables to the input parameters.
We perform a parameter study of the SN rate, diffusion coefficient, and rigidity-dependent diffusion power index to investigate the observables: secondary to primary and unstable to stable isotope ratios.
 
The plan of the paper is as follows:
Section~\ref{sec:prior-art} introduces the formal description of multiple CR hadronic species within the two-moment approach and the piece-wise power-law approximation, including the production and propagation of primary and secondary CRs.
Section~\ref{sect:sim_setup} % Simulation setup
introduces the numerical simulation setup, including the magnetized adiabatic ISM, perturbed by randomly occurring supernovae injecting CR protons, and describes CR spectrum properties. We also describe the spallation channels producing secondaries.
In Section~\ref{sect:dyn_evol} % Dynamical evolution
we analyze the system's dynamic evolution, inspecting the thermal plasma's CR-driven dynamics and the properties of the CR energy spectra of primaries and secondaries.
In Section~\ref{sect:BtoC}, we show predictions of our simulations for the secondary to primary ratio \BtoC and the unstable to stable Berylium isotope \BettoBen and compare our results with the observational data.
In Section~\ref{sect:conclusions}, we conclude this paper.

\section{Physical processes and computational methods}\label{sec:prior-art}

\subsection{Fokker Planck equation for cosmic rays and the two moment approach}

We assume  that CR species, labeled by superscript $a$, are represented by a distribution function $f^a(\vec{r},\vec{p},t)$ in phase space and are described by a system of Fokker-Planck equations \citep{1975MNRAS.172..557S} :
\begin{eqnarray}
\label{eq.1}
    & & (\partial_t + \vec{v} \cdot \nabla - \nabla \cdot (D\nabla))f^a(\vec{r},\vec{p},t)  = \frac{1}{3}\nabla \cdot\vec{v}p\partial_p f^a(\vec{r},\vec{p},t)  \nonumber \\ & &+ \frac{1}{p^2}\partial_p(p^2b_l(p)+D_{pp}\partial_p)f^a(\vec{r},\vec{p},t) - \frac{1}{\gamma^a(p)\tau^a}f^a(\vec{r},\vec{p},t) \nonumber \\ &&+ j^a(\vec{r},\vec{p},t), \nonumber \\
\end{eqnarray}
where $D$ and $D_{pp}$ are diffusion coefficients in real and momentum space, $b_l(p)=-\mathrm{d}p/\mathrm{d}t$ is the energy loss term of microscopic processes such as synchrotron, inverse Compton or Coulomb losses, $\gamma^a(p)\tau^a$ is the lifetime of radioactive species in the observer's reference frame and $j^a(\vec{r},\vec{p},t)$ represents supernovae source and spallation catastrophic loss for primaries, and spallation source for secondaries.
Following \cite{2001CoPhC.141...17M} and \cite{2021ApJS..253...18O},
we divide the momentum space into finite intervals $[p_{\mathrm{L}},p_{\mathrm{R}}]$, referred to as bins, and introduce the following two moments of the distribution function $f^a(\vec{r},p,t)$, the number density $n^a$ and kinetic energy density $e^a$:

\begin{equation}
\label{eq.2}
    n^a(\vec{r},t) = \int_{\pL}^{\pR} 4 \pi p^2  f^a(\vec{r},p,t) \mathrm{d}p,
\end{equation}
\begin{equation}
\label{eq.3}
    e^a(\vec{r},t) = \int_{\pL}^{\pR}  4 \pi p^2  f^a(\vec{r},p,t)T^a(p) \mathrm{d}p,
\end{equation}
where $T^a(p)$ is the kinetic energy. For simplicity we do not include the CR streaming; therefore, it is enough to resolve the equation for number density instead of flux density.

We calculate the corresponding number density of the Fokker-Planck Equation (\ref{eq.1}) to obtain

\begin{eqnarray}
\label{eq.4}
    & & (\partial_{t} - \nabla(\langle D \rangle _n \nabla))n^a(\vec{r},t) + \nabla.(n^a(\vec{r},t)\vec{v}) =  Q^a(\vec{r},t) \nonumber \\ & &  + \left[\left(\frac{1}{3}\nabla.\vec{v}\;p+b_l(p)\right)4\pi p^2 f^a(p)\right]^{p_{\mathrm{R}}}_{p_{\mathrm{L}}} \nonumber \\ & & - \left<\frac{1}{\gamma^a(p)\tau^a}\right> _n n^a(\vec{r},t), \nonumber \\
\end{eqnarray}
where $Q^a(\vec{r},t)$ is the number density source, $\langle D \rangle _n$, and $\left<1/\gamma^a(p)\tau^a\right> _n$ are the bin averaged diffusion tensor and radioactive decay loss rate for number density. Analogously, we calculate the kinetic energy density of (\ref{eq.1}) to get:

\begin{eqnarray}
\label{eq.5}
    & & (\partial_{t} - \nabla( \langle D \rangle _e \nabla))e^a(\vec{r},t) + \nabla.(e^a(\vec{r},t)\vec{v}) = S^a (\vec{r},t) \nonumber \\ && + \left[\left(\frac{1}{3}\nabla.\vec{v}\;p+b_l(p)\right)4\pi p^2 f^a(p)T(p)\right]^{p_{\mathrm{R}}}_{p_{\mathrm{L}}} \nonumber \\ &&   - \int^{p_{\mathrm{R}}}_{p_{\mathrm{L}}} 4\pi p^2\mathrm{d}p  \left(\frac{1}{3}\nabla.\vec{v}\;p+b_l(p)\right)   f^a(p) \beta^a(p)c \nonumber \\ && - \left<\frac{1}{\gamma^a(p)\tau^a}\right> _e e^a(\vec{r},t) \nonumber \\
\end{eqnarray}
where $S^a(\vec{r},t)$ is the kinetic energy density source,
 $\langle D \rangle _e$ and $\left<1/\gamma^a(p)\tau^a\right> _e$ are the bin averaged diffusion tensor and radioactive decay loss rate for the energy density, $\beta^a(p)=v^a/c=p/\sqrt{p^2+m_a^2c^2}$. $Q^a$ and $S^a$ terms include SN injection and spallation loss for primaries and spallation creation for secondaries.
Details of the
averaged diffusion coefficients are given in \cite{2021LRCA....7....2H},
while spallation source terms  $Q^a_{\mathrm{spal}}(\vec{r},t)$, $S^a_\mathrm{spal}(\vec{r},t)$  and radioactive decay rates are presented in Sections
\ref{subs:secondaries} and  (\ref{subs:radioactive_decay})  respectively.

\subsection{Secondary nuclei production}\label{subs:secondaries}

We aim to evolve Equation (\ref{eq.1}) for primary and secondary species by implementing secondary nuclei particle production through spallation processes. We implement nuclear reactions following the physics incorporated in GALPROP (see the Galprop Explanatory Supplement). We consider a secondary species labeled $a$. For its spectral number density function per unit momentum $\mathrm{d}n^a/\mathrm{d}p$ in GALPROP, the source term for secondary nuclei is:
\begin{equation}
\label{eq.6}
     q^{a}_\mathrm{spal}(\vec{r},p,t) = \sum_{i,j} n_i \beta_j c\int_{\mathbb{R}^+} \mathrm{d} p' \; \frac{\mathrm{d}{n}_j}{\mathrm{d}p'}(p')\frac{\mathrm{d}\sigma_{ij}^a}{\mathrm{d}p}(p,p')
\end{equation}

 where $n_i$ represents the density of H and He atoms of the thermal ISM species (in this paper we assume only hydrogen), $n_j$ is the number density of primary species $j$, $p'$ is the momentum of primaries, $p$ the momentum of secondaries, $\mathrm{d}\sigma_{ij}^a/\mathrm{d}p$ is the cross section per unit of momentum for the production of secondary CR element $a$, $\beta_j c$ is the velocity of primary CRs.
 
We compare relations (\ref{eq.1}) and (\ref{eq.6}), and by making a dimensional analysis, we obtain the expression $ 4\pi p^2j^a_\mathrm{spal}(\vec{r},p,t) = q^a_\mathrm{spal}(\vec{r},p,t)$. The calculation of sources for Equation (\ref{eq.4}) gives:
\begin{center}
\begin{eqnarray}
\label{eq.7}
    Q^a_\mathrm{spal}(\vec{r},t) &= &\int_{\pL}^{\pR} 4\pi p^2 j^a(\vec{r},p,t)\mathrm{d}p \\ & =& \sum_{i,j} n_i \beta_j c \int^{p_{\mathrm{R}}}_{p_{\mathrm{L}}}\mathrm{d}p\int_{\mathbb{R}^+} \mathrm{d} p' \; \frac{\mathrm{d}n_j}{\mathrm{d}p'}(p')\frac{\mathrm{d}\sigma_{ij}}{\mathrm{d}p}(p,p') \nonumber
\end{eqnarray}
\end{center}
For Equation (\ref{eq.5}), the source term is:
\begin{eqnarray}
\label{eq.8}
    S^a_\mathrm{spal}(\vec{r},t) &=& \int_{\pL}^{\pR} 4\pi p^2 j^a(\vec{r},p,t)T^a(p)\mathrm{d}p\ \\ & =&  \sum_{i,j} n_i \beta_j c \int^{p_{\mathrm{R}}}_{p_{\mathrm{L}}}\mathrm{d}p\; T^a(p) \nonumber \\ &&\times\int_{\mathbb{R}^+} \mathrm{d} p' \; \frac{\mathrm{d}n_j}{\mathrm{d}p'}(p')\frac{\mathrm{d}\sigma ^a_{ij}}{\mathrm{d}p}(p,p') \nonumber
\end{eqnarray}
We now have the general expression of the source term for energy density and number density.
Following the method in GALPROP, we write the differential cross section as:
\begin{eqnarray}
\label{eq.9}
    \frac{\mathrm{d}\sigma^a_{ij}}{\mathrm{d}p'}(p,p') &=& \sigma^a_{ij}(p') \delta\left(p - \frac{A_{\mathrm{a}}}{A_{\mathrm{j}}}p'\right) \\ &=&\sigma^a_{ij}(p')\frac{A_{\mathrm{j}}}{A_{\mathrm{a}}}\delta\left(p' - \frac{A_{\mathrm{j}}}{A_{\mathrm{a}}}p\right) \nonumber
\end{eqnarray}
Here, $A_{\mathrm{j}}$ and $A_{\mathrm{a}}$ are the atomic number of primary and secondary species. In this study, we only considered constant values for cross sections. Then, the expression of $Q^a_\mathrm{spal}(p)$ and $S^a_\mathrm{spal}(p)$ is :
\begin{eqnarray}
\label{eq.10}
     Q^a_\mathrm{spal} &=& \sum_{i,j} n_i \beta_j c\frac{A_{\mathrm{j}}}{A_{\mathrm{a}}}\int^{p_{\mathrm{R}}}_{p_{\mathrm{L}}}\mathrm{d}p\; \frac{\mathrm{d}n_j}{\mathrm{d}p'}\left(\frac{A_{\mathrm{j}}}{A_{\mathrm{a}}}p\right) \sigma^a_{ij} \\ & =& \sum_{i,j} n_i \beta_j c\sigma^a_{ij} n_{j} \nonumber
\end{eqnarray}
and
\begin{eqnarray}
\label{eq.11}
     S^a_\mathrm{spal} &=& \sum_{i,j} n_i \beta_j c\frac{A_{\mathrm{j}}}{A_{\mathrm{a}}}\int^{p_{\mathrm{R}}}_{p_{\mathrm{L}}}\mathrm{d}p\;T\left(p\right) \frac{\mathrm{d}n_j}{\mathrm{d}p'}\left(\frac{A_{\mathrm{j}}}{A_{\mathrm{a}}}p\right) \sigma^a_{ij} \nonumber \\ & =& \sum_{i,j} n_i \beta_j c\frac{A_{\mathrm{a}}}{A_{\mathrm{j}}}\sigma^a_{ij} e_{j}
\end{eqnarray}
There, we integrated over the quantity $p' = (A_{\mathrm{j}}/A_\mathrm{a})p$, which is the momentum of primaries. The ratio $A_{\mathrm{a}}/A_\mathrm{j} < 1$ shows that the total energy loss of the primary nucleus is not completely converted to the secondary nuclei particle since this fraction is transferred in other particle products such as high-energy photons or pions $\pi$ \citep{2011hea..book.....L}, which are not present in our model. The total momentum is not conserved, but the momentum per nucleon is, $p'/A_{\mathrm{j}} = p/A_\mathrm{a}$.

\subsection{Radioactive decay} \label{subs:radioactive_decay}
In our two-moment approach, the radioactive decay term in Equation (\ref{eq.1}) leads to a bin-averaged catastrophic loss for number and energy density. For $p$ in $[p_{\mathrm{L}},p_{\mathrm{R}}]$, the explicit calculation from equations (\ref{eq.4}) and (\ref{eq.5}) gives:
\begin{equation}
\left<\frac{1}{\gamma^a(p)\tau^a}\right> _n = \frac{1}{n^a(\vec{r},t)} \int^{p_{\mathrm{R}}}_{p_{\mathrm{L}}} 4\pi p^2 \;\frac{f^a(\vec{r},p,t)}{\gamma^a(p)\tau^a}\mathrm{d}p
\label{eq.12}
\end{equation}
\begin{equation}
\left<\frac{1}{\gamma^a(p)\tau^a}\right> _e = \frac{1}{e^a(\vec{r},t)} \int^{p_{\mathrm{R}}}_{p_{\mathrm{L}}} 4\pi p^2 \;\frac{f^a(\vec{r},p,t)T^a(p)}{\gamma^a(p)\tau^a}\mathrm{d}p
\label{eq.13}
\end{equation}
where $\gamma^a(p) = \sqrt{1+(p/m_a c)^2}$. The integration of equations (\ref{eq.12}) and (\ref{eq.13}) can be done using the approach described in the section \ref{subs:piece_wise_power_law}.

\subsection{CR energy losses} \label{subs:energy_losses}

In this simplified model, we compute two relevant energy loss processes for every species. The first one is adiabatic expansion loss, following:
\begin{equation}
    b(p)=-\frac{\mathrm{d}p}{\mathrm{d}t}=\frac{1}{3}\nabla.\vec{v}\:p
\label{eq.14}
\end{equation}

The second is Coulomb energy loss, which is significant for hadronic species at non-relativistic energies (see Appendix B). We use the Bethe-Block formula, following \cite{1972Phy....58..379G} for kinetic energy $T$:
\begin{equation}
-\left(\frac{\mathrm{d}T}{\mathrm{d}t}\right)_\mathrm{C}=\frac{\omega_\mathrm{pl}^2Z^2e^2}{\beta c}\left(\mathrm{ln}\left(\frac{2m_ec^2\gamma \beta^2}{\hbar \omega_\mathrm{pl}}\right)-\frac{\beta^2}{2}\right)
\label{eq.15}
\end{equation}
where $\omega_\mathrm{pl}=\sqrt{4\pi e^2 n_e/m_e}$ is the plasma frequency depending on ionized electron gas density $n_e$.
Adiabatic expansion is incorporated in the CRESP algorithm, which progressively cools the spectrum (see Appendix \ref{App.A},
%\citep{2001CoPhC.141...17M,2021ApJS..253...18O}.
In Appendix \ref{App.B} we describe our implementation of Coulomb losses using the free-cooling method developed in \cite{2020MNRAS.491..993G}.

\subsection{Piece-wise power-law approach in the CRESP algorithm} \label{subs:piece_wise_power_law}
CRESP algorithm in \cite{2021ApJS..253...18O} is based on the piecewise power-law (coarse-grained) method for self-consistent and numerically efficient CR propagation in the magnetized interstellar medium (ISM) of galaxies.
Following  \citep{2001CoPhC.141...17M,2020MNRAS.491..993G,2021ApJS..253...18O,2021LRCA....7....2H,2022MNRAS.tmp.1768H}, we numerically integrate number density, energy density and related energy gain/loss terms by assuming a piece-wise power-law, isotropic distribution function
\begin{equation}
\label{eq.16}
f^a(p) = f^a_{l-1/2} \left(\frac{p}{p_{l-1/2}}\right)^{-q^a_l}
\quad \mbox{for}\quad p \in [p_{l-1/2},p_{l+1/2}],
\end{equation}
\begin{equation}
\label{eq.17}
    q^a_l = -\left. \mathrm{ln}\left(\frac{f^a_{l+1/2}}{f^a_{l-1/2}}\right)\right/\mathrm{ln}\left(\frac{p_{l+1/2}}{p_{l-1/2}}\right)
\end{equation}
in an arbitrarily chosen range of momentum space spanning $(p_\mathrm{min}, p_\mathrm{max})$,
where $q^a_l$ is the power index of the bin $l$ for the $a$-th species, $f^a_{l-1/2}$ is the corresponding distribution function amplitude at the left bin edge $p_{l-1/2}$.

The CRESP algorithm so far has been implemented with the ultra-relativistic limit for particle energy, i.e., $T^a(p) = cp$. A new purpose of our work is to implement CRs in the trans-relativistic limit ($cp \approx m_a c^2$) and non-relativistic limit ($cp \ll m_a c^2$), implying the use of the more general relation:
\begin{equation}
\label{eq.18}
    T^a(p)=\sqrt{c^2p^2 + m_a^2c^4} - m_a c^2
\end{equation}
To integrate kinetic energy density (\ref{eq.3}) we also use the piece-wise power-law approximation for kinetic energy \citep[see][]{2021LRCA....7....2H}:

\begin{equation}
\label{eq.19}
    T^a(p) = T^a_{l-1/2}\left(\frac{p}{p_{l-1/2}}\right)^{s^a_l}
\quad \mbox{for}\quad p \in [p_{l-1/2},p_{l+1/2}],
\end{equation}
\begin{equation}
\label{eq.20}
    s^a_l = \left. \mathrm{ln}\left(\frac{T^a_{l+1/2}}{T^a_{l-1/2}}\right)\right/\mathrm{ln}\left(\frac{p_{l+1/2}}{p_{l-1/2}}\right)
\end{equation}
The last equation gives $s^a_l \approx 1$ for ultra-relativistic CRs and $s^a_l \approx 2$ for non-relativistic CRs. From (\ref{eq.16}), $\mathrm{d}T^a/\mathrm{d}p = cp/(\gamma^a(p)m_a c)$ and, writing $\gamma^a$ in terms of $p$ and $\mathrm{d}T^a/\mathrm{d}p$, compute radioactive decay terms for equations (\ref{eq.12}) and (\ref{eq.13}) only using power-laws in (\ref{eq.14}) and (\ref{eq.17}) (see Appendix A for explicit calculations).

Having attributed quantities $f^a_{l-1/2}$, $q^a_l$, $T^a_{l-1/2}$ and $s^a_l$ in each spatial cell and each momentum bin $l$ of every species to compute $e^a_l$ and $n^a_l$, we have to retrieve those quantities at each time step in the evolution of the system. Each CR species is described with two different sets:
\begin{equation}
    (f^a_{l-1/2},q^a_l) \leftrightarrow (n^a_l, e^a_l)
\label{eq.21}
\end{equation}
We compute new $q^a_l$ at each time step in CRESP by solving, for each species:
\begin{equation}
    \frac{e^a_l}{n^a_l T^a_{l-1/2}} = \mathrm{func} (q^a_l,s^a_l)
    \label{eq.22}
\end{equation}
corresponding to Equation (\ref{eq.A30}) and (\ref{eq.A31}) in appendix A. The right-hand side is only a function of $s^a_l$, fixed for every bin, and the unknown quantity $q^a_l$ to find after each time step. For electrons, a Newton-Raphson algorithm was used for this purpose. To model several species simultaneously, we switched to an interpolation method: by noting $\alpha^a_l$ the left-hand side of the Equation (\ref{eq.20}), we construct an array of possible values of $\alpha^a$ and their corresponding power-law $q^a$ for each species. We identify in which range $[\alpha^a_i, \alpha^a_j]$ of the array the value $\alpha^a_l$ in the code belongs, and make a linear interpolation to find $q^a_l$ in the range $[q^a_i, q^a_j]$.

\section{Description of the simulation setup} \label{sect:sim_setup}
To validate our algorithm's new capabilities, we perform a series of gravity-stratified box simulations that validate the code by reproducing physical predictions of the ISM near the Solar System that can be compared with the observational data. This chapter details the numerical setup used for this purpose, the CR species computed, the physical processes, and the general properties of the CR spectra.

\subsection{PIERNIK MHD code}

PIERNIK is a grid-based multifluid MHD code, using conservative numerical schemes, available from the public repository at: \url{https://github.com/AntoineBaldacchino/piernik}.
The functionality of PIERNIK includes the modeling of multiple fluids: gas,  dust, magnetic field, CRs, and their mutual interactions. PIERNIK is parallelized on the basis of the MPI  library, and its dataIO communication utilizes parallel HDF5 output.
The MHD algorithm is based on the standard set of resistive MHD equations. PIERNIK code is equipped with the HLLD Riemann solver \cite{2005JCoPh.208..315M}  combined with the divergence cleaning \cite{2002JCoPh.175..645D} algorithm introduced in  \cite{2023mnras.522.5529p}, together with thermal cooling using the exact integration scheme by \cite{2009ApJS..181..391T}, and the star formation feedback algorithm based on \cite{2013ApJ...770...25A}. The code also includes the Adaptive Mesh Refinement (AMR) technique, multigrid (MG) Poisson solver, multigrid diffusion solver, and an N-body particle-mesh solver for a large number of point masses representing stellar and dark matter components of galaxies.

The CR propagation is dynamically coupled to the MHD system. CRs can be included in numerical simulations as single-bin (spectrally unresolved), diffusive relativistic fluids, as e.g. in \cite{2023mnras.522.5529p}, or spectrally-resolved, multiple-bin systems processed with the CRESP algorithm. A combination of spectrally unresolved protons with other spectrally-resolved components (electrons, heavier primary and secondary CR nuclei) is also available.

\subsection{3D Gravity-stratified box setup}

\begin{table*}[ht]
\centering
\begin{tabular}{llllllll}
\hline
Species    &  A & Z & spectral & Primary/Secondary & Lifetime & SN abundance ($N_i/N_p$) \\ \hline
Proton &  $1$                                                                                     & $1$                                                                               & No    & Primary & Stable    & $1$                                                                          \\
$^7\mathrm{Li}$  & $7$                                                                                     & $3$                                                                               & Yes    & Secondary & Stable   & None                                                                            \\
$^9\mathrm{Be}$  & $9$                                                                                     & $4$                                                                               & Yes   &      Secondary &  Stable      & None                                                                  \\
$^{10}\mathrm{Be}$ &  $10$                                                                                     & $4$                                                                               & Yes   & Secondary & $1.6\; \mathrm{Myr}$ & None                                                                               \\
$^{10}\mathrm{B}$  & $10$                                                                                     & $5$                                                                               & Yes    & Secondary & Stable    & None                                                                           \\
$^{11}\mathrm{B}$  & $11$                                                                                     & $5$                                                                               & Yes   & Secondary & Stable    & None                                                                            \\
$^{12}\mathrm{C}$ &  $12$                                                                                     & $6$                                                                               & Yes   & Primary & Stable  & $4.5 \times 10^{-3}$ \\
$^{14}\mathrm{N}$  & $14$                                                                                     & $7$                                                                               & Yes    & Primary & Stable   & $1.0 \times 10^{-3}$                                                                             \\
$^{16}\mathrm{O}$  & $16$                                                                                     & $8$                                                                               & Yes   & Primary & Stable    & $4.0 \times 10^{-3}$                                                                                \\

\end{tabular}

\caption{Cosmic Ray species included in the simulations and their different parameters: mass number $A$, charge number $Z$, lifetime (if radioactive), initial SN relative abundance to protons for primaries, which species are spectral or not, which are primaries or secondaries. }
\label{table_1}
\end{table*}
To demonstrate the capabilities of CRESP in PIERNIK code in simulations of multiple hadronic spectral CR components, we construct a test model for local simulations of galactic ISM. We assume a local rectangular patch of the ISM in initial hydrostatic equilibrium of dimension $L_x \times L_y \times  L_z = 0.5 \;\mathrm{kpc} \times 1 \;\mathrm{kpc} \times 8 \;\mathrm{kpc} $ in $x$, $y$, $z$ directions, with minimum cell size $\Delta x=\Delta y=\Delta z= 62.5\;\mathrm{pc}$. The box is stratified by vertical gravity. Following \cite{1998ApJ...497..759F}, we choose physical conditions typical for the neighborhood of the Solar orbit in the Galaxy, with the thermal gas density $ n_{\mathrm{ISM}}\approx 1 \;\mathrm{cm}^{-3}$ and temperature $T \approx 7000 \;\mathrm{K}$ at the galactic mid-plane. We assume an initial magnetic field of $3 \;\mathrm{\mu G}$ along the $y$ axis. Periodic boundary conditions are chosen for the $x$-axis and $y$-axis and outer for the $z$-axis, meaning that CR flux and other fluids are set to zero at the vertical edges.

Having defined the initial equilibrium state, we inject primary CRs in randomly located supernovae \citep{2004ApJ...605L..33H}. We assume that each supernova remnant provides CRs with 10\% of $E_{\mathrm{SN}}=10^{51} \mathrm{erg}$ energy in the proton component. We inject spectrally unresolved proton energy density coupled to the magnetized thermal gas of the ISM through the CR pressure gradient and driving the ISM. We treat CR protons as a relativistic fluid in a single-bin (grey) approximation. Adiabatic energy losses are included for physical consistency, while we omit hadronic (pionic) losses, as this work does not aim to model proton spectra or gamma-ray emissivities. In the present model, CR protons serve as a driver of the ISM dynamics. We will address the impact of hadronic losses on proton distributions, together with gamma-ray observables, in future work. Table (\ref{table_1}) lists all the CR species injected in the simulations. There is three primary nuclei components and their relative abundances. We assume, for simplicity, that no thermal or kinetic energy is input from supernovae. The spallation process of the primaries with thermal gas injects secondary CRs.

\subsection{Primary/secondary spectrum properties and initial/boundary conditions}
\label{sect:spectrum_properties}

\begin{figure}[!htb]
\includegraphics[scale=0.55]{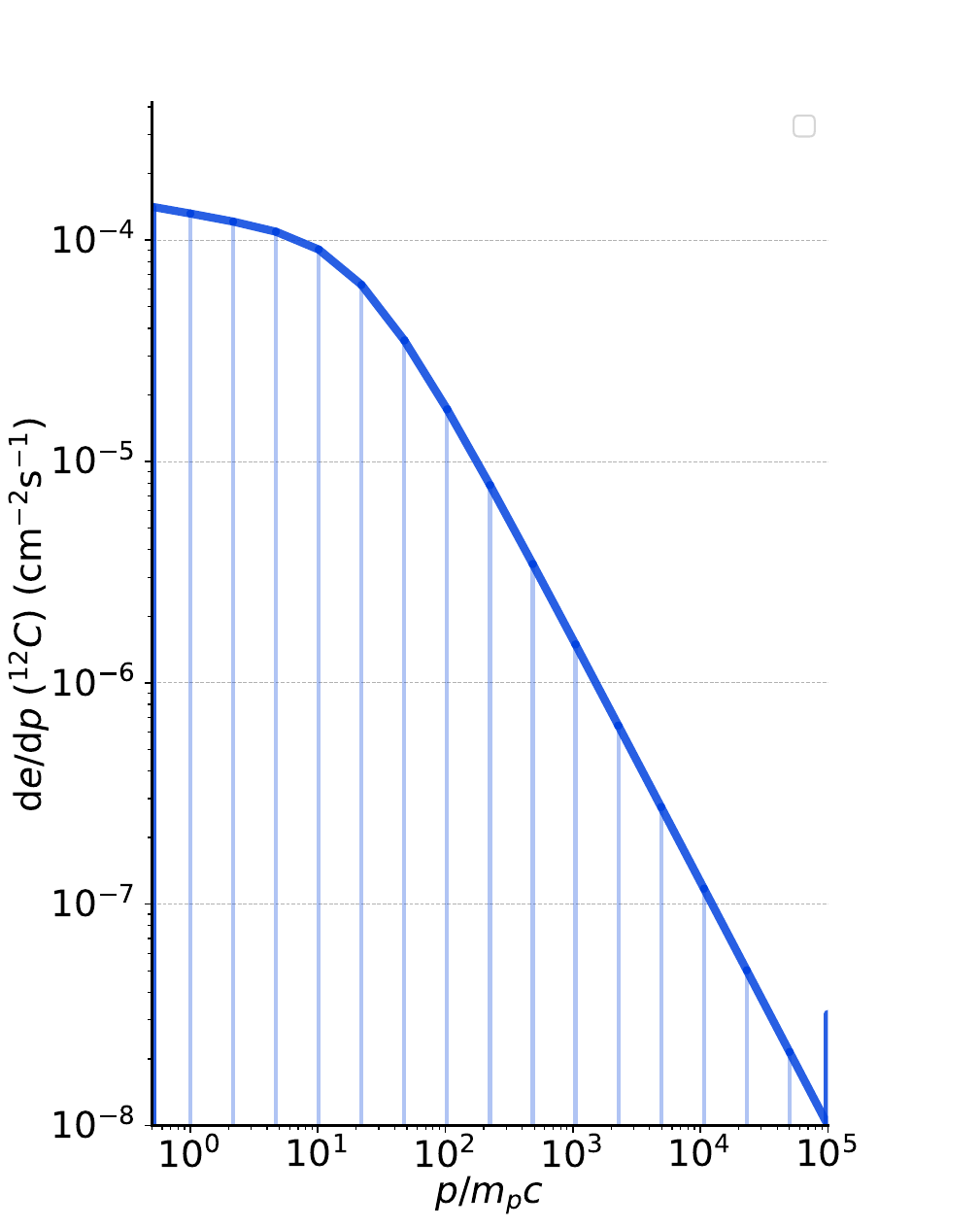}
\caption{Example of \Ct injection differential kinetic energy density spectrum $\mathrm{d}e/\mathrm{d}p$ at $t=0$ from a one cell simulation test. In the first and fourth bins (non-relativistic and trans-relativistic), the spectrum follows $p^{2 + s - q}\approx p^{-2.1}$. Above $p/m_pc=A_C$ ($A_C=12$ for Carbon), the ultra-relativistic bins exhibit a flatter power law $p^{-1.1}$.
}
\label{fig.1}
\end{figure}

The spectrally resolved CRs are energetically distributed in 16 bins in the range $p = (5\times10^{-1}-10^5)m_p c $ for every species, where $m_p$ is the proton mass. The left boundary value of the momentum is chosen in the non-relativistic regime, while applying minimal substeping constraints for Coulomb energy loss (see Appendix \ref{App.B} for details of Coulomb cooling algorithm) to prevent the simulation runs to slow down. The left and right edge bins are chosen wider, each covering half of a full momentum decade, acting like a particle reservoir providing sufficient CRs. Momentum boundary conditions consist of a power-law cutoff as described in \cite{2021ApJS..253...18O}, but, unlike CR electrons, the momentum boundary values are fixed for the whole simulations. The minimum numerical value for $n^a_l$ and $e^a_l$ is set to $10^{-15}$ in the normalized units presented in Appendix (\ref{App.A2}), for each bin $l$. For the present simulations, we choose the initial spectral index to be $q = 4.1$ for all bins, including cutoff bins, and all species. All those features are present in Figure (\ref{fig.1}).

We set a fixed value of the initial slope, $q=4.1$. This choice is motivated by diffusive shock acceleration theory \citep{2020LRCA....6....1M} in SN remnants. In the discussion on the spectra in section (\ref{sect:cr_propagation}), we focus on the slope deviation from the SN injection spectrum, and between primaries and secondaries, to demonstrate the self-consistency of the spectral CR transport.

Random SNe inject primary CRs, while secondary CRs are produced by spallation. The secondary injection spectrum involves all reactions with primary species and momentum-dependent reaction rates. Following relations (\ref{eq.10}) and (\ref{eq.11}), the secondary number and energy density injection spectrum for a single reaction $\mathrm{N}_j + (\mathrm{p},\mathrm{He})_i \rightarrow \mathrm{N}'_a$ read:
\begin{equation}
\label{eq.23}
     Q^a_{ij} = \Gamma^a_{ij}(p)n_j(p) \propto (\Gamma^a_{ij})^{(0)} \beta_j(p)p^{3-q_j}
\end{equation}
\begin{equation}
\label{eq.24}
     S^a_{ij} =  \frac{A_a}{A_j} \Gamma^a_{ij}(p)e_j(p) \propto (\Gamma^a_{ij})^{(0)} \beta_j(p)T_j(p)p^{3-q_j}
\end{equation}
where $\Gamma^a_{ij}=(\Gamma^a_{ij})^{(0)} \beta_j(p) = n_i\sigma^a_{ij}c\beta_j(p)$ is the reaction rate, and $\sigma^a_{ij}$ cross sections are constant for all spallation reactions. This choice of constant, non-energy dependent cross sections limits the quantitative accuracy at non-relativistic energies below $1\,\GeV\,n^{-1}$, but does not affect the qualitative trends and correlations of secondary-to-primary ratios explored in the latest sections. Our analysis in the following sections considers only primary and secondary particles, and we do not consider tertiary products in this work.

\begin{figure}[ht]
    \centerline{
    \includegraphics[width=1.\linewidth]{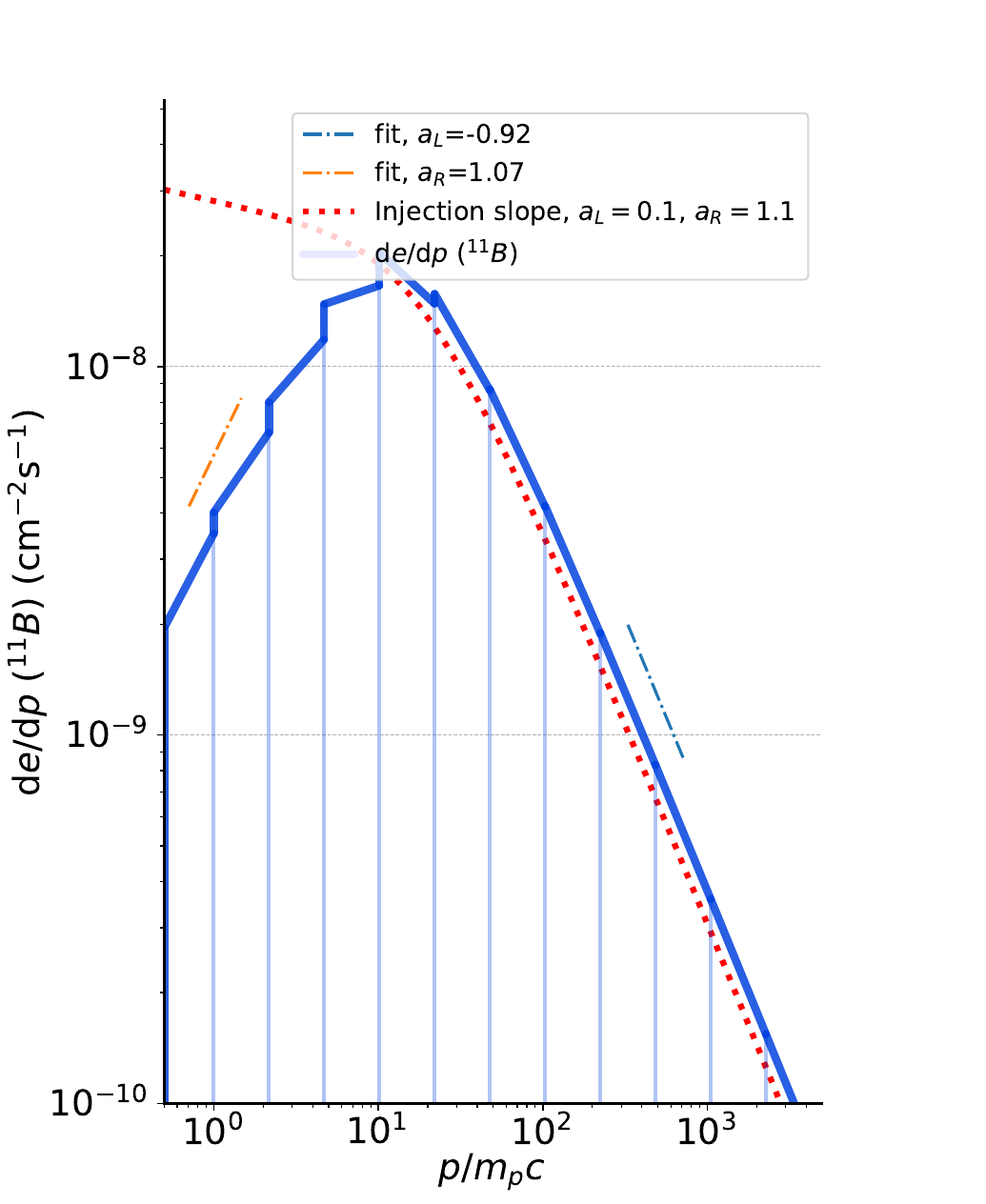}}
    \caption{CR differential kinetic energy density spectrum for \Ct after one step in a test simulation with primary \Ct and secondary \Bel only. The red curve is the primary injection spectrum of \Ct (the amplitude is changed to compare both curves), as seen in Figure (\ref{fig.1}). We observe the differences between primaries and secondaries below $p/m_p c\leq A_C$, showing how the momentum-dependent spallation rate changes the slope. A power-law fitting method applied to a few bins estimates the slope of the \Bel spectrum (blue curve). The corresponding curves and slopes are displayed. As expected, the slope of \Bel and \Ct differ by one for $p/m_p c \leq A_C$ and are equal for $p/m_p c \geq A_C$. }
    \label{fig.2}
\end{figure}

All the reaction channels and corresponding cross section values are displayed in Table (\ref{table_2}). For $p\geq m_jc$, $\beta_j\approx 1$, but for $p\leq m_jc$, $\beta_j \approx p/(m_jc)$ and the injection slope changes. We expect that the low-energy part of the secondary spectrum differs from the primaries. At non-relativistic energies, the velocity of particles is much less than the speed of light $c$, and collisions are sporadic. This phenomenon is shown in Figure (\ref{fig.2}): the comparison between the Carbon injection spectrum and Boron spectrum after a one-time step highlights how the left part of the spectrum for $p\leq A_Cm_pc$ is modified. A power-law fitting method estimates the slope of the secondary \Bel spectrum at non-relativistic energy $a_L$ and relativistic energy $a_R$. Those slopes adopt the same sign convention as $q$, implying that the displayed spectra scale as $p^{-a_L}$ below $10\,\GeV\,c^{-1}$, and $p^{-a_R}$ above $10\,\GeV\,c^{-1}$. Its estimation gives $a_R$(\Ct)$\approx a_R$(\Bel)$ $, and $a_L$(\Bel)$\approx a_L$(\Ct)$-1$ within a few \% error estimation, which is satisfying and confirms the consistency of the method.

\begin{table}
\centering
\begin{tabular}{llllllll}
\hline
Channel    &  $\sigma^a_{ij}\;(\mathrm{mb})$ \\ \hline
\Ct $+ \mathrm{p} \rightarrow $\Lis   &  $6.8$                                                                                                                                                             \\
\Ct $+ \mathrm{p} \rightarrow $\Ben   &  $6.8$                                                                                                                                                                \\
\Ct $+ \mathrm{p} \rightarrow $\Bet   &  $4$                                                                                                                                                     \\
\Ct $+ \mathrm{p} \rightarrow $\Bt  &  $12.3$                                                                                                                                                          \\
\Ct $+ \mathrm{p} \rightarrow $\Bel   &  $30$
\\
\Nf $+ \mathrm{p} \rightarrow $\Lis   &  $9.3$                                                                                                                                                                 \\
\Nf $+ \mathrm{p} \rightarrow $\Ben   &  $2.1$
\\
\Nf $+ \mathrm{p} \rightarrow $\Bt   &  $10.3$
\\
\Nf $+ \mathrm{p} \rightarrow $\Bel   &  $17.3$
\\
\Os $+ \mathrm{p} \rightarrow $\Lis   &  $11.2$
\\
\Os $+ \mathrm{p} \rightarrow $\Ben   &  $3.7$
\\
\Os $+ \mathrm{p} \rightarrow $\Bet  &  $2.2$
\\
\Os $+ \mathrm{p} \rightarrow $\Bt   &  $10.9$
\\
\Os $+ \mathrm{p} \rightarrow $\Bel   &  $18.2$
\\

\end{tabular}
\caption{Table presenting all the spallation channels in our model with the corresponding cross sections $\sigma^a_{ij}$ in $\mathrm{mb}$. The values are taken from \cite{2018PhRvC..98c4611G}.}
\label{table_2}
\end{table}

\subsection{CR transport}
For CR transport, we take into account two processes: advection with thermal gas represented by the term $\vec{v}\cdot \nabla f^a$ in the Fokker-Planck Equation (\ref{eq.1}), and energy-dependent,  magnetic field-aligned anisotropic diffusion. We compute the diffusion of nuclei along magnetic field  lines with their rigidity $R_\mathrm{N}=cp/Ze\; (\mathrm{GV})$ scaled with the fixed rigidity of protons $R^0_\mathrm{p} = cp^0/e = 10 \; \mathrm{GV}$:
\begin{eqnarray}
\label{eq.25}
    D_\parallel(R_\mathrm{N}) &=& D_\parallel^0\beta(p) \left(\frac{R_\mathrm{N}}{R^0_\mathrm{p}}\right)^\delta \nonumber \\ &=&D_\parallel^0\frac{v(p)}{c}\left(\frac{p}{Z p^0}\right)^\delta
\end{eqnarray}
where $D_\parallel^0 = 3\times10^{28} \mathrm{cm}^2 \mathrm{s}^{-1}$ at $p^0 =  10 \; \mathrm{GeV}\,c^{-1}$, $\delta$ is a parameter we can modulate. We also include a diffusion perpendicular to the magnetic field lines, $D_\perp(R_\mathrm{N}) = 1 \%$ of $D_\parallel(R_\mathrm{N})$ \citep{2007ARNPS..57..285S}.
For each momentum bin $l$ of range $[p_{l-1/2},p_{l+1/2}]$, the rigidity-dependent diffusion coefficient is computed with the centered value $p_{\mathrm{mid}} = \sqrt{(p_{l-1/2})(p_{l+1/2})}$ so that $D_{\parallel,l}\propto\beta(p_{l,\mathrm{mid}})(p_{l,\mathrm{mid}})^\delta $.
\section{Dynamical evolution of the system} \label{sect:dyn_evol}
In this section, we describe the dynamical evolution of the different components of the stratified boxes: ISM gas, magnetic field, CR protons, and the different spectrally resolved CR nuclei. We describe in detail how the live MHD environment acts on spectrally-resolved nuclei propagation in the stratified box. For this purpose, we investigate the spectra of different CR nuclei at $t= 500\;\mathrm{Myr}$ evolution.
\begin{table*}
\centering
\begin{tabular}{llllllll}
\hline
Model    &  SN rate ($\mathrm{kpc}^{-2}\mathrm{Myr}^{-1}$) & $D^0_{\parallel}$ $(\mathrm{cm}^2 \mathrm{s}^{-1})$ & $\delta$ & $L_x \times L_y \times L_z$ ($\mathrm{kpc}^3$) \\ \hline

A1 & $80$                                                                                     & $3\times 10^{28}$                                                                               & $0.3$    & $0.5 \times 1 \times 8$                                                                                \\
A2  & $60$                                                                                     & $3\times 10^{28}$                                                                               & $0.3$    & $0.5 \times 1 \times 8$                                                                                \\
A3 & $20$                                                                                     & $3\times 10^{28}$                                                                               & $0.3$    & $0.5 \times 1 \times 8$                                                                                \\
B1 &  $80$                                                                                     & $6\times 10^{28}$                                                                               & $0.3$    & $0.5 \times 1 \times 8$                                                                                \\
C1 & $80$                                                                                     & $3\times 10^{28}$                                                                               & $0.5$      & $0.5 \times 1 \times 8$                                                                                \\
\end{tabular}
\caption{The different simulation models tested with the relevant parameters for CR transport (from left to right): random SN rate, diffusion coefficient, diffusion power-law, and spatial dimensions of the 3D stratified box. A1 is the fiducial case, from which we change the different parameter values in the other simulations (see discussion in Section \ref{sect:BtoC}).}
\label{table_3}
\end{table*}

\begin{figure*}
    \centerline{
    \includegraphics[width=.5\linewidth]{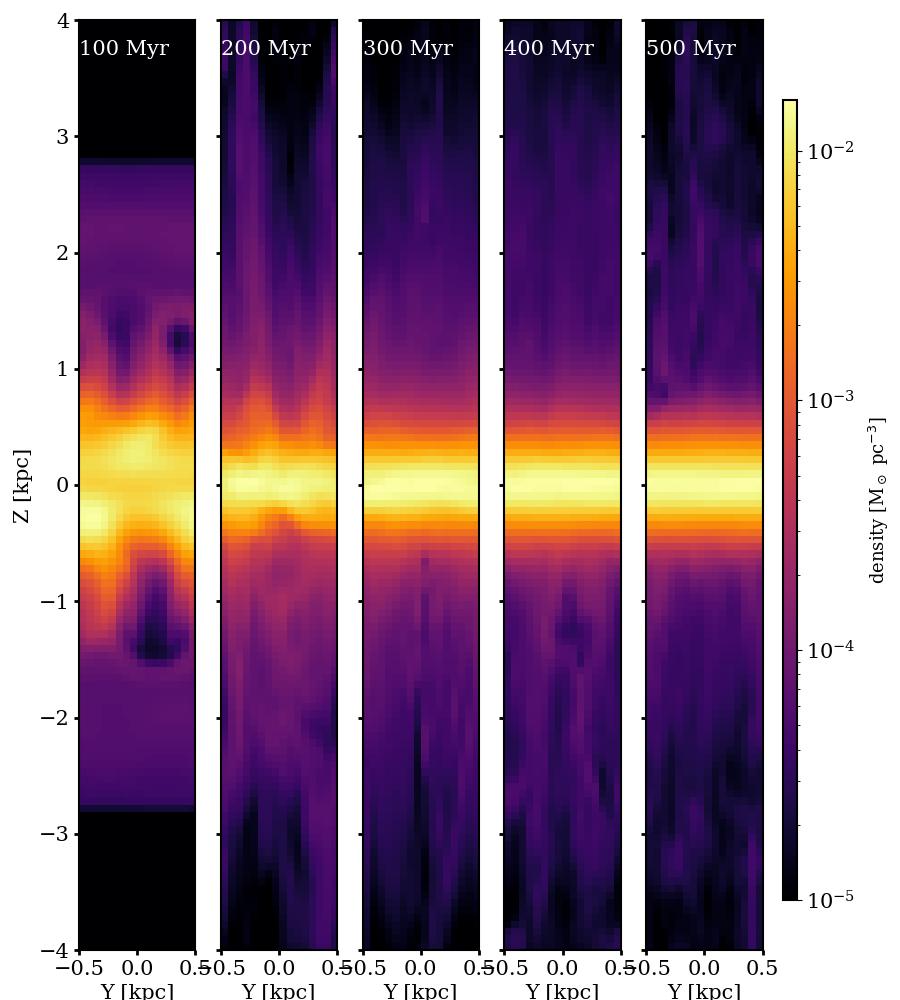}
    \includegraphics[width=.5\linewidth]{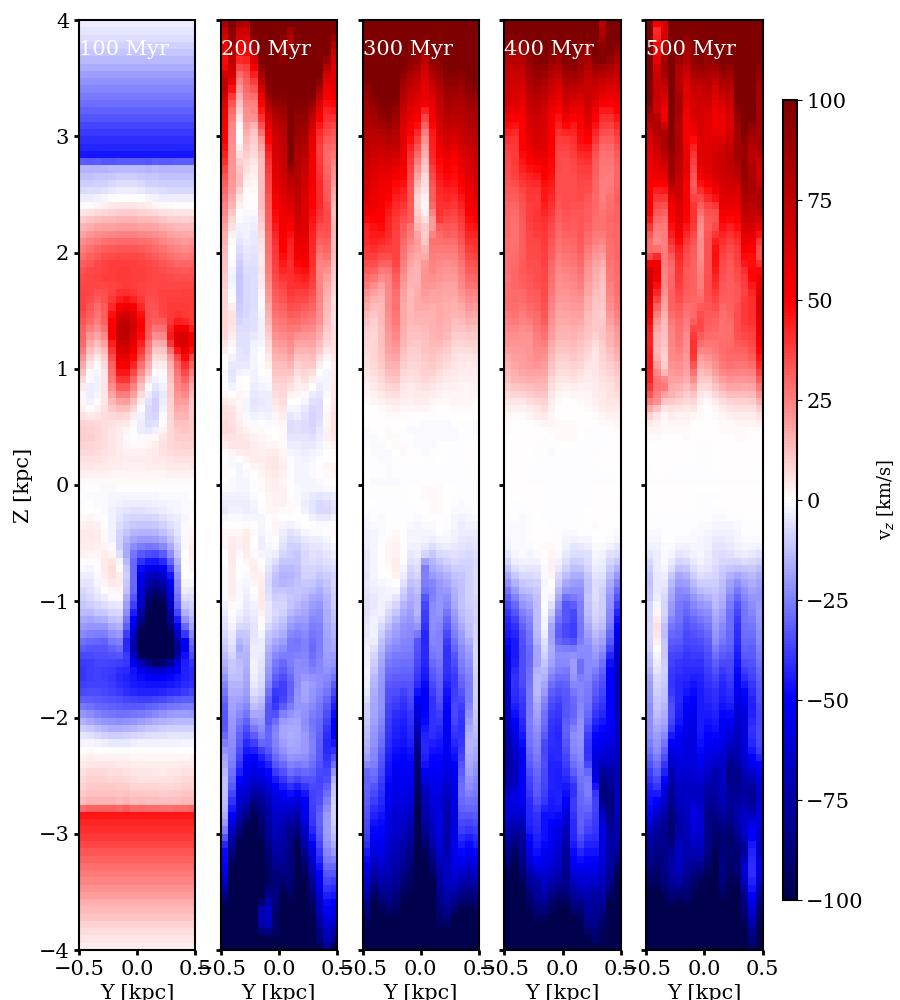}}
    \caption{Snapshots of the ISM density evolution (left) and vertical velocity evolution (right) in the stratified box over $500 \; \mathrm{Myr}$ in the $y-z$ plane at $x = 0 $ for the model case A1 with SN rate $=80\;\mathrm{kpc}^{-2}\mathrm{Myr}^{-1}$, $D_{\parallel}^0 = 3\times 10^{28} \mathrm{cm}^2\mathrm{s}^{-1}$. The ISM presents CR-driven buoyancy structures, generating outflows in the vertical direction. The vertical velocity snapshots indicate the presence of outflows, meaning that advection for CRs is present.}
    \label{fig.3}
\end{figure*}
\subsection{Evolution of ISM gas, magnetic fields and CR protons}\label{sect:ISM_evolution}
\begin{figure}[ht]
    \centerline{
    \includegraphics[width=.9\linewidth]{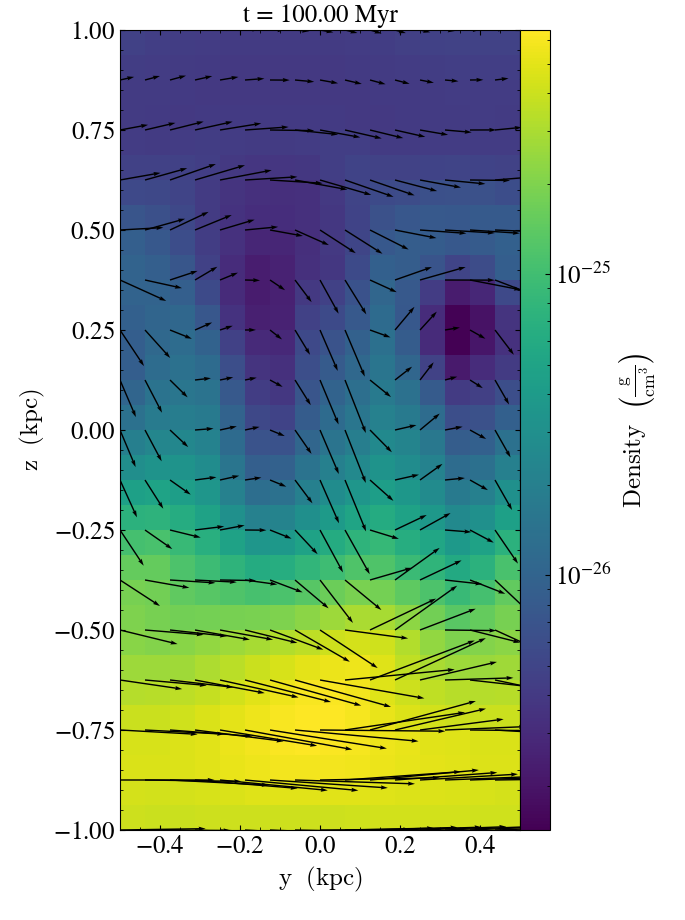}}
    \centerline{
    \includegraphics[width=.9\linewidth]{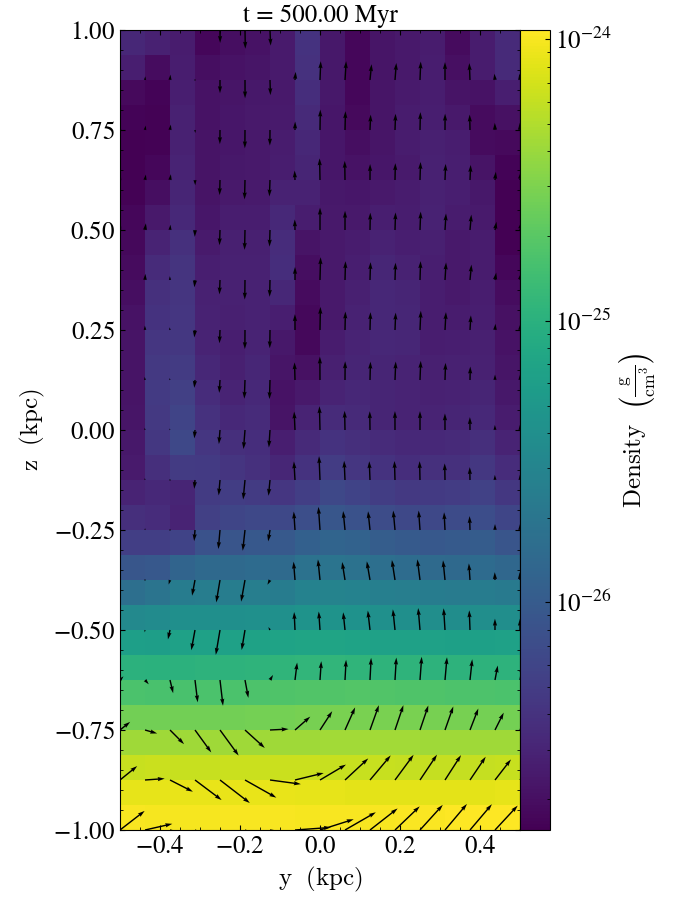}}
    \caption{Top: snapshot of the gas density within the section $z \in [0,2]\,\mathrm{kpc}$ for A1 run at $t=100\;\mathrm{Myr}$, where magnetic fields vectors are displayed. Bottom: same snapshot as top, but at $t=500\;\mathrm{Myr}$. 
    }
    \label{fig.4}
\end{figure}

We perform five simulations listed in Table \ref{table_3} assuming adiabatic thermal ISM gas, magnetic field, CR protons in the grey approximation, and selected spectrally-resolved CR nuclei. We let the simulations run for $500\; \mathrm{Myr}$ to let the system achieve a quasi-stationary state. Figure \ref{fig.3} displays a time sequence of snapshots showing the physical evolution of the ISM gas density and velocity in time intervals of $100 \Myr$ for the whole vertical extent of the computational domain. Figure \ref{fig.4} shows gas density and magnetic field structure in the region $z\in[0,2]\,\kpc$ for $t=100$ and $500\,\mathrm{Myr}$.

At  $t = 100\;\mathrm{Myr}$, we observe the effects of CR buoyancy,  leading to structures generated by Parker instability \citep{1966ApJ...145..811P}, apparent as the vertical undulation of the magnetic field vectors in the left panel of Figure~\ref{fig.4}. The effects of CR injection thicken the disc and initiate a vertical gas outflow \citep{2003A&A...412..331H}. At $200\;\mathrm{Myr}$, the buoyant loop-like structures extend in the vertical direction and form outflows reaching the computational domain's lower and upper boundaries. The disc achieves a more homogeneous form after $300\;\mathrm{Myr}$.
After $200\;\mathrm{Myr}$, the velocity near the outer domain boundaries ($z = \pm 4 \mathrm{kpc}$) reaches values of $\pm100\;\mathrm{km} \,\mathrm{s}^{-1}$ showing that the ISM gas is accelerated over the extent of the vertical box. The vertical gas velocity remains relatively small in the direct disk vicinity ($ z \in [-1 \mathrm{kpc}, 1\mathrm{kpc}]$).
CR protons drive gas outflows, and the vertical propagation of CRs is partially due to the advection with the vertically accelerated gas and partially due to diffusion along vertically stretched magnetic field lines. The vertical non-uniform outflows generating vertical magnetic fields of alternating orientation are pronounced at the end of the simulation.

\begin{figure*}
    \centerline{
    \includegraphics[width=.4\linewidth]{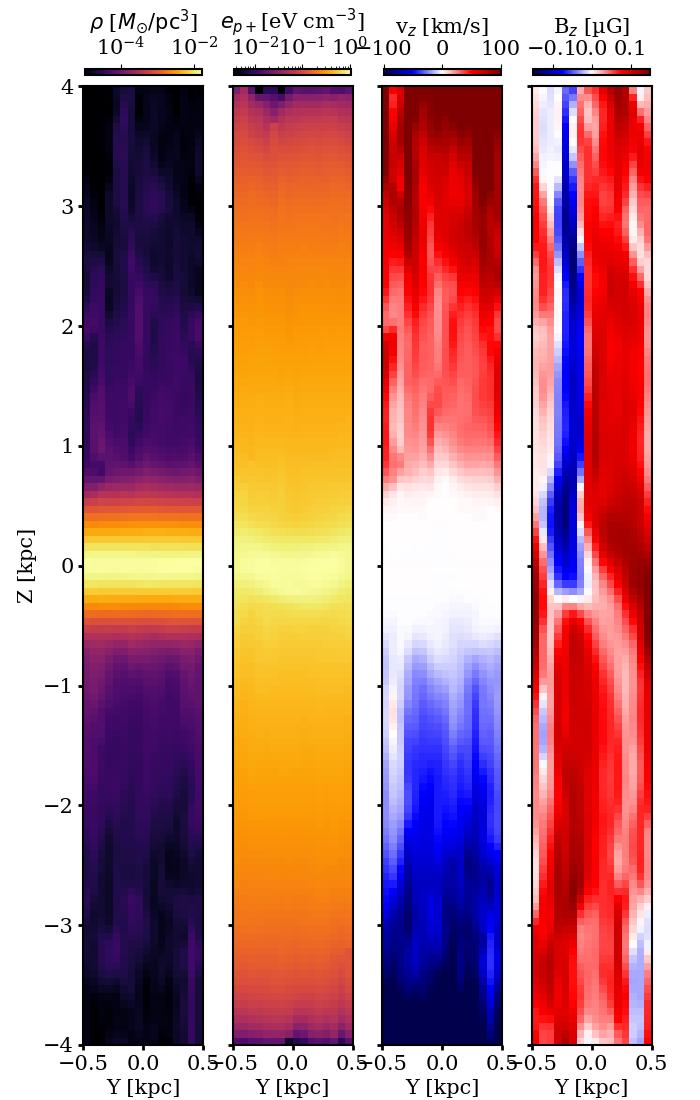}
    \includegraphics[width=.4\linewidth]{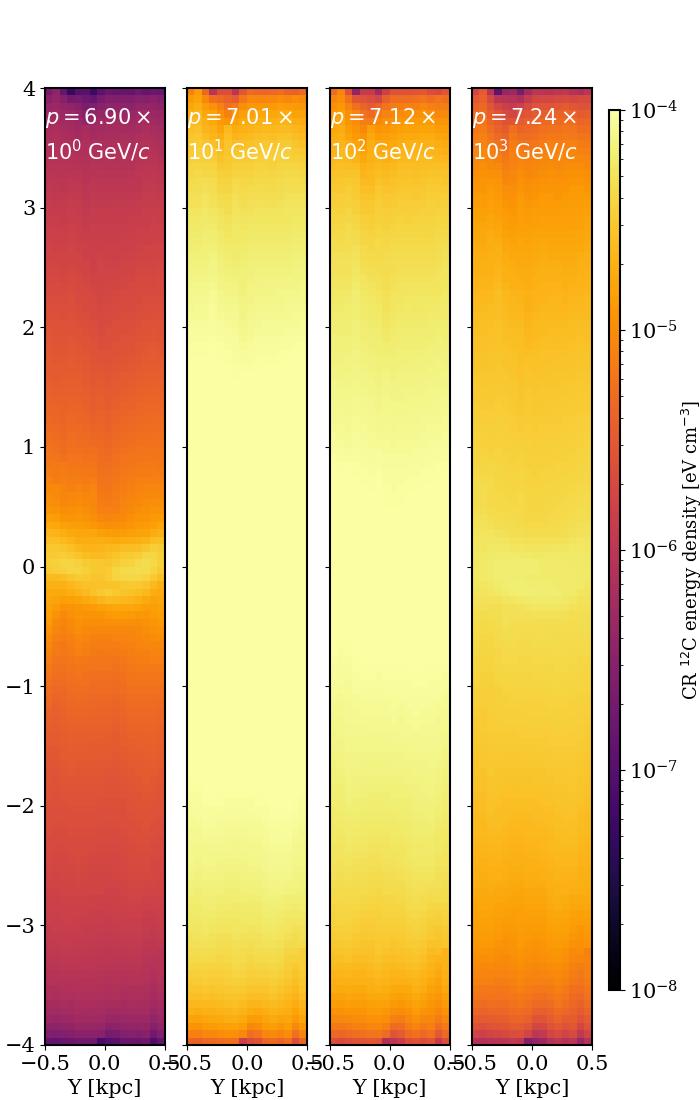}}
    \centerline{
    \includegraphics[width=.4\linewidth]{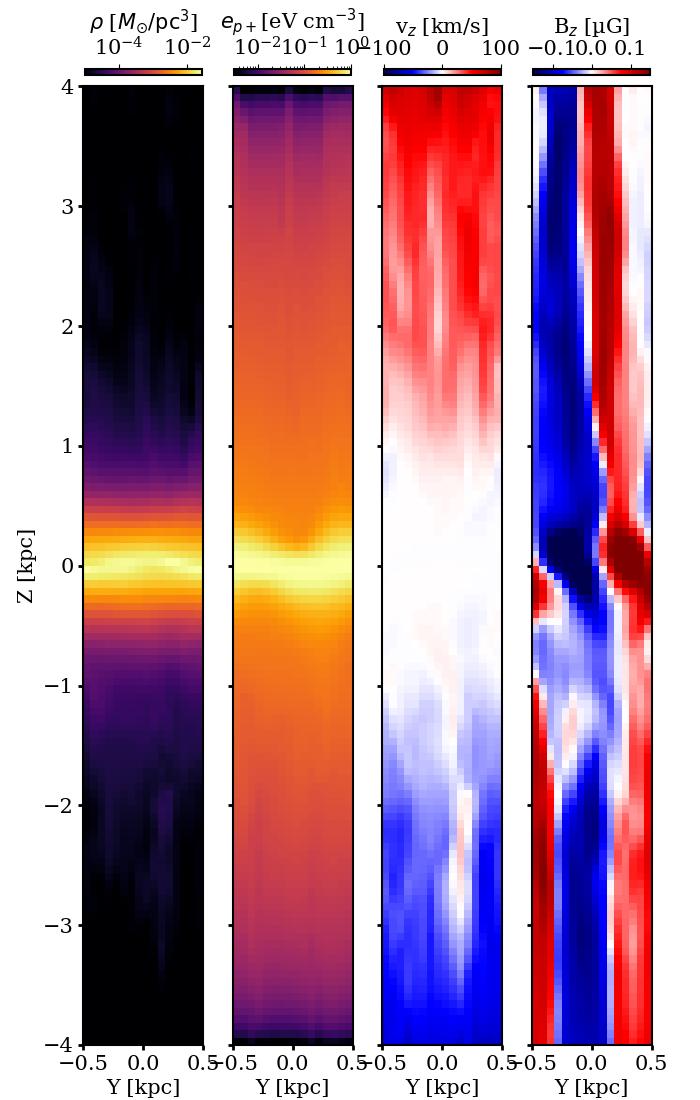}
    \includegraphics[width=.4\linewidth]{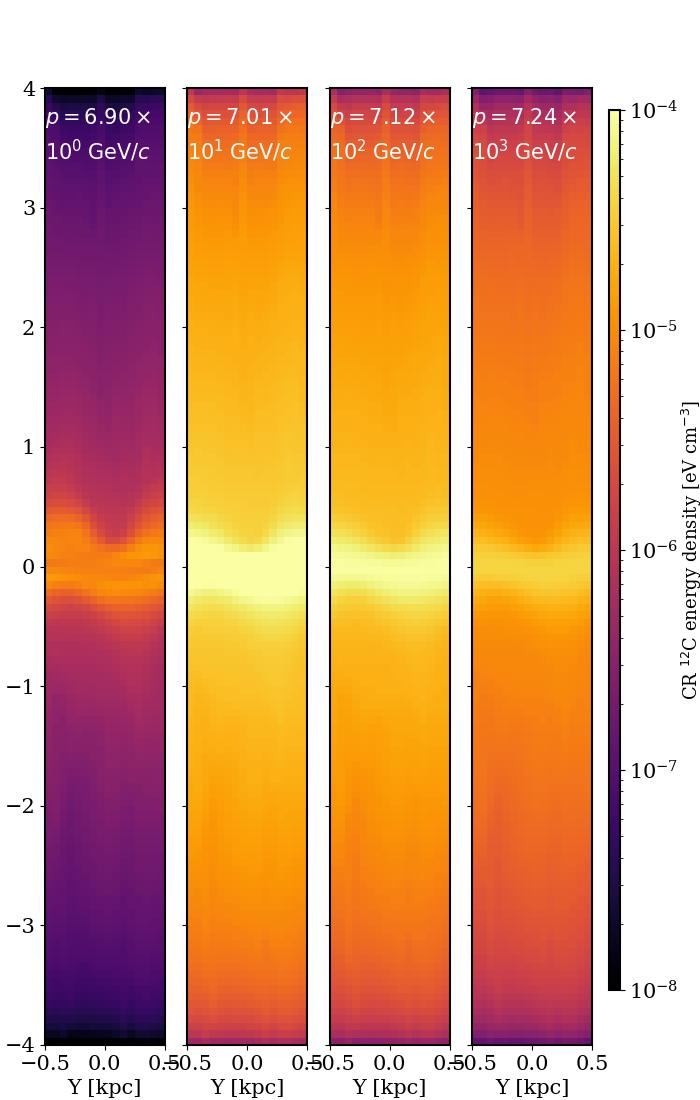}}
    \caption{Snapshots of the stratified box from A1 (top row) and A3 (bottom row) simulations at $t = 500\; \mathrm{Myr}$ with, from left to right: gas density, CR protons energy density, the vertical velocity of the gas, vertical magnetic field, and four CR \Ct bins at middle-valued momentum $p = 6.90\;\mathrm{GeV} \, \mathrm{c}^{-1}$, $p = 7.01\times10^{1}\;\mathrm{GeV} \, \mathrm{c}^{-1}$, $p = 7.12\times10^{2}\;\mathrm{GeV} \,\mathrm{c}^{-1}$ and $p = 7.24\times10^{3}\;\mathrm{GeV} \,\mathrm{c}^{-1}$. We observe a correlation between the magnetic field structure, outflows of gas, and outflows of CR protons and Carbon. Carbon at relativistic energies is more spread in the vertical direction, showing the action of rigidity-dependent diffusion.
    %\note[MH]{Remember to produce new plots for the new simulations for consistency! }
    }
    \label{fig.5}
\end{figure*}

In addition to Figures \ref{fig.3} and \ref{fig.4}, the gas dynamics is also shown in Figure \ref{fig.5}, where the following quantities are displayed in the snapshots at $t=500\;\mathrm{Myr}$: ISM gas density, velocity $v_z$, the vertical magnetic field  $B_z$, the CR proton energy density, and the spectrally resolved CR \Ct energy density in four different momentum bins. Those snapshots are from A1 and A3 models, which differ in the value of the SN rate (see Table \ref{table_3}). Both models present similarities in their evolution: the structure of $B_z$ corroborates with the alternating orientation of the magnetic field displayed in Figure \ref{fig.4}.

The CR proton energy density spreads all along the vertical direction, providing a non-thermal pressure gradient driving the outflows. However, for the A3 case, where the SN rate is much lower than for A1, $v_z$ has a lower maximal value, and the gas is more concentrated near the disk midplane. We then have a model that presents a thinner disk and weaker outflows. The CR proton energy density also has lower values all along the box due to lower SN injection and less efficient transport.  All those observations are expected since CR-driven gas outflows come from the coupling between the gas and CR pressure gradient. With a lower SN rate, the total CR pressure decreases, which injects less energy of the gas to create outflows and CR advection.

Although different parameters impact the dynamics of gas and CRs, the models listed in Table \ref{table_3} present a similar global behavior. With this simulation set, we can analyze in detail which transport modes rule the propagation of the spectral CRs and assess how much the spectrum and observable properties are sensitive to variations of the model parameters.

\subsection{Propagation of spectrally-resolved CR nuclei}\label{sect:cr_propagation}
\begin{figure*}

    \centerline{

    \includegraphics[width=0.5\linewidth]{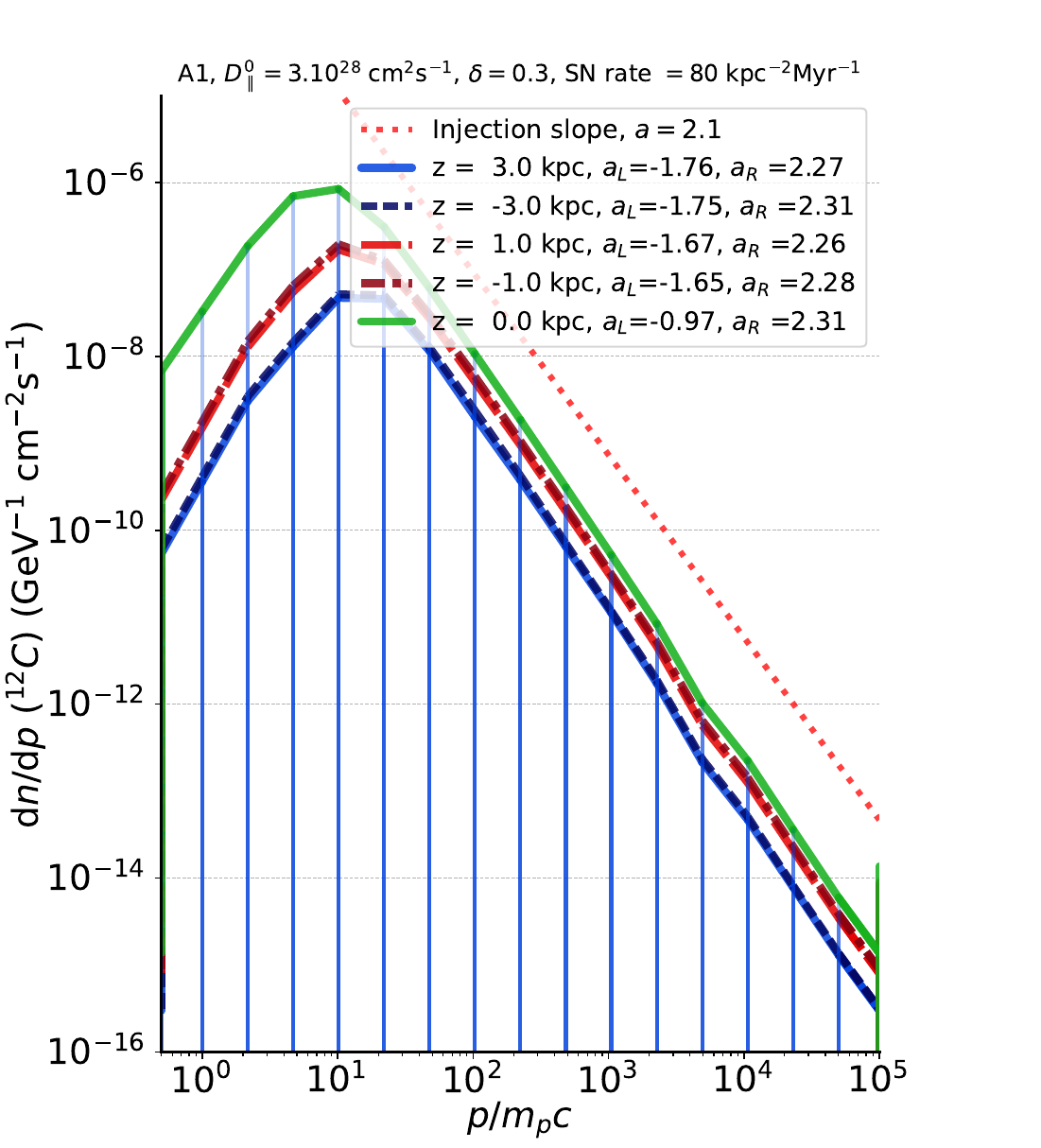}

    \includegraphics[width=0.5\linewidth]{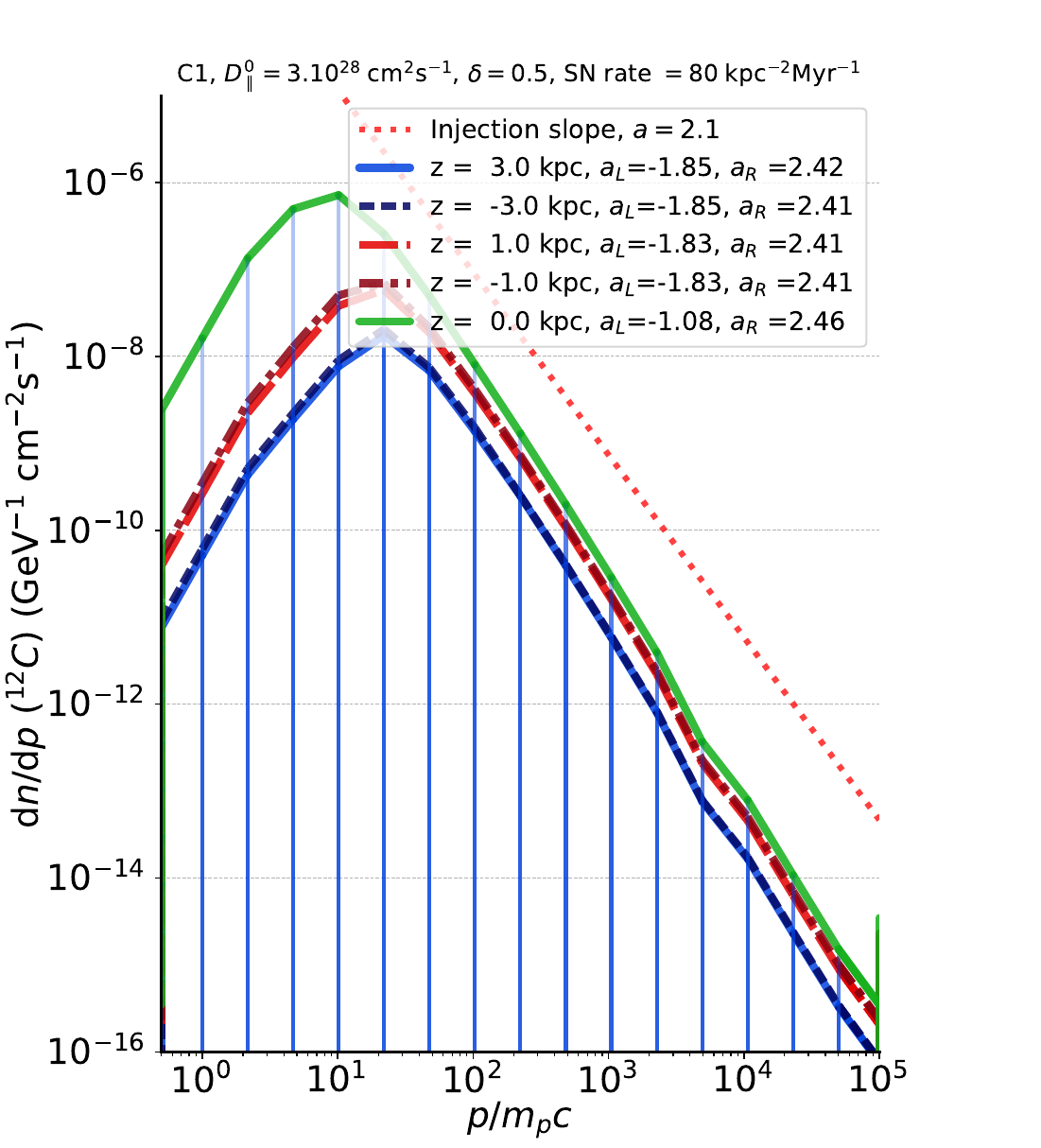}}
    \centerline{

    \includegraphics[width=0.5\linewidth]{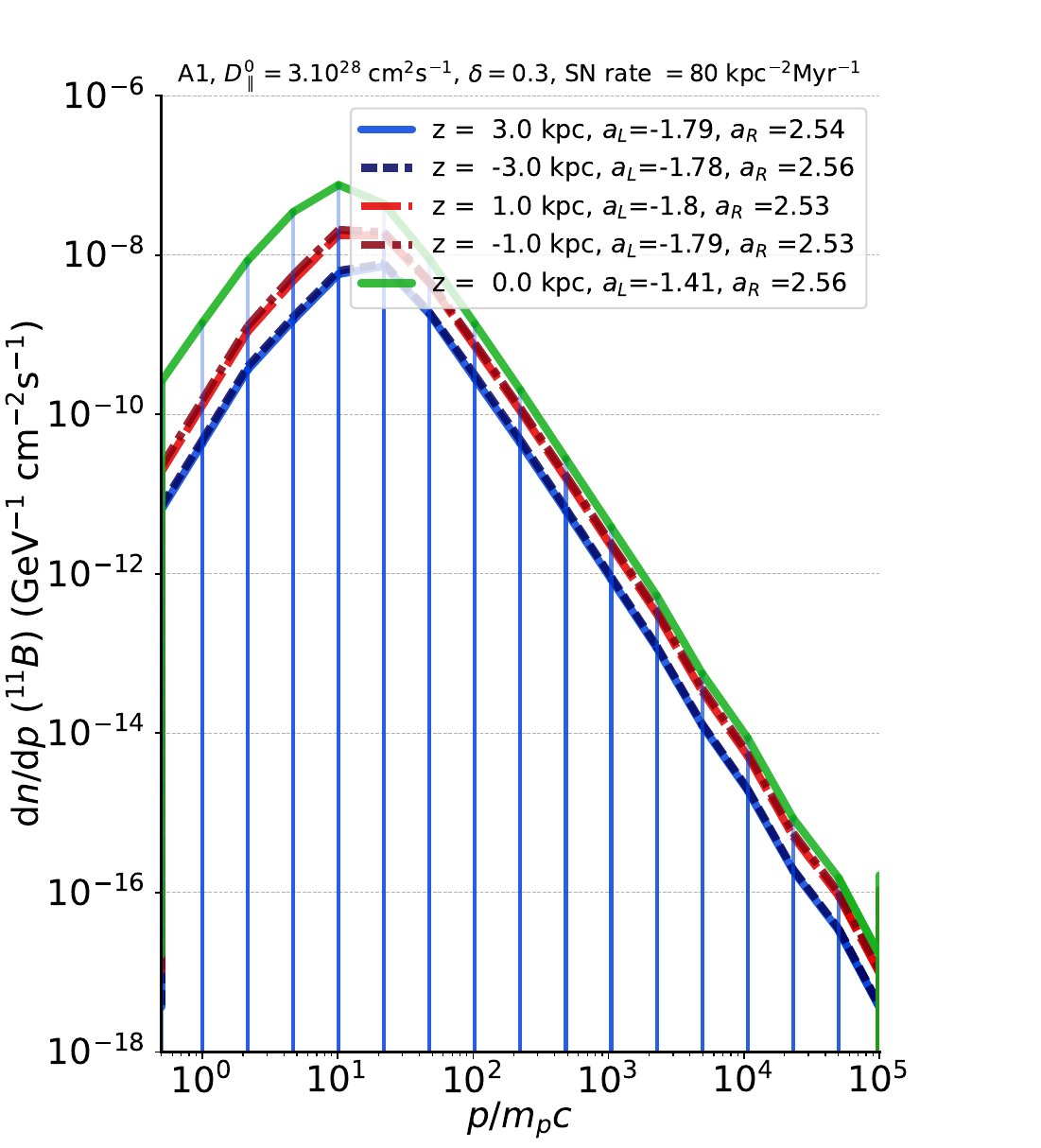}

    \includegraphics[width=0.5\linewidth]{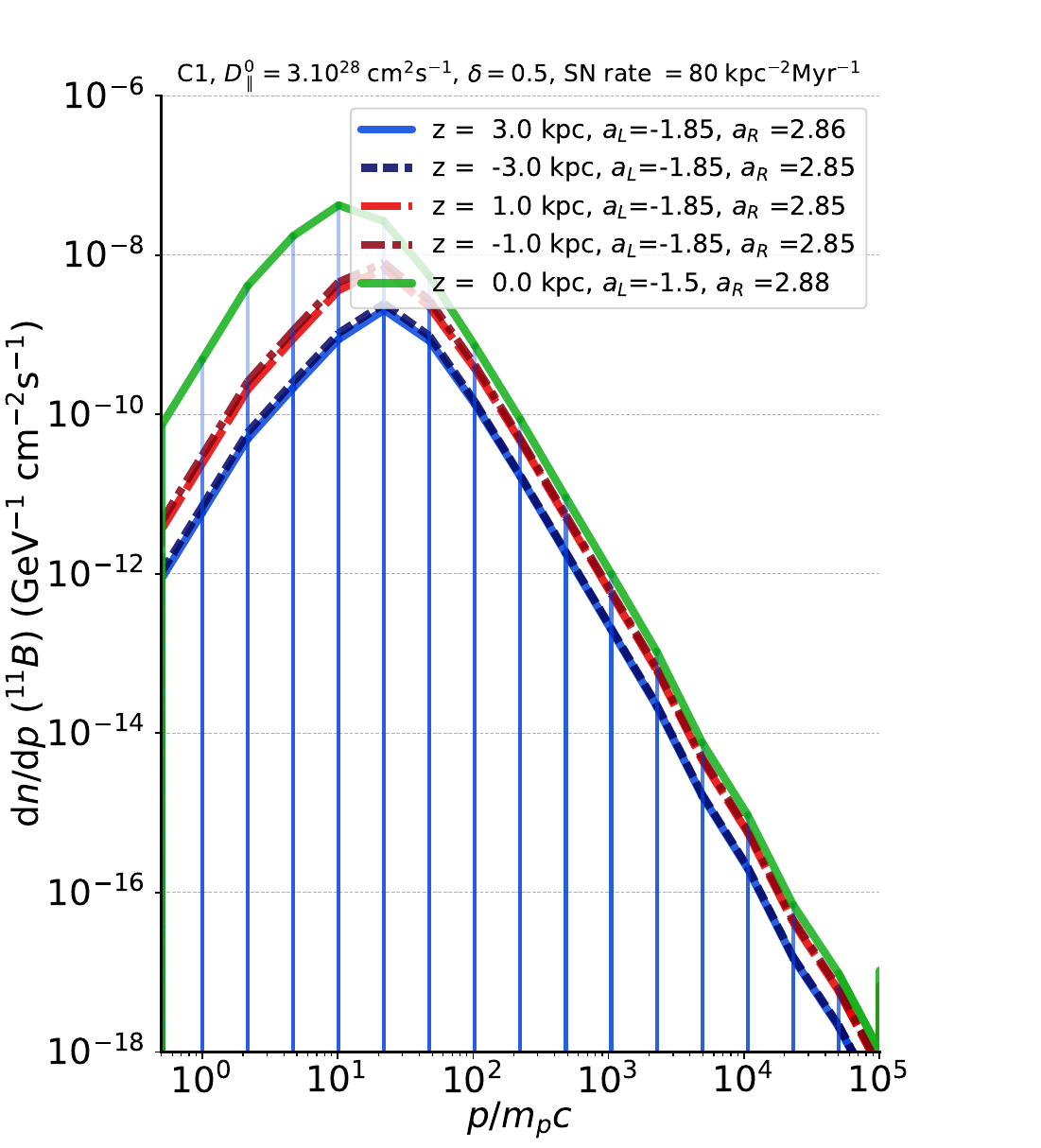}}
    \caption{CR differential number density spectrum $\mathrm{d}n/\mathrm{d}p$ for \Ct (top row) and \Bel isotope (bottom row) in model A1 (left) and model C1 (right) after $500\; \mathrm{Myr}$ evolution taken at five different altitudes points in the box (with $x=y=0$). The red dashed lines represent the initial slope $a$ of the primary injection spectrum. For each spectrum, a fit slope estimation is given, with $a_L$ the slope at non-relativistic energy and $a_R$ the slope at relativistic energy. The slope differs in those energy ranges because of the different energy-dependent spallation, Coulomb energy loss and diffusion processes.}
    \label{fig.6}
\end{figure*}

\begin{figure*}

    \centerline{

    \includegraphics[width=0.5\linewidth]{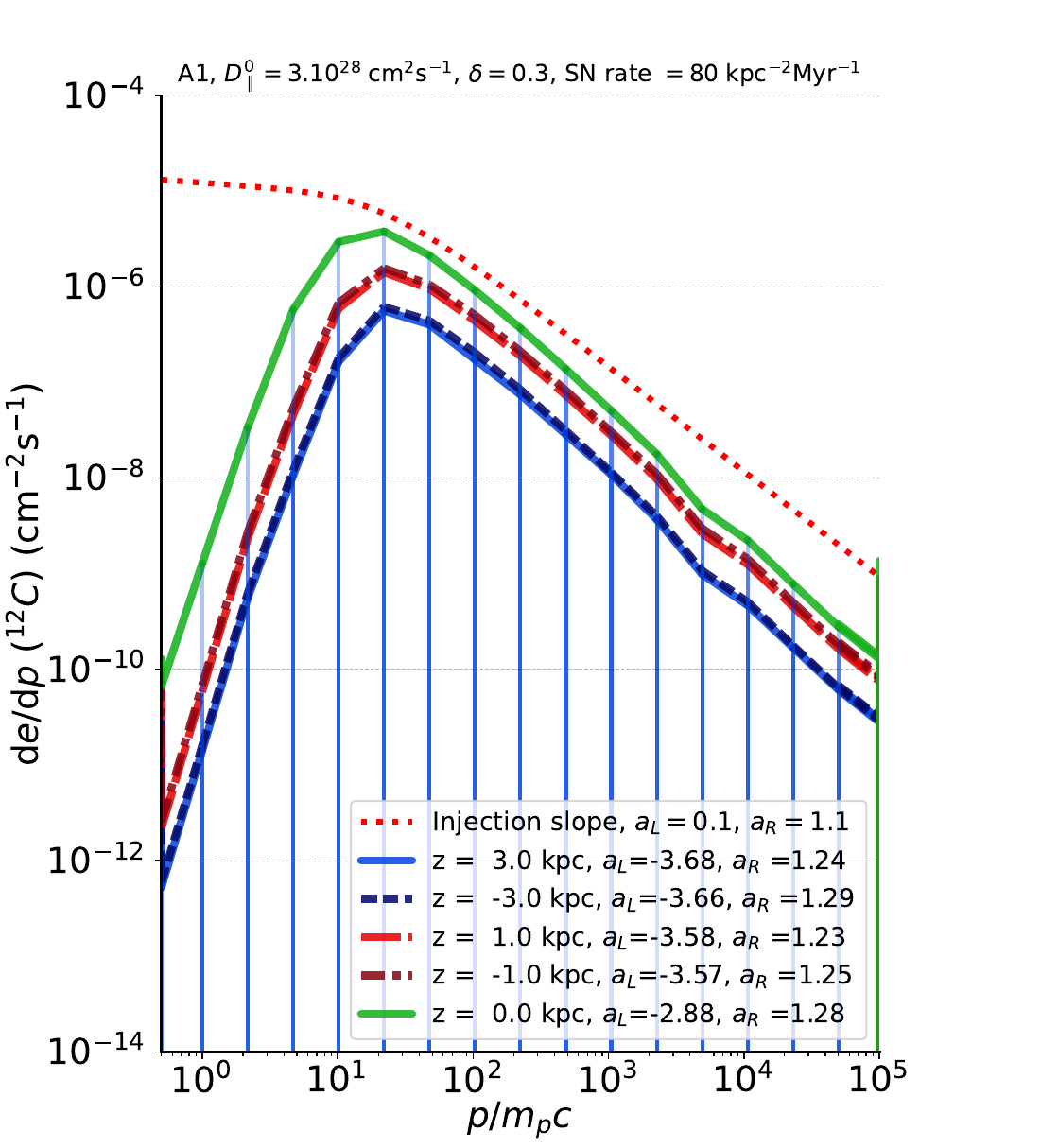}

    \includegraphics[width=0.51\linewidth]{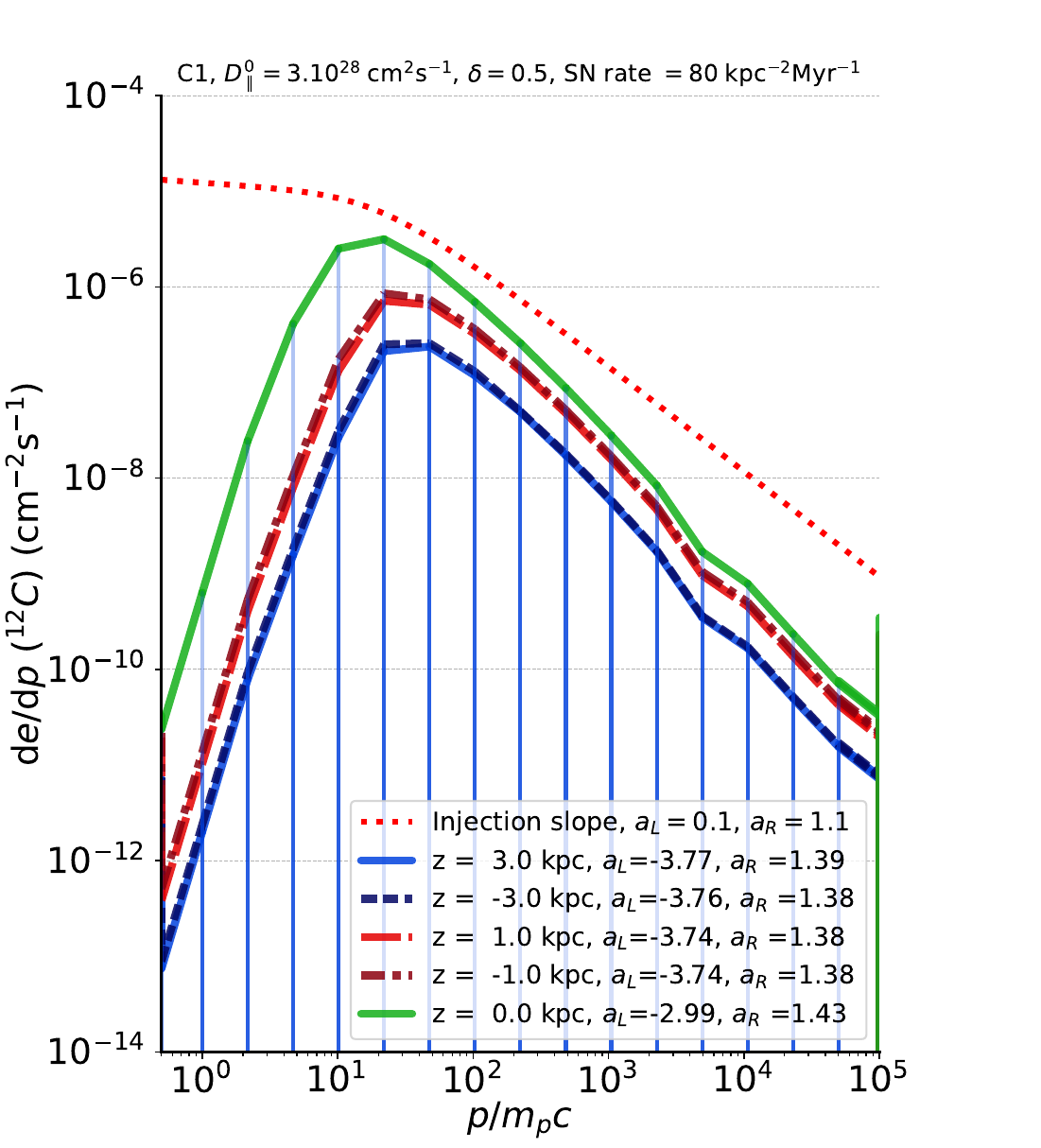}}
    \centerline{

    \includegraphics[width=0.5\linewidth]{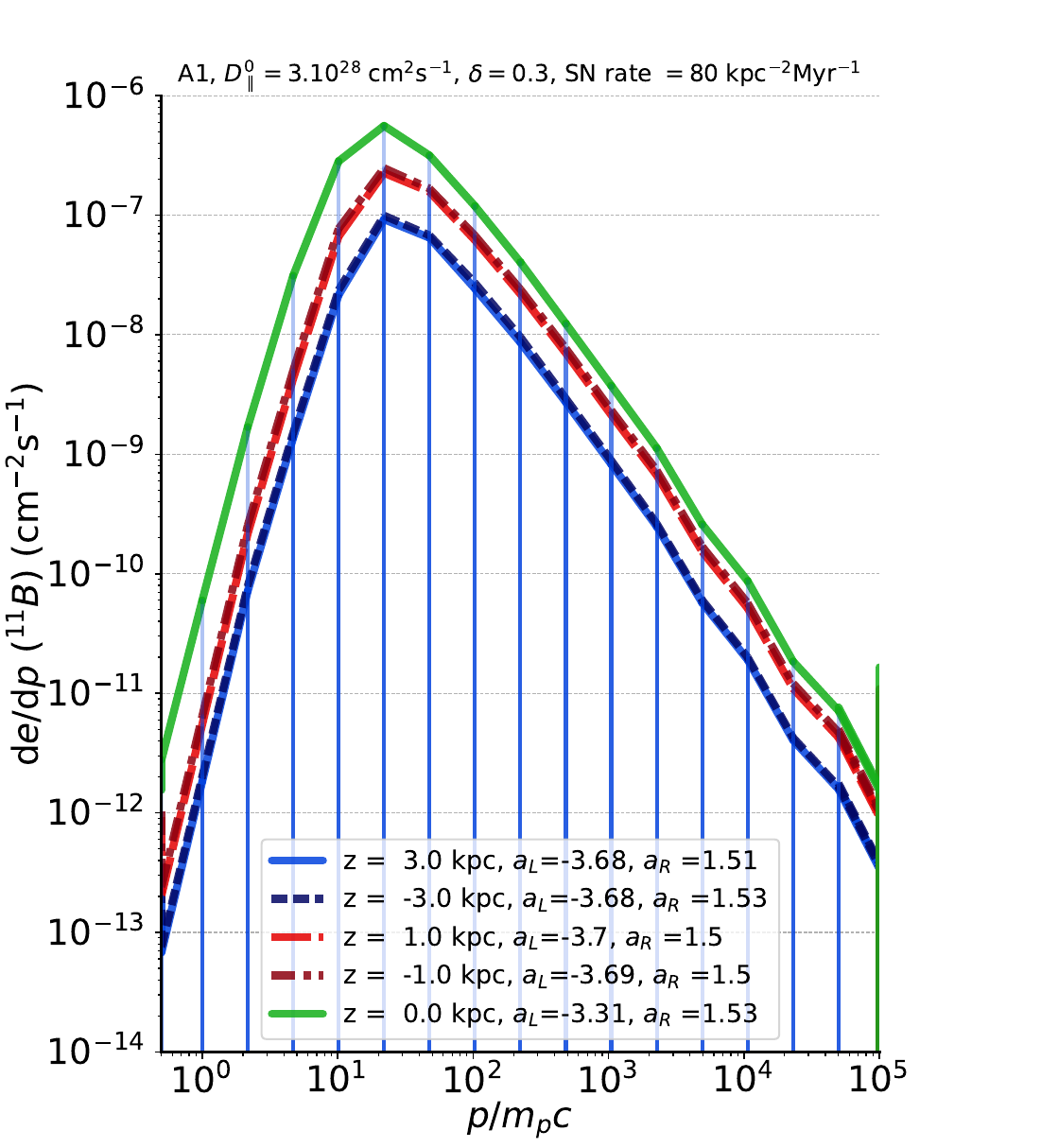}

    \includegraphics[width=0.51\linewidth]{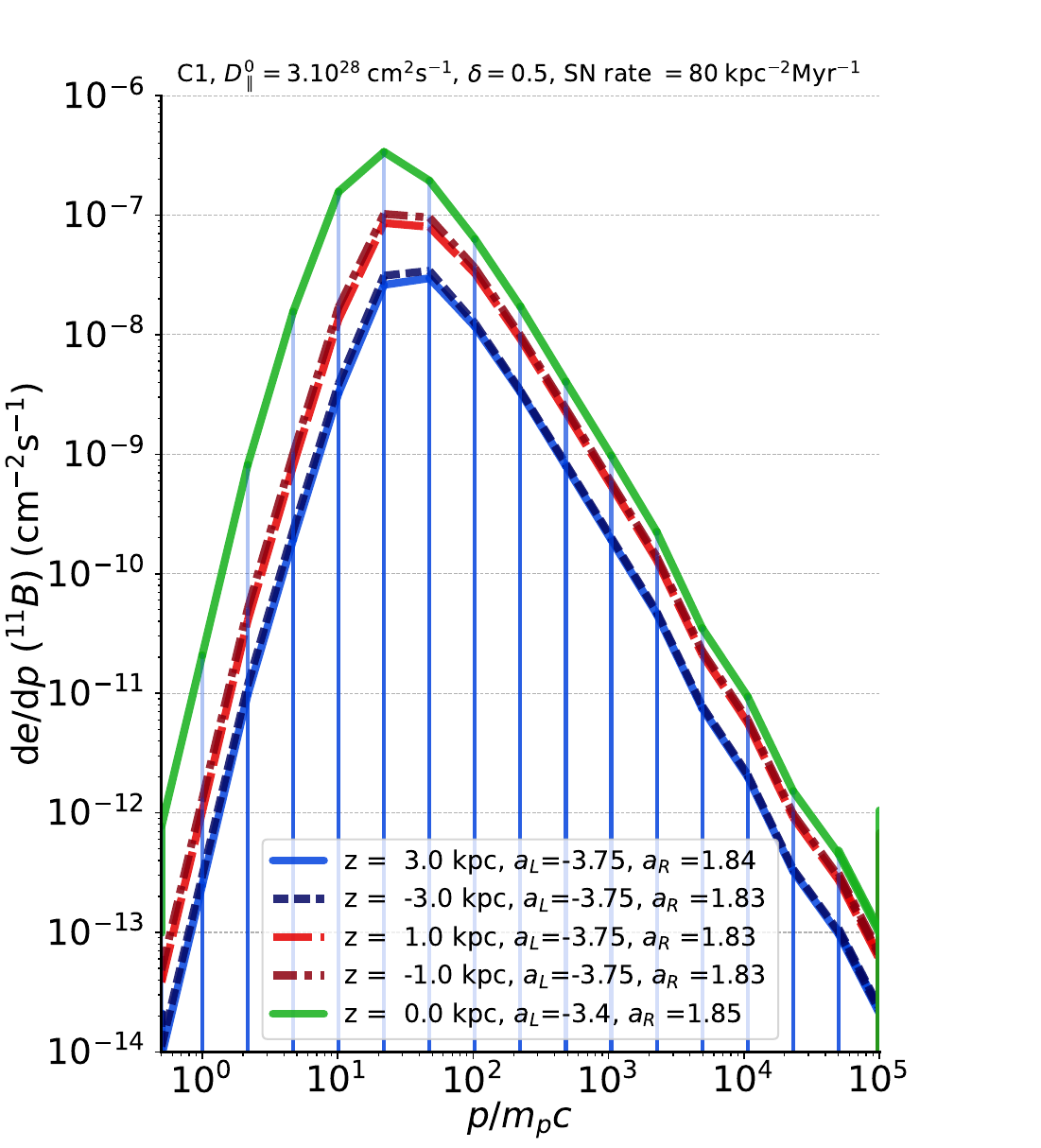}}
    \caption{CR differential energy density spectrum $\mathrm{d}e/\mathrm{d}p$ for \Ct (top row) and \Bel isotope (bottom row) in A1 model (left) and C1 model (right) after $500\; \mathrm{Myr}$ evolution taken at the same altitude points than the previous Figure. As for Figure \ref{fig.6}, the red dashed lines represent the initial slopes $a_L$ and $a_R$ of the primary injection spectrum. For each spectrum, a fit slope estimation is given, with $a_L$ the slope at non-relativistic energy and $a_R$ the slope at relativistic energy. The change of slope in the low-energy range is more visible because of the trans-relativistic transition, the spallation process and the Coulomb energy loss.}
    \label{fig.7}
\end{figure*}
On the right part of Figure \ref{fig.5},  the spectrally resolved \Ct kinetic energy density for four different momentum bins are shown for models A1 and A3 at $t=500\;\mathrm{Myr}$. As for protons, the global distribution of \Ct shows the influence of the underlying evolution of the thermal gas and magnetic field. The \Ct outflows are broader in the box with a higher SN rate. This property, also observed for the spectrally unresolved CR protons, can be attributed to stronger winds generated at higher SN rates.

The most energetic bins components spread more than the less energetic ones: trans-relativistic CRs in the left column (with momentum $p=6.90\;\mathrm{GeV}\,c^{-1}$)  remain confined in the disk while ultra-relativistic CRs vertically spread in the right column ($p=7.24\times10^{3}\;\mathrm{GeV}\,c^{-1}$). The two other columns with intermediate energy display a progressive vertical broadening. This difference in CR propagation across four displayed orders of magnitude in momentum results from the rigidity-dependent diffusion process (see Equation \ref{eq.25}). The second column ($p=7.10\times10^{1};\mathrm{GeV},c^{-1}$) contains the highest abundance of cosmic rays, which matches the maximum seen in the spectra presented in the next section.

Comparison of the small and high SN rate runs presented in Figure~\ref{fig.5} indicates that the higher advection rate corresponding to the high SN rate with the same diffusion coefficients, leads to the broader distribution of CR particles in corresponding momentum bins. The differences in the vertical distribution of CRs for the varying SN rate (and vertical outflow velocities) contradict the potential conjecture of the dominant role of diffusion in the overall vertical transport of CRs because the slope difference is insensitive to the speed of the advection.

To understand better the CR evolution in our simulations, we analyze the CR spectrum for the different primary and secondary species (an analysis of \BtoC and \BettoBen is presented in the next section). The number density and energy density spectrum of primary \Ct and secondary \Bel \   are shown respectively in Figures \ref{fig.6} and \ref{fig.7} for the A1 and C1 models for $t = 500\:\mathrm{Myr}$, which only differ in parameters by rigidity-dependent power index $\delta$.

We observe the presence of small wiggles above $p\approx10^3\,\GeV\,c^{-1}$ for all spectra. Those artefacts, induced by the Coulomb loss routine, remain stable all along the runs, and do not interfer with the physical discussion.

To analyze the different spectra, we focus on the fitted slope values $a_L$ and $a_R$ (introduced in section \ref{sect:spectrum_properties}) in Figures \ref{fig.6} and \ref{fig.7} for non-relativistic and ultra-relativistic ranges respectively. We adopt the convention that a positive slope corresponds to a negative $a_L$ or $a_R$ index, and vice versa. We compare primaries to the initial SN injection slope (the red dot curve). For secondaries, we compare it to the primary slopes, which are the secondary injection slopes. In both cases,  the variation of slope $\Delta a$ is different for $cp  \leq m_\mathrm{prim}c^2$ (momentum range of non-relativistic particles) and above, at $cp  \geq m_\mathrm{prim}c^2$ (ultra-relativistic particles), i.e. $\Delta a_R \neq \Delta a_L$. At relativistic energies, the variation in the slope is due to the rigidity-dependent diffusion. For primary CRs, this is visible in the difference between the number density injection slope and the spectrum of the evolved CR population. A similar effect is observed between primary \Ct and secondary \Bel in Figures \ref{fig.6} and \ref{fig.7}. For the A1 run (left panels), the difference of diffusion power index between the Carbon spectrum (the injection spectrum of secondaries) and the Boron spectrum is $\Delta a_R \approx 0.3$, which is equal to the value of $\delta$ in this run. For the C1 run (right panels), the difference is $\Delta a_R \approx 0.5$, again equal to $\delta$. Rigidity-dependent diffusion along magnetic fields is then a non-negligible transport process.

The slope variation at relativistic energy between the primary injection spectrum and the actual primary spectra is $\Delta a < \delta$. This effect is due to mixing old particles diffusing in the ISM and fresh injected particles from SN events. This effect can be explained by simply estimating the number density spectrum of a primary species. Using the numerical scheme  (formula \ref{eq.A5}) for a given momentum $p$ within a relativistic bin $[p_{L},p_{R}]$ we find after one-time step:
\begin{eqnarray}
    n^\mathrm{p}_l(t+\Delta t)&=&n^\mathrm{p}_l(t)|_\mathrm{old}  -\Gamma^{\mathrm{p}}_l n^\mathrm{p}_l(t)|_\mathrm{old}\Delta t  + n^\mathrm{\mathrm{prim}}_l(\Delta t)|_\mathrm{inj} \nonumber \\ &=& n^{\mathrm{p},0}_l(t)|_\mathrm{old}(1  -\Gamma^{\mathrm{p}}_l \Delta t)\left(\frac{p}{p_{L}}\right)^{-(a_\mathrm{inj}+\delta)} \nonumber\\&&
    +n^\mathrm{\mathrm{p},0}_l(\Delta t)|_\mathrm{inj}\left(\frac{p}{p_{L}}\right)^{-a_\mathrm{inj}} \propto \left(\frac{p}{p_{L}}\right)^{-a'}
\label{eq.26}
\end{eqnarray}

where advection steps are neglected for clarity, and 'inj' terms refer to the fresh population injected in SN remnants. $\Gamma^\mathrm{p}_l$ results from the summation of the different spallation channel reaction rates. Here, $a'$, corresponding to $a_R$ in Figure \ref{fig.6}, results from mixing two CR populations with different spectral indexes. Spallation hardens the spectra while SN events soften them. We have $a_\mathrm{inj}<a'<a_\mathrm{inj}+\delta$ in the general case. The argument is similar for the energy density spectrum. This mixing does not act for secondaries that are injected only by spallation.

At non-relativistic energy, the slope $a_L$ is set by spallation (for secondaries), rigidity-dependent diffusion, and Coulomb energy losses.
At non-relativistic energies, the slopes are  {\bf positive (negative $a_L$)} and steep.  For \Ct with diffusion index $\delta = 0.3$, Figures \ref{fig.6} and \ref{fig.7} show, in the disc, $a_L\approx-1$ for $\mathrm{d}n/\mathrm{d}p$, and $a_L\approx-2.9$ for $\mathrm{d}e/\mathrm{d}p$. Since $a_L=q-2$ for $\mathrm{d}n/\mathrm{d}p$ and $a_L=q-4$ for $\mathrm{d}e/\mathrm{d}p$, $q\approx1$ in the disc ($z\approx0$). The same analysis gives $q\approx0.2-0.3$ in the halo ($z\approx3$).

In a one zone test where only Coulomb losses act, the spectral index evolves to the limit $q=0.1$ \citep{2020MNRAS.491..993G}, independently of the initial slope. However, $q>0.1$ in our simulations, due to particle mixing between old, cooled-down particles and fresh, new particles.
New particles diffuse with a power-law index $1+\delta$ (see equation \ref{eq.25}, with $\beta = p/mc$) at non-relativistic energies.
Moreover, the displayed slope is calculated between the third and fifth momentum bins, in a region approaching the mid-relativistic range, so transition effects influence the measurement.

The spectrum exhibits distinct peaks: for the number density, between $p=10$ and $20\,\GeV\,c^{-1}$; for the energy density, between $p=20$ and $30\,\GeV\,c^{-1}$, above the natural mid-relativistic momentum range $p=Am_pc\approx 10\,\GeV\,c^{-1}$. Particles continuously cool with propagation time, so even relativistic particles start to lose energy.

Figures \ref{fig.6} and \ref{fig.7} present notable differences between spectra at $z=0$ and spectra at high altitudes: the CR number density and energy density are more abundant in the disc area.
At relativistic energies, $cp\gg m_\mathrm{prim}c^2$, the slope index $a_R$ is the same in the disc ($z\approx0$)and the halo ($z\approx3$), being consistent with a CR system undergoing diffusion. At non-relativistic energies, the slopes $a_L$ are systematically harder in the disc.
For primaries, \Ct losses due to spallation become less relevant towards the non-relativistic energy end of the spectrum. Particle acceleration in SN remnants dominates over spallation losses, and diffusion is slow in the non-relativistic energy range, causing non-relativistic primaries to stay longer in the disc. For secondaries, the explanation is the following: suppose primary CR particles from SN remnants are more abundant near the disk midplane. In that case, non-relativistic spallation products are more abundant in the disc compared to high altitudes, even if their production efficiency is low, due to the longer residence time of their primaries near the disk mid-plane.

The subtle effects of primary and secondary CR spectra derived from our discussed simulations are consistent with the diffusion advection propagation model.

At relativistic energies, the slope index $a_R$ for $\mathrm{d}e/\mathrm{d}p$ is $1.25$ for primaries, $1.5$ for secondaries. Experimental measurements lie between $1.8$ and $2.2$ \citep{2015ARA&A..53..199G,2015arXiv150507601N, 2017A&A...606A..22N}. This difference can be attributed to several factors: the power-law $q$ set from SN injection may be too hard to evolve toward the observed value. In addition, the diffusion power index $\delta$ may be underestimated. In the run with $\delta=0.5$, increasing the slope index to $q=4.5$ would raise $a_R$ toward the observed range.

\section{Boron to Carbon and unstable to stable Beryllium isotope ratios} \label{sect:BtoC}

CR observables such as \BtoC=\BBtoC and \BettoBen have been strongly experimentally and theoretically studied in the past years. The leaky-box model \citep[see, e.g.,][for theoretical details]{2011hea..book.....L,2016crpp.book.....G}, in which the solutions of Equation (\ref{eq.1}) represent steady states, introduces the escape time $\tau_{\mathrm{esc}}$ and escape length $\lambda_\mathrm{esc}$, which are respectively the time and length required for particles to escape the Galaxy. The unstable isotope \Bet abundance (also known as Cosmic ray clocks) depends primarily on the decay time $\gamma \tau$. In the leaky-box model, the secondary to primary ratio $f_\mathrm{S}/f_\mathrm{P}=\BtoC$ and unstable to stable isotope ratio \BettoBen explicitly depend on the escape time or escape length and can be then used as a tool to estimate those escape parameters. Including the diffusion process, theoretical works \citep{2002cra..book.....S}
show that $\lambda_\mathrm{esc}\propto D_\parallel^{-1}\propto R^{-\delta}$ and the diffusion coefficient and power index can be retrieved from those ratios.

\begin{figure}
        \includegraphics[scale=.53]{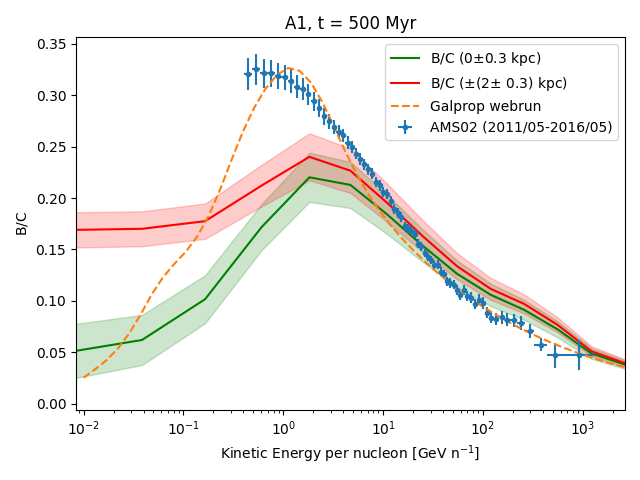}

        \includegraphics[scale=.53]{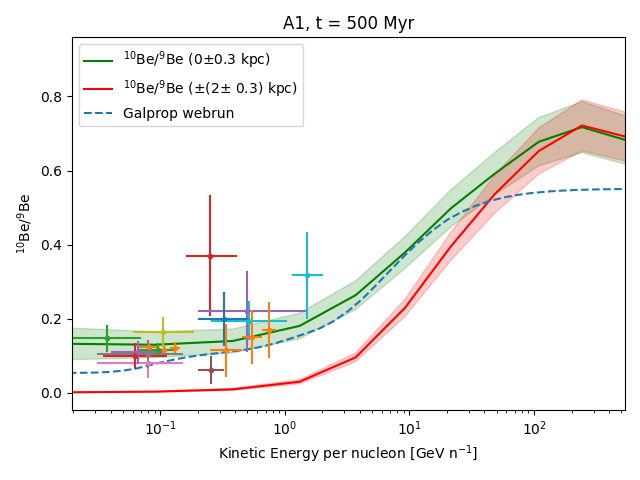}
    \caption{Top: Secondary to primary ratio \BtoC from the A1 run in Table \ref{table_3} vs. Kinetic energy per nucleon, compared with the fiducial Galprop Webrun simulation and AMS-02 experimental data. Bottom: Unstable to stable isotope ratio \BettoBen from the same run vs Kinetic energy per nucleon, also compared with the Galprop Webrun model and with different experimental data extracted from Cosmic Ray Data Base. The green curves are the mean points calculated in a 3D region of $0.625\;\kpc$ vertical size centered in the disc at $z=0$, and the red curves are the mean points calculated in a 3D region of the same vertical size, far from the disc at $\pm 2\;\kpc$. The red and green colored bands are the respective standard deviations.}
    \label{fig.8}
\end{figure}

The AMS-02 experiment has released several new observational data \citep{2016PhRvL.117w1102A,2018PhRvL.120b1101A}, and many models attempted to reproduce and interpret those results \citep{2019PhRvD..99j3023E,2020PhRvD.101b3013E,2023MNRAS.526..160J,2023APh...14402776T}.
In this section, we investigate how the choice of model parameters influences \BtoC and \BettoBen observables in our simulations and compare it to the data of \href{https://galprop.stanford.edu/webrun.php}{Galprop WebRun} \citep{2011CoPhC.182.1156V} and the observational data extracted in \href{https://lpsc.in2p3.fr/crdb/}{Cosmic Ray Data Base (CRDB)} \citep{2014A&A...569A..32M,2020Univ....6..102M,2023EPJC...83..971M}.

In Figure \ref{fig.8} we present \BtoC and \BettoBen vs. kinetic energy per nucleon obtained from the A1 run with $D_\parallel^0=3\times10^{28}\mathrm{cm}^2\mathrm{s}^{-1}$, $\delta=0.3$, and SN rate $=80\; \kpc^{-2}\Myr^{-1}$, displayed together with AMS-02 observational data and GALPROP WebRun simulation data. 					We use the GALPROP WebRun data as a simple but qualitatively accurate reference for comparison, while acknowledging that it is less precise than a full GALPROP setup. Our primary validation is based on AMS-02 data; however, we also note that low-energy Voyager measurements provide additional constraints. In particular, the recent work by \cite{2025arXiv251203385P} shows that Voyager data span energies per nucleon in the range $10^{-2}-10^{-1}\GeV$, with measured B/C ratios between $0.1-0.17$. Our models A2 and B1 (see Figures \ref{fig.9} and \ref{fig.11}) fit well to these data within the standard deviation uncertainties.

The green curves are obtained from data points in a 3D region centered in the disc plane, and red ones from a 3D region at $\pm 2 \;\kpc$ far above the disc. In both plots, the deviation between our simulations and the data is satisfying, except for \BettoBen at $\pm 2 \kpc$, where primary CR nuclei are much less abundant: radioactive decay depletes more efficiently the low-energy radioactive isotopes, so the curve at $z\simeq 2\,\kpc$ is significantly below the other data. Therefore, we do not show the lines of \BettoBen for $\pm2\,\mathrm{kpc}$ in the next plots.

The maximum of \BtoC is $\approx 0.24$, below the experimental value of $0.3$. The fiducial model is a reference point for exploring the parameter space. We will show in the next section how varying the discussed parameters brings the results closer to the observations. We note that \BettoBen is well fitting the data points below $1\,\GeV\,n^{-1}$ for this fiducial run.

We compare the other models to our fiducial one in Figure \ref{fig.8}. We vary the SN rate, diffusion coefficient, and power index and report the results in the following section. The parameter study in this section does not provide a best-fit or make a formal $\chi^2$ comparison with the data. We demonstrate the internal consistency of the physics presented in the different runs, and show that the trends of the observables respond systematically to variations of the chosen input parameters, thereby validating the robustness of the method and the potential of the model to fit the experimental data.

\subsection{Impact of supernova rate}

\begin{figure*}[ht!]

    \centerline{

        \includegraphics[scale=.43]{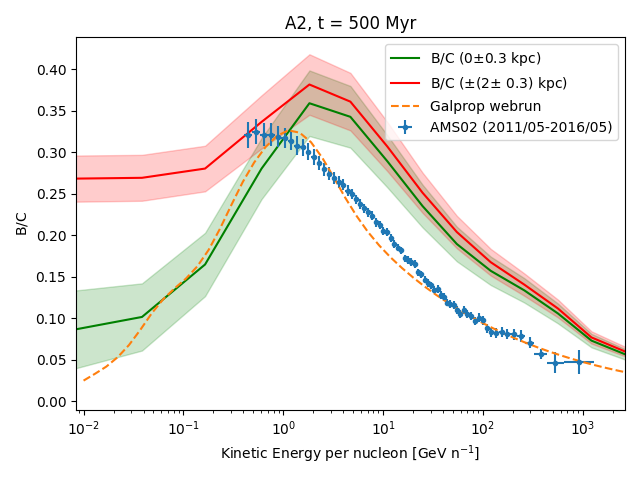}

        \includegraphics[scale=.43]{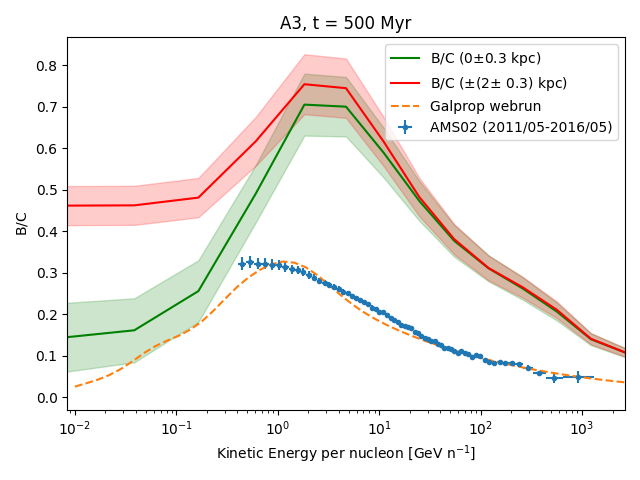}}

    \centerline{

        \includegraphics[scale=.43]{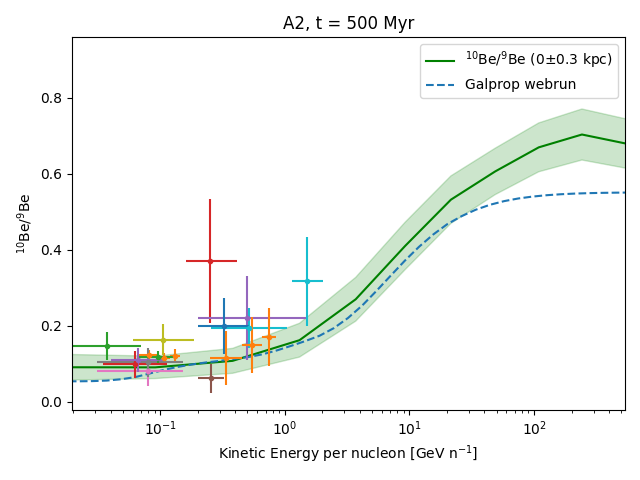}

        \includegraphics[scale=.43]{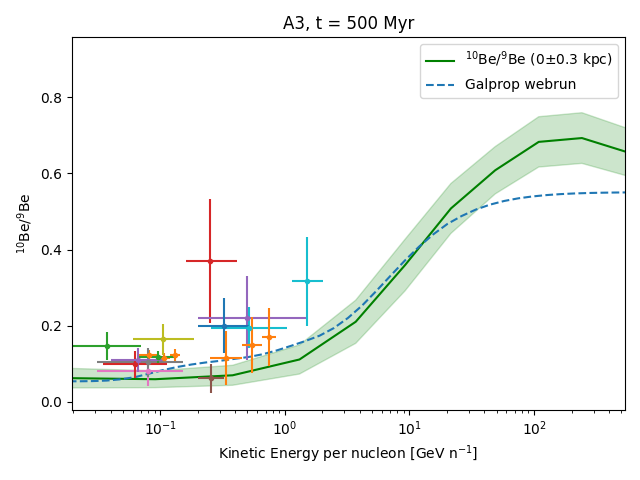}}

    \caption{Same \BtoC and \BettoBen plots as the previous Figure \ref{fig.8}, for A2 and A3 models, only differing from the fiducial case by SN rate. The left panels have SN rate $=60\;\kpc^{-2}\mathrm{Myr}^{-1}$, the right panels $20\;\kpc^{-2}\mathrm{Myr}^{-1}$.}
    \label{fig.9}
\end{figure*}

\begin{figure*}[ht!]
    \centerline{
        \includegraphics[scale=.37]{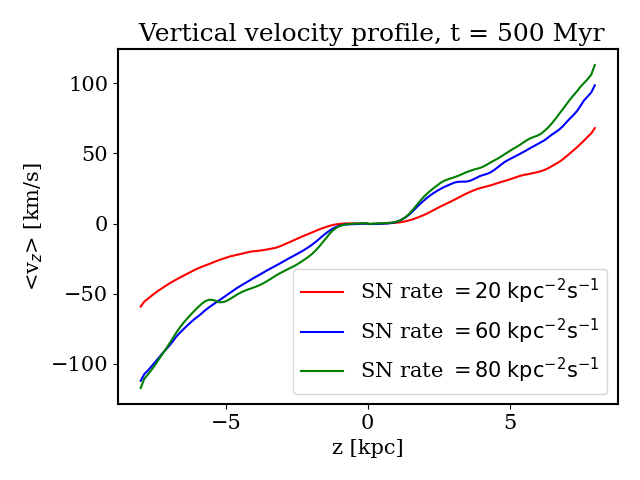}

        \includegraphics[scale=.37]{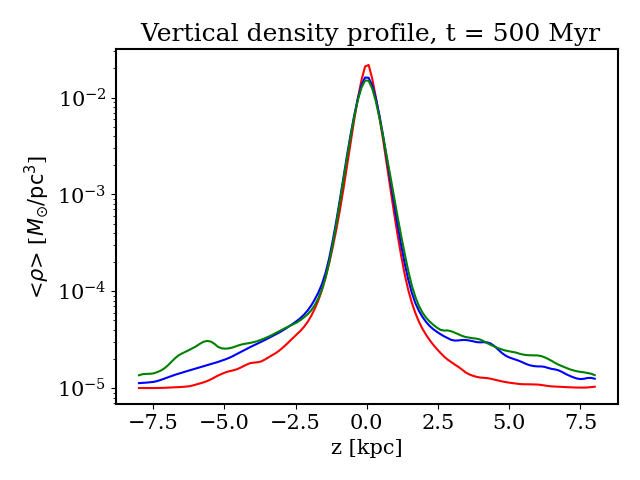}

        \includegraphics[scale=.37]{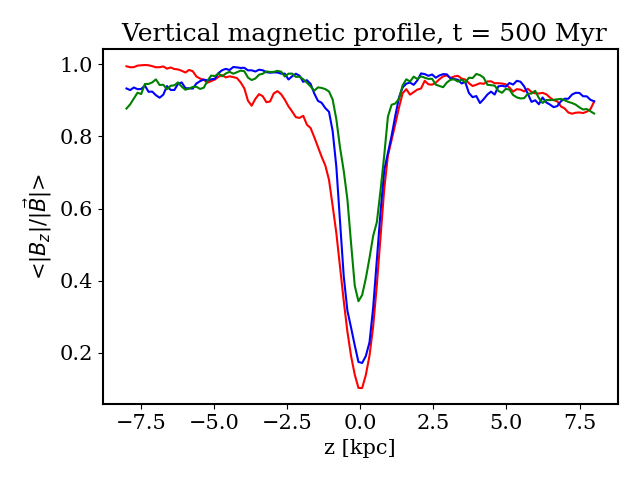}}
    \centerline{
        \includegraphics[scale=.37]{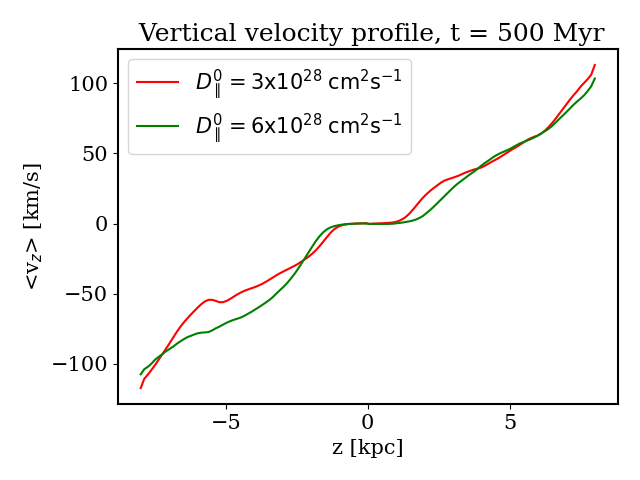}
        \includegraphics[scale=.37]{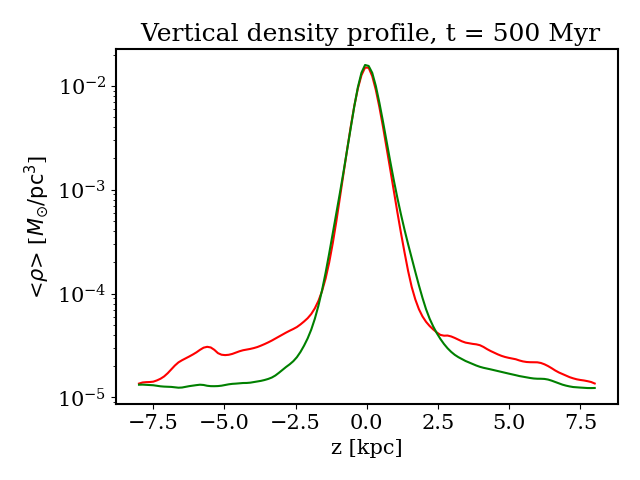}
        \includegraphics[scale=.37]{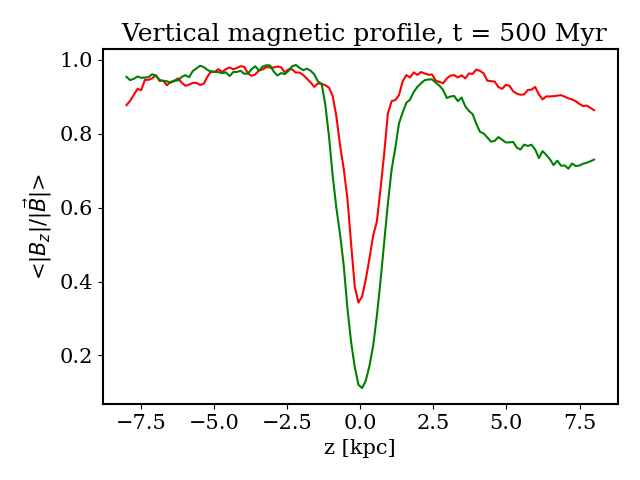}}
    \caption{Top: profile of vertical ISM gas velocity $\langle v_z\rangle$ (left), gas density $\langle \rho \rangle$ (middle) and $\langle|B_z|/|\vec{B}|\rangle$ (right) averaged in $x-y$ plan vs altitude $z$ in $\kpc$ for three runs having different SN rate values, displayed in the caption. Bottom: profile of vertical ISM gas velocity $\langle v_z\rangle$ (left), gas densiry $\rho$ (middle) and $\langle|B_z|/|\vec{B}|\rangle$ (right) averaged in $x-y$ plan vs altitude $z$ in $\kpc$ for two runs having different diffusion coefficient $D_\parallel^0$ values, displayed in the caption. The captions in the left panels apply to all the plots in the row.}
    \label{fig.10}
\end{figure*}

Figure \ref{fig.9} presents the \BtoC and \BettoBen plots vs. kinetic energy per nucleon (in $\mathrm{GeV}\,n^{-1}$) from the models A2 (SN rate $=60\;\kpc^{-2}\mathrm{Myr}^{-1}$) and A3 (SN rate $= 20\;\kpc^{-2}\mathrm{Myr}^{-1}$). Comparing the models in Figures \ref{fig.8} and \ref{fig.9}, we notice a different response of the system to the SN rate in each simulation: the maximum of $\BtoC$ reaches higher values if the SN rate is lower. As discussed in Section \ref{sect:cr_propagation}, a lower SN rate generates weaker outflows in which CR advection in the vertical direction is less efficient. CR proton pressure from a higher SN rate in the ISM drives stronger outflows, strengthening advection transport and leading the secondary nuclei to deplete parallel to diffusive transport. This is confirmed in Figure \ref{fig.10} where the top row shows the vertical velocity $\langle v_z\rangle$, gas density $\langle\rho\rangle$, and vertical magnetic profile over $z$ direction for the three runs. The vertical magnetic profile is shown using the quantity $\langle|B_z|/|\vec{B}|\rangle=\langle\mathrm{sin}(\alpha)\rangle$ which specifies the mean angle between $B_z$ component and the whole vector $\vec{B}$. $\langle\mathrm{sin}(\alpha)\rangle$ has values between $0$ (no vertical magnetic field) and $1$ (only vertical magnetic field).  The velocity $v_z$ and density $\rho$ in the disc region ($-1\, \kpc \leq z \leq 1\, \kpc$) are identical for all models. Still, at several kpc altitudes, they have higher values if the SN rate is higher, because of more efficient vertical outflows. In the disc, $\langle|B_z|/|\vec{B}|\rangle$ increases if the SN rate increases. CRs propagating along magnetic field lines escape more efficiently outside of the disc region due to a higher magnetic field inclination (see discussion of Figure \ref{fig.4} in Section \ref{sect:ISM_evolution}). Consequently, fewer secondary CRs are produced in the central disc, and \BtoC is decreasing. At high altitudes, $B_z$ dominates for all the runs.  This is expected since all discussed models possess efficient, spatially non-uniform outflows.

For \BettoBen, at the lowest SN rate = $20\;\mathrm{kpc}^{-2}\mathrm{Myr}^{-1}$ the ratio \BettoBen is the lowest in the subrelativistic limit.
The depletion of the unstable secondary isotope \Bet still occurs while less efficient advection leads to higher abundances of the stable \Ben in the disk.

\subsection{Impact of diffusion coefficient and its power index}

\begin{figure*}[ht!]

    \centerline{
        \includegraphics[scale=.43]{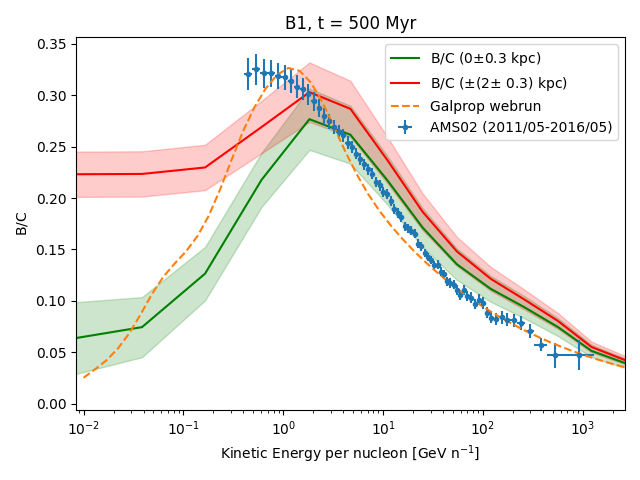}

        \includegraphics[scale=.43]{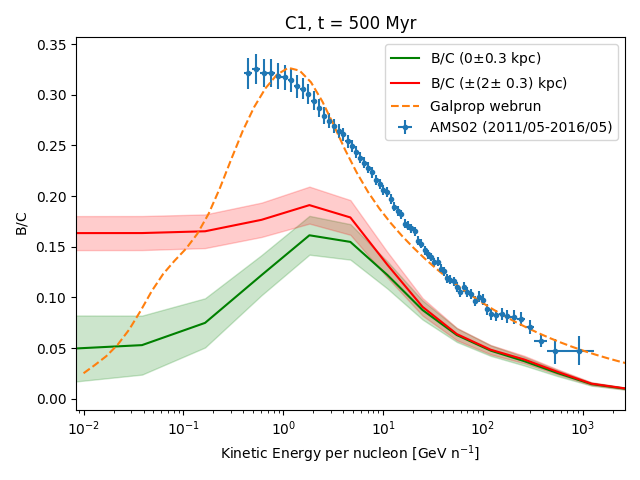}}

    \centerline{

        \includegraphics[scale=.43]{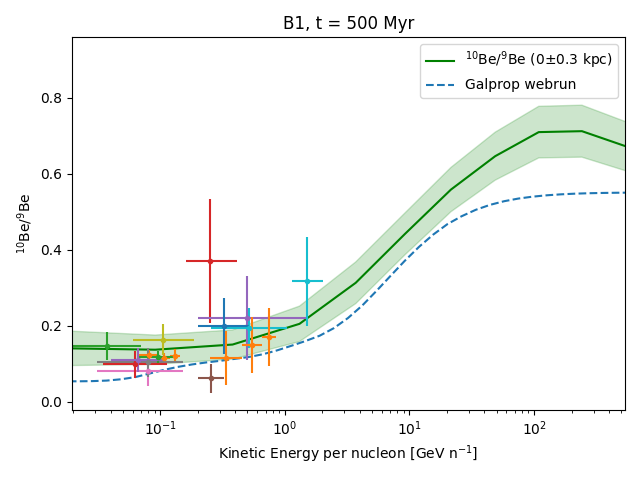}

        \includegraphics[scale=.43]{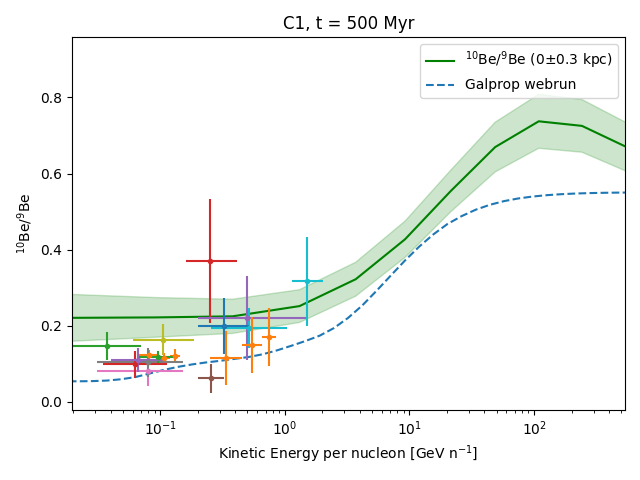}}
    \caption{Same \BtoC and \BettoBen plots as the previous Figures \ref{fig.8} and \ref{fig.9} for B1 and C1 models, only differing from the fiducial case by diffusion coefficient and diffusion power index. Left panels have $D_\parallel^0 = 6\times 10^{28}\;\mathrm{cm}^2\mathrm{s}^{-1}$ and $\delta = 0.3$, and right panels have $D_\parallel^0 = 3\times 10^{28}\;\mathrm{cm}^2\mathrm{s}^{-1}$ and $\delta = 0.5$.}
    \label{fig.11}
\end{figure*}

In Figure \ref{fig.11}, we analyze \BtoC and \BettoBen for models differing in diffusion  $D^0_\parallel$ and $\delta$ (see Equation \ref{eq.25}).  Comparing Figure \ref{fig.8} and the left panel of Figure \ref{fig.11},
the maximum of \BtoC near $1\;\mathrm{GeV}\,n^{-1}$  is higher if $D^0_\parallel$ is greater, meaning a longer residence
(or escape) time of CRs in the dense disk regions.

This result contrasts with the predictions of the leaky box and the phenomenological (GALPROP-type) stationary models. 
From the Leaky-box model, \BtoC depends on the diffusion coefficient:
in the relativistic energy range, and assuming isotropic diffusion coefficient $D_\parallel$, increasing
\BtoC coincides with a smaller diffusion coefficient and greater escape time:
\begin{equation}
\frac{\mathrm{B}}{\mathrm{C}}\left(\gg1\,\mathrm{GeV}\,n^{-1}\right) \approx \tau_\mathrm{esc}=\frac{H^2}{D_\parallel}
\label{eq.27}
\end{equation}
where $H$ is the scale height of the disk \citep{2023arXiv230900298E}. The system does not follow the relation (\ref{eq.27}) in our simulations.

To explain this qualitative difference between the stationary approach and our self-consistent model, we analyze the basic dynamics of the ISM driven by supernovae through the coupling of CRs with the thermal ISM. In Figure \ref{fig.10}, we show the vertical velocity $\langle v_z\rangle$, gas density $\langle\rho\rangle$ and vertical magnetic profile of $\langle|B_z|/|\vec{B}|\rangle$ along the $z$-direction for the two runs discussed here. We find that density is lower at $|z| >1\,\kpc$  for a greater diffusion coefficient  $D_\parallel^0=6\times10^{28}\;\mathrm{cm}^2\mathrm{s}^{-1}$. The lower density coincides with the higher outflow velocity, in the statistical sense, since the vertical velocity fluctuates and is asymmetric.
The higher outflow velocity and lower density in the halo favor lower values of \BtoC for the greater diffusion coefficient. The opposite effect that we observe motivates us to investigate the differences in magnetic field structure.

We find that in the dense layer of the disk, $\langle|B_z|/|\vec{B}|\rangle$ is lower if $D^0_\parallel$ is higher.
This difference points us to define the effective vertical diffusion coefficient $D_{\mathrm{vert}} \equiv D_\parallel \langle \sin (\alpha )\rangle $, which depends on $D_\parallel$ and the average inclination of the magnetic field perturbed with CR buoyancy effects. In our models, $D_{\mathrm{vert}}$ has to replace $D_\parallel$ in Expression (\ref{eq.27}).

The lower right panel of Figure~\ref{fig.10} demonstrates that twice bigger parallel diffusion coefficient of $D_\parallel^0=6\times10^{28}\;\mathrm{cm}^2\mathrm{s}^{-1}$ lowers the value of $\langle\mathrm{sin}(\alpha)\rangle$ from $\sim0.4$ to $\sim0.1$ near the disc midplane, making vertical diffusion almost four times slower and increasing the escape time by same factor. Consequently, more secondaries are produced in the disc with a higher diffusion coefficient, and \BtoC peak value becomes higher. A difference in $\langle|B_z|/|\vec{B}|\rangle$  is also present in the upper halo at high altitudes, where this factor decreases below a value of $0.8$ above $1\,\kpc$ for $D_\parallel^0=6\times10^{28}\;\mathrm{cm}^2\mathrm{s}^{-1}$, while it is close to $1$ for $D_\parallel^0 =  3\times10^{28}\;\mathrm{cm}^2\mathrm{s}^{-1}$.

Those results demonstrate the essential difference between the stationary models and our self-consistent CRMHD approach, in which CRs, gas, and magnetic fields are dynamically coupled. Varying parameters, such as diffusion coefficient, change the distribution and geometry of the magnetic fields, making the CR transport along the field lines significantly different between the different models. In GALPROP, one should simultaneously fix $D_\parallel$ with the value of the vertical wind velocity. In PIERNIK, the velocity is directly determined by the CR-ISM dynamics. Moreover, the escape time we deduce from \BtoC and \BettoBen depends not only on $D_\parallel$, but also on the MF geometry via the $\mathrm{sin}(\alpha)$ parameter.

Looking at the bottom row of Figure \ref{fig.11}, \BettoBen is identically close to the Galprop prediction for $D^0_\parallel = 3\times10^{28}\;\mathrm{cm}^2\mathrm{s}^{-1}$ (left panel) and for $6\times10^{28}\;\mathrm{cm}^2\mathrm{s}^{-1}$ (right panel). The \BettoBen differences between A1 and B1 models are marginal.

A comparison of A1 and C1 models presented in Figure \ref{fig.8} and the right-hand part of Figure \ref{fig.11} shows the dependency of \BtoC on the diffusion power index $\delta$. For $\delta=0.5$, \BtoC significantly decreases at relativistic energies. The non-relativistic part of the spectrum (see section \ref{sect:cr_propagation}) is modified by the change of $\delta$. However, we did not include solar modulation effects on CRs, so no definitive conclusion concerning the ratios in the non-relativistic range can be drawn regarding possible observational data.

From Equation \ref{eq.25}, we expect that, at relativistic energy, \BtoC follows a power-law behavior in $p^{-\delta}$ and the slope differs for $\delta=0.3$ (model A1) and $\delta=0.5$ (model C1). The observational data from \cite{2016PhRvL.117w1102A} fit with a power-law $\delta = 1/3$, corresponding to the Kolmogorov theory of turbulence \citep{1991RSPSA.434....9K}. In model A1, shown in Figure \ref{fig.8}, where $\delta=0.3$, the slope at relativistic energy is not accurately close to the data points. In model C1  ($\delta = 0.5$) shown in the right panel of Figure \ref{fig.11}, the simulation results show a strong agreement with the data beyond  $10\,\GeV\,n^{-1}$, even if the amplitude is different. \BtoC is then sensitive to the power index $\delta$ in our simulations, as expected, but it also suggests to search for a more optimal combination of $\delta$, $D_\parallel$ and SN rate to reproduce the correct slope and amplitude.

\BettoBen does significantly change between the two models, approaching $0.2$ at non-relativistic energy, which is in tension with the data and GALPROP results.

\section{Conclusions} \label{sect:conclusions}

Following the implementation of spectrally resolved propagation of CR electrons within the Cosmic Ray Energy SPectrum (CRESP) algorithm of PIERNIK MHD code presented in Paper I \citep{2021ApJS..253...18O}, in this paper, we presented an extension of this algorithm to the cases of a high number of primary and secondary CR nuclei, which are subject to spallation reactions and radioactive losses. Similarly to the previous paper, we have used the two-moment piece-wise power-law approach to solve the Fokker-Planck Equation (\ref{eq.1}) and compute CR number and energy density (Relations (\ref{eq.2}) and (\ref{eq.3})) in a relatively low number (typically less than 20) of momentum bins at each step of the simulations.  CR primary and secondary nuclei are coupled to the live MHD environment. The whole dynamics of the present system are in the present simplified setup entirely driven by spectrally unresolved CR proton. The other spectrally resolved nuclei are subject to advection and rigidity-dependent diffusion in the magnetized ISM environment shaped by CR protons. Momentum-dependent spallation collisions of C, N, and O nuclei against the ISM hydrogen nuclei generate secondary particles: Li, Be, and B, including the unstable isotope \Bet. Those processes influence the spectra of primaries and secondaries for both non-relativistic and ultra-relativistic ranges. We tested this model using a set of 3D gravity-stratified box simulations with ISM gas, magnetic field, and spectrally unresolved CR protons  (grey approximation) driving the ISM dynamics. We performed a parameter study of that system by investigating the impact of three parameters: supernova  (SN) rate, CR parallel diffusion coefficient $D_\parallel$, and the parameter $\delta$ representing the power-law dependence of the diffusion coefficient on rigidity.
We verified that the differences between the slopes of primary CR injection and evolved spectra and between the slopes of primary and secondary CR components are consistent with the standard interpretation of the impact of the diffusion power-law index $\delta$.
We analyzed the evolution of CR spectra and compared the energy-dependent \BtoC and \BettoBen ratios with the observational data and Galprop Webrun results. 

The main conclusions of our investigations are as follows:
\begin{enumerate}
\item The CR spectrum, \BtoC, and \BettoBen are sensitive to model input parameters: SN rate, diffusion coefficient, and diffusion power index. While the AMS-02 experiment well constrains the $\delta$ parameter, tuning the SN rate and parallel diffusion coefficient is needed to reproduce the  \BtoC and \BettoBen consistent with experimental data.

\item The SN rate controls the vertical outflow velocity through the dynamical coupling of CR protons to the thermal ISM components and magnetic field. The higher the SN rate, the more CR energy is injected into the system and the faster the CR-driven winds are. This relation implies a shorter residence time of primary CR nuclei in the dense disk region and, therefore, lower values for the \BtoC ratio. The effect is pronounced since variations of SN rate by a factor of a few changes \BtoC by a similar factor.
 
\item  A novel result of our investigation is that CRs dynamically coupled to the magnetized ISM can drastically change these observables' response to diffusion coefficient variations. Contrary to the previous models,  our model predicts that \BtoC can grow when the diffusion coefficient grows. We attribute this property to the magnetic field structure reaction to diffusion coefficient variations. Higher CR diffusion coefficients lead to weaker vertical fields resulting from the CR buoyancy effects. 
A plausible explanation is that a more efficient diffusion distributes CRs more evenly along horizontal field lines of the disk. Consequently, buoyancy forces due to discrete local CR injection events are more evenly distributed along a horizontal flux tube. This leads to less differentiation of the flux tube elevation and less vertical magnetic field. Such a field configuration traps the primary CR particles for longer near the disk midplane, leading to higher abundancies of secondary CRs. 

\item Our results show that \BtoC and \BettoBen observables do not provide the exact CR escape time deducible from the diffusion input parameters, but instead a practical value that indicates how the coupling of CRs to the ISM influences their transport in a non-trivial manner.

\end{enumerate}

However, we note some caveats in our model. We mention that:

- A simple narrow vertical patch of galactic disk is represented in our simulation, and our current resolution $\Delta x = 62.5\;\mathrm{pc}$ we are using is far from other state-of-the-art CRMHD simulations such as in \cite{2018MNRAS.479.3042G}, for which $\Delta x = 4\;\mathrm{pc}$, or \citep{2021ApJ...922...11A,2022ApJ...929..170A,2024ApJ...964...99A}, having $\Delta x = \mathcal{O}( \mathrm{1\;pc})$ cell size.

- Consequently, to the previous point, the geometry of the stratified box and its topology (periodic boundary conditions in x and y directions) may influence CR propagation and the system's response to input parameter variations. This circumstance may lead to a significant bias in our results, rendering them qualitatively similar, but quantitatively distinct from what a comprehensive galactic halo model can predict.

- Our approach is simplified in some aspects that might be essential to consider for tuning the model against observational data:
the model does not include the streaming propagation mode of CRs \citep[see][for a numerical approach example]{2019MNRAS.485.2977T}. Also, we assume a simple ideal thermal ISM gas, with adiabatic equation of state, and we do not resolve the thermal structure nor small-scale turbulence dynamics that form the background for CR propagation. A multiphase ISM with thermal feedback from different sources also impacts CR transport; therefore, the results presented in this paper may change significantly in such a system.

We plan to address these issues in the next steps of our study.
\section{Acknowledgements}

This work was supported by the Polish National Science Center through the OPUS grant no. 2015/19/B/ST9/02959.
Calculations were carried out at the Centre of Informatics – Tricity Academic Supercomputer \& networK (TASK) and on the HYDRA cluster at the Institute of Astronomy of Nicolaus Copernicus University in Torun.

The authors thank the anonymous referee for their helpful report and suggestions. MH would like to thank the Aspen Physics Center for its hospitality during the 2024 workshop “Cosmic Ray Feedback in Galaxy Evolution,”  and the participants of that workshop, in particular Isabelle Grenier, for inspiring discussions. PG acknowledges financial support from the European Research Council via the ERC Synergy Grant ``ECOGAL'' (project ID 855130). AB acknowledges the kind hospitality of the Hamburg Sternwarte and the valuable discussions during his visit.
\section{Data Availability}

The data underlying this paper will be shared on reasonable request to the corresponding author.

\bibliography{paper}{}

@Article{	  sharma_2010,
  title		= {Numerical Implementation of Streaming Down the Gradient:
		  Application to Fluid Modeling of Cosmic Rays and Saturated
		  Conduction},
  volume	= {32},
  issn		= {1095-7197},
  url		= {http://dx.doi.org/10.1137/100792135},
  doi		= {10.1137/100792135},
  number	= {6},
  journal	= {SIAM Journal on Scientific Computing},
  publisher	= {Society for Industrial & Applied Mathematics (SIAM)},
  author	= {Sharma, Prateek and Colella, Phillip and Martin, Daniel
		  F.},
  year		= {2010},
  month		= {Jan},
  pages		= {3564–3583}
}

@Article{	  2018apj...854....5j,
  author	= {{Jiang}, Yan-Fei and {Oh}, S. Peng},
  title		= "{A New Numerical Scheme for Cosmic-Ray Transport}",
  journal	= {\apj},
  year		= "2018",
  month		= "Feb",
  volume	= {854},
  number	= {1},
  eid		= {5},
  pages		= {5},
  doi		= {10.3847/1538-4357/aaa6ce},
  archiveprefix	= {arXiv},
  eprint	= {1712.07117},
  primaryclass	= {astro-ph.HE},
  adsurl	= {https://ui.adsabs.harvard.edu/abs/2018ApJ...854....5J}
}

@Article{	  2019MNRAS.485.2977T,
  author	= {{Thomas}, T. and {Pfrommer}, C.},
  title		= "{Cosmic-ray hydrodynamics: Alfv{\'e}n-wave regulated
		  transport of cosmic rays}",
  journal	= {\mnras},
  year		= "2019",
  month		= "May",
  volume	= {485},
  number	= {3},
  pages		= {2977-3008},
  doi		= {10.1093/mnras/stz263},
  archiveprefix	= {arXiv},
  eprint	= {1805.11092},
  primaryclass	= {astro-ph.HE},
  adsurl	= {https://ui.adsabs.harvard.edu/abs/2019MNRAS.485.2977T}
}

@ARTICLE{2022MNRAS.510.3917G,
       author = {{Girichidis}, Philipp and {Pfrommer}, Christoph and {Pakmor}, R{\"u}diger and {Springel}, Volker},
        title = "{Spectrally resolved cosmic rays - II. Momentum-dependent cosmic ray diffusion drives powerful galactic winds}",
      journal = {\mnras},
         year = 2022,
        month = mar,
       volume = {510},
       number = {3},
        pages = {3917-3938},
          doi = {10.1093/mnras/stab3462},
archivePrefix = {arXiv},
       eprint = {2109.13250},
 primaryClass = {astro-ph.GA},
       adsurl = {https://ui.adsabs.harvard.edu/abs/2022MNRAS.510.3917G}
}

@ARTICLE{2015ARA&A..53..199G,
       author = {{Grenier}, Isabelle A. and {Black}, John H. and {Strong}, Andrew W.},
        title = "{The Nine Lives of Cosmic Rays in Galaxies}",
      journal = {\araa},
         year = 2015,
        month = aug,
       volume = {53},
        pages = {199-246},
          doi = {10.1146/annurev-astro-082214-122457},
       adsurl = {https://ui.adsabs.harvard.edu/abs/2015ARA&A..53..199G}
}

@ARTICLE{2022MNRAS.tmp.1768H,
       author = {{Hopkins}, Philip F. and {Butsky}, Iryna S. and {Panopoulou}, Georgia V. and {Ji}, Suoqing and {Quataert}, Eliot and {Faucher-Gigu{\`e}re}, Claude-Andr{\'e} and {Kere{\v{s}}}, Du{\v{s}}an},
        title = "{First predicted cosmic ray spectra, primary-to-secondary ratios, and ionization rates from MHD galaxy formation simulations}",
      journal = {\mnras},
         year = 2022,
        month = jul,
          doi = {10.1093/mnras/stac1791},
archivePrefix = {arXiv},
       eprint = {2109.09762},
 primaryClass = {astro-ph.HE},
       adsurl = {https://ui.adsabs.harvard.edu/abs/2022MNRAS.tmp.1768H}
}

@ARTICLE{2021LRCA....7....2H,
       author = {{Hanasz}, Micha{\l} and {Strong}, Andrew W. and {Girichidis}, Philipp},
        title = "{Simulations of cosmic ray propagation}",
      journal = {Living Reviews in Computational Astrophysics},
         year = 2021,
        month = dec,
       volume = {7},
       number = {1},
          eid = {2},
        pages = {2},
          doi = {10.1007/s41115-021-00011-1},
archivePrefix = {arXiv},
       eprint = {2106.08426},
 primaryClass = {astro-ph.HE},
       adsurl = {https://ui.adsabs.harvard.edu/abs/2021LRCA....7....2H}
}

@ARTICLE{2021ApJS..253...18O,
       author = {{Ogrodnik}, Mateusz A. and {Hanasz}, Micha{\l} and {W{\'o}lta{\'n}ski}, Dominik},
        title = "{Implementation of Cosmic Ray Energy Spectrum (CRESP) Algorithm in PIERNIK MHD Code. I. Spectrally Resolved Propagation of Cosmic Ray Electrons on Eulerian Grids}",
      journal = {\apjs},
         year = 2021,
        month = mar,
       volume = {253},
       number = {1},
          eid = {18},
        pages = {18},
          doi = {10.3847/1538-4365/abd16f},
archivePrefix = {arXiv},
       eprint = {2009.06941},
 primaryClass = {astro-ph.HE},
       adsurl = {https://ui.adsabs.harvard.edu/abs/2021ApJS..253...18O}
}

@ARTICLE{2020MNRAS.491..993G,
       author = {{Girichidis}, Philipp and {Pfrommer}, Christoph and {Hanasz}, Micha{\l} and {Naab}, Thorsten},
        title = "{Spectrally resolved cosmic ray hydrodynamics - I. Spectral scheme}",
      journal = {\mnras},
         year = 2020,
        month = jan,
       volume = {491},
       number = {1},
        pages = {993-1007},
          doi = {10.1093/mnras/stz2961},
archivePrefix = {arXiv},
       eprint = {1909.12840},
 primaryClass = {astro-ph.HE},
       adsurl = {https://ui.adsabs.harvard.edu/abs/2020MNRAS.491..993G}
}

@article{1966ApJ...145..811P,
	Adsurl = {http://adsabs.harvard.edu/abs/1966ApJ...145..811P},
	Author = {{Parker}, E.~N.},
	Doi = {10.1086/148828},
	Journal = {\apj},
	Month = sep,
	Pages = {811-+},
	Title = {{The Dynamical State of the Interstellar Gas and Field}},
	Volume = 145,
	Year = 1966}

@ARTICLE{2005JCoPh.208..315M,
   author = {{Miyoshi}, T. and {Kusano}, K.},
    title = "{A multi-state HLL approximate Riemann solver for ideal magnetohydrodynamics}",
  journal = {Journal of Computational Physics},
     year = 2005,
    month = sep,
   volume = 208,
    pages = {315-344},
      doi = {10.1016/j.jcp.2005.02.017},
   adsurl = {http://adsabs.harvard.edu/abs/2005JCoPh.208..315M}
}

@ARTICLE{2002JCoPh.175..645D,
       author = {{Dedner}, A. and {Kemm}, F. and {Kr{\"o}ner}, D. and {Munz}, C. -D. and {Schnitzer}, T. and {Wesenberg}, M.},
        title = "{Hyperbolic Divergence Cleaning for the MHD Equations}",
      journal = {Journal of Computational Physics},
         year = 2002,
        month = jan,
       volume = {175},
       number = {2},
        pages = {645-673},
          doi = {10.1006/jcph.2001.6961},
       adsurl = {https://ui.adsabs.harvard.edu/abs/2002JCoPh.175..645D}
}

@ARTICLE{1975MNRAS.172..557S,
   author = {{Skilling}, J.},
    title = "{Cosmic ray streaming. I - Effect of Alfven waves on particles}",
  journal = {\mnras},
     year = 1975,
    month = sep,
   volume = 172,
    pages = {557-566},
   adsurl = {http://adsabs.harvard.edu/abs/1975MNRAS.172..557S}
}

@ARTICLE{2001CoPhC.141...17M,
   author = {{Miniati}, F.},
    title = "{COSMOCR: A numerical code for cosmic ray studies in computational cosmology}",
  journal = {Computer Physics Communications},
   eprint = {astro-ph/0105447},
     year = 2001,
    month = nov,
   volume = 141,
    pages = {17-38},
      doi = {10.1016/S0010-4655(01)00293-4},
   adsurl = {http://adsabs.harvard.edu/abs/2001CoPhC.141...17M}
}

@ARTICLE{Strong&Moskalenko1998,
  author = {{Strong}, A.~W. and {Moskalenko}, I.~V.},
   title = "{Propagation of Cosmic-Ray Nucleons in the Galaxy}",
 journal = {\apj},
  eprint = {arXiv:astro-ph/9807150},
    year = 1998,
   month = dec,
  volume = 509,
   pages = {212-228},
     doi = {10.1086/306470},
  adsurl = {http://esoads.eso.org/abs/1998ApJ...509..212S}
}

@ARTICLE{2018MNRAS.479.3042G,
       author = {{Girichidis}, Philipp and {Naab}, Thorsten and {Hanasz}, Micha{\l} and {Walch}, Stefanie},
        title = "{Cooler and smoother - the impact of cosmic rays on the phase structure of galactic outflows}",
      journal = {\mnras},
         year = 2018,
        month = sep,
       volume = {479},
       number = {3},
        pages = {3042-3067},
          doi = {10.1093/mnras/sty1653},
archivePrefix = {arXiv},
       eprint = {1805.09333},
 primaryClass = {astro-ph.GA},
       adsurl = {https://ui.adsabs.harvard.edu/abs/2018MNRAS.479.3042G}
}

@ARTICLE{2013ApJ...777L..38H,
   author = {{Hanasz}, M. and {Lesch}, H. and {Naab}, T. and {Gawryszczak}, A. and
	{Kowalik}, K. and {W{\'o}lta{\'n}ski}, D.},
    title = "{Cosmic Rays Can Drive Strong Outflows from Gas-rich High-redshift Disk Galaxies}",
  journal = {\apjl},
archivePrefix = "arXiv",
   eprint = {1310.3273},
 primaryClass = "astro-ph.GA",
     year = 2013,
    month = nov,
   volume = 777,
      eid = {L38},
    pages = {L38},
      doi = {10.1088/2041-8205/777/2/L38},
   adsurl = {http://adsabs.harvard.edu/abs/2013ApJ...777L..38H}
}

@INPROCEEDINGS{2012EAS....56..367H,
   author = {{Hanasz}, M. and {Kowalik}, K. and {W{\'o}lta{\'n}ski}, D. and
   {Paw{\l}aszek}, R.},
    title = "{PIERNIK MHD code - a multi-fluid, non-ideal extension of the relaxing-TVD scheme (IV)}",
booktitle = {EAS Publications Series},
     year = 2012,
   series = {EAS Publications Series},
   volume = 56,
archivePrefix = "arXiv",
   eprint = {0901.0104},
 primaryClass = "astro-ph.GA",
   editor = {{de Avillez}, M.~A.},
    month = sep,
    pages = {367-370},
      doi = {10.1051/eas/1256060},
   adsurl = {http://adsabs.harvard.edu/abs/2012EAS....56..367H}
}

@inproceedings{2012EAS....56..363H,
	Adsurl = {http://adsabs.harvard.edu/abs/2012EAS....56..363H},
	Author = {{Hanasz}, M. and {Kowalik}, K. and {W{\'o}lta{\'n}ski}, D. and {Paw{\l}aszek}, R.},
	Booktitle = {EAS Publications Series},
	Doi = {10.1051/eas/1256059},
	Editor = {{de Avillez}, M.~A.},
	Month = sep,
	Pages = {363-366},
	Series = {EAS Publications Series},
	Title = {{PIERNIK MHD code - a multi-fluid, non-ideal extension of the relaxing-TVD scheme (III)}},
	Volume = 56,
	Year = 2012}

@inproceedings{2010EAS....42..281H,
	Adsurl = {http://adsabs.harvard.edu/abs/2010EAS....42..281H},
	Archiveprefix = {arXiv},
	Author = {{Hanasz}, M. and {Kowalik}, K. and {W{\'o}lta{\'n}ski}, D. and {Paw{\l}aszek}, R. and {Kornet}, K.},
	Booktitle = {EAS Publications Series},
	Doi = {10.1051/eas/1042030},
	Editor = {{K.~Go{\'z}dziewski, A.~Niedzielski, \& J.~Schneider}},
	Eprint = {0812.2799},
	Month = apr,
	Pages = {281-285},
	Series = {EAS Publications Series},
	Title = {{The PIERNIK MHD code - a multi-fluid, non-ideal extension of the relaxing-TVD scheme (II)}},
	Volume = 42,
	Year = 2010}

@inproceedings{2010EAS....42..275H,
	Adsurl = {http://adsabs.harvard.edu/abs/2010EAS....42..275H},
	Archiveprefix = {arXiv},
	Author = {{Hanasz}, M. and {Kowalik}, K. and {W{\'o}lta{\'n}ski}, D. and {Paw{\l}aszek}, R.},
	Booktitle = {EAS Publications Series},
	Doi = {10.1051/eas/1042029},
	Editor = {{K.~Go{\'z}dziewski, A.~Niedzielski, \& J.~Schneider}},
	Eprint = {0812.2161},
	Month = apr,
	Pages = {275-280},
	Series = {EAS Publications Series},
	Title = {{The PIERNIK MHD code - a multi-fluid, non-ideal extension of the relaxing-TVD scheme (I)}},
	Volume = 42,
	Year = 2010}

@ARTICLE{2004ApJ...605L..33H,
   author = {{Hanasz}, M. and {Kowal}, G. and {Otmianowska-Mazur}, K. and
	{Lesch}, H.},
    title = "{Amplification of Galactic Magnetic Fields by the Cosmic-Ray-driven Dynamo}",
  journal = {\apjl},
   eprint = {arXiv:astro-ph/0402662},
     year = 2004,
    month = apr,
   volume = 605,
    pages = {L33-L36},
      doi = {10.1086/420697},
   adsurl = {http://adsabs.harvard.edu/abs/2004ApJ...605L..33H}
}

@ARTICLE{2003A&A...412..331H,
   author = {{Hanasz}, M. and {Lesch}, H.},
    title = "{Incorporation of cosmic ray transport into the ZEUS MHD code. Application for studies of Parker instability in the ISM}",
  journal = {\aap},
   eprint = {arXiv:astro-ph/0309660},
     year = 2003,
    month = dec,
   volume = 412,
    pages = {331-339},
      doi = {10.1051/0004-6361:20031433},
   adsurl = {http://adsabs.harvard.edu/abs/2003A%26A...412..331H}
}

@BOOK{2011hea..book.....L,
   author = {{Longair}, M.~S.},
    title = "{High Energy Astrophysics}",
booktitle = {High Energy Astrophysics},
  PUBLISHER = {Cambridge, UK: Cambridge University Press},
     year = 2011,
    month = feb,
   adsurl = {http://adsabs.harvard.edu/abs/2011hea..book.....L}
}

@ARTICLE{1998ApJ...497..759F,
   author = {{Ferriere}, K.},
    title = "{Global Model of the Interstellar Medium in Our Galaxy with New Constraints on the Hot Gas Component}",
  journal = {\apj},
     year = 1998,
    month = apr,
   volume = 497,
    pages = {759-+},
      doi = {10.1086/305469},
   adsurl = {http://adsabs.harvard.edu/abs/1998ApJ...497..759F}
}

@ARTICLE{2007ARNPS..57..285S,
   author = {{Strong}, A.~W. and {Moskalenko}, I.~V. and {Ptuskin}, V.~S.
	},
    title = "{Cosmic-Ray Propagation and Interactions in the Galaxy}",
  journal = {Annual Review of Nuclear and Particle Science},
   eprint = {arXiv:astro-ph/0701517},
     year = 2007,
    month = nov,
   volume = 57,
    pages = {285-327},
      doi = {10.1146/annurev.nucl.57.090506.123011},
   adsurl = {http://adsabs.harvard.edu/abs/2007ARNPS..57..285S}
}

@ARTICLE{1994A&A...286..983M ,
   author = { {Mannheim}, K. and {Schlickeiser}, R. },
    title = "{Interactions of cosmic ray nuclei }",
  journal = {
    Astronomy and Astrophysics},
archivePrefix = "arXiv",
 primaryClass = "astro-ph.IM",
     year = 1994,
    month = jan,
   volume = 286,
    pages = {983-996},
   adsurl = {https://articles.adsabs.harvard.edu/pdf/1994A%26A...286..983M }
}

@ARTICLE{2011CoPhC.182.1156V,
   author = { {Vladimirov}, A. E. and {Digel}, S. W. and {Jóhannesson}, G. and {Michelson}, P. F. and {Moskalenko}, I. V. and {Nolan}, P. L. and {Orlando}, E. and {Porter}, T. A. and {Strong}, A. W. },
    title = "{GALPROP WebRun: an internet-based service for calculating galactic cosmic ray propagation and associated photon emissions}",
  journal = {Computer Physics Communications},
archivePrefix = "arXiv",
 primaryClass = "astro-ph.IM",
     year = 2011,
    month = may,
      doi = "10.1016/j.cpc.2011.01.017",
   volume = 182,
    pages = {1156-1161},
   adsurl = {https://ui.adsabs.harvard.edu/abs/2011CoPhC.182.1156V/abstract},
}

@ARTICLE{2023A&ARv..31....4R ,
   author = {  {Ruszkowski}, M. and {Pfrommer}, C. },
    title = "{Cosmic ray feedback in galaxies and galaxy clusters }",
  journal = {The Astronomy and Astrophysics Review},
archivePrefix = "arXiv",
 primaryClass = "astro-ph.IM",
     year = 2023,
    month = dec,
      doi = "10.1007/s00159-023-00149-2 ",
   volume = "31",
   issue = "1",
   eid = "4",
   adsurl = {https://ui.adsabs.harvard.edu/abs/2023A%26ARv..31....4R/abstract},
}

@ARTICLE{2023ecrs.confE.138B,
   author = { {Baldacchino-Jordan}, A. and {Hanasz}, M. and {Ogrodnik}, M. and {Wóltański}, D. and {Gawryszczak}, A. },
    title = "{ Propagation of CR secondary species and gamma ray emission in MHD simulations of galaxies }",
  journal = {ECRS 2022},
 primaryClass = "astro-ph.IM",
     year = 2023,
    month = feb,
      doi = "10.22323/1.423.0138",
   adsurl = {https://ui.adsabs.harvard.edu/abs/2023ecrs.confE.138B/abstract},
}

@ARTICLE{2014A&A...569A..32M,
       author = {{Maurin}, D. and {Melot}, F. and {Taillet}, R.},
        title = "{A database of charged cosmic rays}",
      journal = {\aap},
         year = 2014,
        month = sep,
       volume = {569},
          eid = {A32},
        pages = {A32},
          doi = {10.1051/0004-6361/201321344},
archivePrefix = {arXiv},
       eprint = {1302.5525},
 primaryClass = {astro-ph.HE},
       adsurl = {https://ui.adsabs.harvard.edu/abs/2014A&A...569A..32M}
}

@ARTICLE{2020Univ....6..102M,
       author = {{Maurin}, David and {Dembinski}, Hans Peter and {Gonzalez}, Javier and {Mari{\c{s}}}, Ioana Codrina and {Melot}, Fr{\'e}d{\'e}ric},
        title = "{Cosmic-Ray Database Update: Ultra-High Energy, Ultra-Heavy, and Antinuclei Cosmic-Ray Data (CRDB v4.0)}",
      journal = {Universe},
         year = 2020,
        month = jul,
       volume = {6},
       number = {8},
          eid = {102},
        pages = {102},
          doi = {10.3390/universe6080102},
archivePrefix = {arXiv},
       eprint = {2005.14663},
 primaryClass = {astro-ph.HE},
       adsurl = {https://ui.adsabs.harvard.edu/abs/2020Univ....6..102M}
}

@ARTICLE{2023EPJC...83..971M,
       author = {{Maurin}, David and {Ahlers}, Markus and {Dembinski}, Hans and {Haungs}, Andreas and {Mangeard}, Pierre-Simon and {Melot}, Fr{\'e}d{\'e}ric and {Mertsch}, Philipp and {Wochele}, Doris and {Wochele}, J{\"u}rgen},
        title = "{A cosmic-ray database update: CRDB v4.1}",
      journal = {European Physical Journal C},
         year = 2023,
        month = oct,
       volume = {83},
       number = {10},
          eid = {971},
        pages = {971},
          doi = {10.1140/epjc/s10052-023-12092-8},
archivePrefix = {arXiv},
       eprint = {2306.08901},
 primaryClass = {astro-ph.HE},
       adsurl = {https://ui.adsabs.harvard.edu/abs/2023EPJC...83..971M}
}

@ARTICLE{2016PhRvL.117i1103A,
       author = {{Aguilar}, M. and {Ali Cavasonza}, L. and {Alpat}, B. and {Ambrosi}, G. and {Arruda}, L. and {Attig}, N. and {Aupetit}, S. and {Azzarello}, P. and {Bachlechner}, A. and {Barao}, F. and {Barrau}, A. and {Barrin}, L. and {Bartoloni}, A. and {Basara}, L. and {Ba{\c{s}}e{\c{C}}{\textsection}mez-du Pree}, S. and {Battarbee}, M. and {Battiston}, R. and {Bazo}, J. and {Becker}, U. and {Behlmann}, M. and {Beischer}, B. and {Berdugo}, J. and {Bertucci}, B. and {Bindi}, V. and {Boella}, G. and {de Boer}, W. and {Bollweg}, K. and {Bonnivard}, V. and {Borgia}, B. and {Boschini}, M.~J. and {Bourquin}, M. and {Bueno}, E.~F. and {Burger}, J. and {Cadoux}, F. and {Cai}, X.~D. and {Capell}, M. and {Caroff}, S. and {Casaus}, J. and {Castellini}, G. and {Cernuda}, I. and {Cervelli}, F. and {Chae}, M.~J. and {Chang}, Y.~H. and {Chen}, A.~I. and {Chen}, G.~M. and {Chen}, H.~S. and {Cheng}, L. and {Chou}, H.~Y. and {Choumilov}, E. and {Choutko}, V. and {Chung}, C.~H. and {Clark}, C. and {Clavero}, R. and {Coignet}, G. and {Consolandi}, C. and {Contin}, A. and {Corti}, C. and {Coste}, B. and {Creus}, W. and {Crispoltoni}, M. and {Cui}, Z. and {Dai}, Y.~M. and {Delgado}, C. and {Della Torre}, S. and {Demirk{\"o}z}, M.~B. and {Derome}, L. and {Di Falco}, S. and {Dimiccoli}, F. and {D{\'\i}az}, C. and {von Doetinchem}, P. and {Dong}, F. and {Donnini}, F. and {Duranti}, M. and {D'Urso}, D. and {Egorov}, A. and {Eline}, A. and {Eronen}, T. and {Feng}, J. and {Fiandrini}, E. and {Finch}, E. and {Fisher}, P. and {Formato}, V. and {Galaktionov}, Y. and {Gallucci}, G. and {Garc{\'\i}a}, B. and {Garc{\'\i}a-L{\'o}pez}, R.~J. and {Gargiulo}, C. and {Gast}, H. and {Gebauer}, I. and {Gervasi}, M. and {Ghelfi}, A. and {Giovacchini}, F. and {Goglov}, P. and {G{\'o}mez-Coral}, D.~M. and {Gong}, J. and {Goy}, C. and {Grabski}, V. and {Grandi}, D. and {Graziani}, M. and {Guerri}, I. and {Guo}, K.~H. and {Habiby}, M. and {Haino}, S. and {Han}, K.~C. and {He}, Z.~H. and {Heil}, M. and {Hoffman}, J. and {Hsieh}, T.~H. and {Huang}, H. and {Huang}, Z.~C. and {Huh}, C. and {Incagli}, M. and {Ionica}, M. and {Jang}, W.~Y. and {Jinchi}, H. and {Kang}, S.~C. and {Kanishev}, K. and {Kim}, G.~N. and {Kim}, K.~S. and {Kirn}, Th. and {Konak}, C. and {Kounina}, O. and {Kounine}, A. and {Koutsenko}, V. and {Krafczyk}, M.~S. and {La Vacca}, G. and {Laudi}, E. and {Laurenti}, G. and {Lazzizzera}, I. and {Lebedev}, A. and {Lee}, H.~T. and {Lee}, S.~C. and {Leluc}, C. and {Li}, H.~S. and {Li}, J.~Q. and {Li}, J.~Q. and {Li}, Q. and {Li}, T.~X. and {Li}, W. and {Li}, Z.~H. and {Li}, Z.~Y. and {Lim}, S. and {Lin}, C.~H. and {Lipari}, P. and {Lippert}, T. and {Liu}, D. and {Liu}, Hu and {Lu}, S.~Q. and {Lu}, Y.~S. and {Luebelsmeyer}, K. and {Luo}, F. and {Luo}, J.~Z. and {Lv}, S.~S. and {Majka}, R. and {Ma{\~n}{\'a}}, C. and {Mar{\'\i}n}, J. and {Martin}, T. and {Mart{\'\i}nez}, G. and {Masi}, N. and {Maurin}, D. and {Menchaca-Rocha}, A. and {Meng}, Q. and {Mo}, D.~C. and {Morescalchi}, L. and {Mott}, P. and {Nelson}, T. and {Ni}, J.~Q. and {Nikonov}, N. and {Nozzoli}, F. and {Nunes}, P. and {Oliva}, A. and {Orcinha}, M. and {Palmonari}, F. and {Palomares}, C. and {Paniccia}, M. and {Pauluzzi}, M. and {Pensotti}, S. and {Pereira}, R. and {Picot-Clemente}, N. and {Pilo}, F. and {Pizzolotto}, C. and {Plyaskin}, V. and {Pohl}, M. and {Poireau}, V. and {Putze}, A. and {Quadrani}, L. and {Qi}, X.~M. and {Qin}, X. and {Qu}, Z.~Y. and {R{\"a}ih{\"a}}, T. and {Rancoita}, P.~G. and {Rapin}, D. and {Ricol}, J.~S. and {Rodr{\'\i}guez}, I. and {Rosier-Lees}, S. and {Rozhkov}, A. and {Rozza}, D. and {Sagdeev}, R. and {Sandweiss}, J. and {Saouter}, P.},
        title = "{Antiproton Flux, Antiproton-to-Proton Flux Ratio, and Properties of Elementary Particle Fluxes in Primary Cosmic Rays Measured with the Alpha Magnetic Spectrometer on the International Space Station}",
      journal = {\prl},
         year = 2016,
        month = aug,
       volume = {117},
       number = {9},
          eid = {091103},
        pages = {091103},
          doi = {10.1103/PhysRevLett.117.091103},
       adsurl = {https://ui.adsabs.harvard.edu/abs/2016PhRvL.117i1103A}
}

@ARTICLE{2016PhRvL.117w1102A,
       author = {{Aguilar}, M. and {Ali Cavasonza}, L. and {Ambrosi}, G. and {Arruda}, L. and {Attig}, N. and {Aupetit}, S. and {Azzarello}, P. and {Bachlechner}, A. and {Barao}, F. and {Barrau}, A. and {Barrin}, L. and {Bartoloni}, A. and {Basara}, L. and {Ba{\c{s}}e{\v{g}}mez-du Pree}, S. and {Battarbee}, M. and {Battiston}, R. and {Becker}, U. and {Behlmann}, M. and {Beischer}, B. and {Berdugo}, J. and {Bertucci}, B. and {Bindel}, K.~F. and {Bindi}, V. and {Boella}, G. and {de Boer}, W. and {Bollweg}, K. and {Bonnivard}, V. and {Borgia}, B. and {Boschini}, M.~J. and {Bourquin}, M. and {Bueno}, E.~F. and {Burger}, J. and {Cadoux}, F. and {Cai}, X.~D. and {Capell}, M. and {Caroff}, S. and {Casaus}, J. and {Castellini}, G. and {Cervelli}, F. and {Chae}, M.~J. and {Chang}, Y.~H. and {Chen}, A.~I. and {Chen}, G.~M. and {Chen}, H.~S. and {Cheng}, L. and {Chou}, H.~Y. and {Choumilov}, E. and {Choutko}, V. and {Chung}, C.~H. and {Clark}, C. and {Clavero}, R. and {Coignet}, G. and {Consolandi}, C. and {Contin}, A. and {Corti}, C. and {Creus}, W. and {Crispoltoni}, M. and {Cui}, Z. and {Dai}, Y.~M. and {Delgado}, C. and {Della Torre}, S. and {Demakov}, O. and {Demirk{\"o}z}, M.~B. and {Derome}, L. and {Di Falco}, S. and {Dimiccoli}, F. and {D{\'\i}az}, C. and {von Doetinchem}, P. and {Dong}, F. and {Donnini}, F. and {Duranti}, M. and {D'Urso}, D. and {Egorov}, A. and {Eline}, A. and {Eronen}, T. and {Feng}, J. and {Fiandrini}, E. and {Finch}, E. and {Fisher}, P. and {Formato}, V. and {Galaktionov}, Y. and {Gallucci}, G. and {Garc{\'\i}a}, B. and {Garc{\'\i}a-L{\'o}pez}, R.~J. and {Gargiulo}, C. and {Gast}, H. and {Gebauer}, I. and {Gervasi}, M. and {Ghelfi}, A. and {Giovacchini}, F. and {Goglov}, P. and {G{\'o}mez-Coral}, D.~M. and {Gong}, J. and {Goy}, C. and {Grabski}, V. and {Grandi}, D. and {Graziani}, M. and {Guo}, K.~H. and {Haino}, S. and {Han}, K.~C. and {He}, Z.~H. and {Heil}, M. and {Hoffman}, J. and {Hsieh}, T.~H. and {Huang}, H. and {Huang}, Z.~C. and {Huh}, C. and {Incagli}, M. and {Ionica}, M. and {Jang}, W.~Y. and {Jinchi}, H. and {Kang}, S.~C. and {Kanishev}, K. and {Kim}, G.~N. and {Kim}, K.~S. and {Kirn}, Th. and {Konak}, C. and {Kounina}, O. and {Kounine}, A. and {Koutsenko}, V. and {Krafczyk}, M.~S. and {La Vacca}, G. and {Laudi}, E. and {Laurenti}, G. and {Lazzizzera}, I. and {Lebedev}, A. and {Lee}, H.~T. and {Lee}, S.~C. and {Leluc}, C. and {Li}, H.~S. and {Li}, J.~Q. and {Li}, J.~Q. and {Li}, Q. and {Li}, T.~X. and {Li}, W. and {Li}, Y. and {Li}, Z.~H. and {Li}, Z.~Y. and {Lim}, S. and {Lin}, C.~H. and {Lipari}, P. and {Lippert}, T. and {Liu}, D. and {Liu}, Hu and {Lordello}, V.~D. and {Lu}, S.~Q. and {Lu}, Y.~S. and {Luebelsmeyer}, K. and {Luo}, F. and {Luo}, J.~Z. and {Lv}, S.~S. and {Machate}, F. and {Majka}, R. and {Ma{\~n}{\'a}}, C. and {Mar{\'\i}n}, J. and {Martin}, T. and {Mart{\'\i}nez}, G. and {Masi}, N. and {Maurin}, D. and {Menchaca-Rocha}, A. and {Meng}, Q. and {Mikuni}, V.~M. and {Mo}, D.~C. and {Morescalchi}, L. and {Mott}, P. and {Nelson}, T. and {Ni}, J.~Q. and {Nikonov}, N. and {Nozzoli}, F. and {Oliva}, A. and {Orcinha}, M. and {Palmonari}, F. and {Palomares}, C. and {Paniccia}, M. and {Pauluzzi}, M. and {Pensotti}, S. and {Pereira}, R. and {Picot-Clemente}, N. and {Pilo}, F. and {Pizzolotto}, C. and {Plyaskin}, V. and {Pohl}, M. and {Poireau}, V. and {Putze}, A. and {Quadrani}, L. and {Qi}, X.~M. and {Qin}, X. and {Qu}, Z.~Y. and {R{\"a}ih{\"a}}, T. and {Rancoita}, P.~G. and {Rapin}, D. and {Ricol}, J.~S. and {Rosier-Lees}, S. and {Rozhkov}, A. and {Rozza}, D. and {Sagdeev}, R. and {Sandweiss}, J. and {Saouter}, P. and {Schael}, S. and {Schmidt}, S.~M.},
        title = "{Precision Measurement of the Boron to Carbon Flux Ratio in Cosmic Rays from 1.9 GV to 2.6 TV with the Alpha Magnetic Spectrometer on the International Space Station}",
      journal = {\prl},
         year = 2016,
        month = dec,
       volume = {117},
       number = {23},
          eid = {231102},
        pages = {231102},
          doi = {10.1103/PhysRevLett.117.231102},
       adsurl = {https://ui.adsabs.harvard.edu/abs/2016PhRvL.117w1102A}
}

@ARTICLE{2018PhRvL.120b1101A,
       author = {{Aguilar}, M. and {Ali Cavasonza}, L. and {Ambrosi}, G. and {Arruda}, L. and {Attig}, N. and {Aupetit}, S. and {Azzarello}, P. and {Bachlechner}, A. and {Barao}, F. and {Barrau}, A. and {Barrin}, L. and {Bartoloni}, A. and {Basara}, L. and {Ba{\c{s}}e{\v{g}}mez-du Pree}, S. and {Battarbee}, M. and {Battiston}, R. and {Becker}, U. and {Behlmann}, M. and {Beischer}, B. and {Berdugo}, J. and {Bertucci}, B. and {Bindel}, K.~F. and {Bindi}, V. and {de Boer}, W. and {Bollweg}, K. and {Bonnivard}, V. and {Borgia}, B. and {Boschini}, M.~J. and {Bourquin}, M. and {Bueno}, E.~F. and {Burger}, J. and {Burger}, W.~J. and {Cadoux}, F. and {Cai}, X.~D. and {Capell}, M. and {Caroff}, S. and {Casaus}, J. and {Castellini}, G. and {Cervelli}, F. and {Chae}, M.~J. and {Chang}, Y.~H. and {Chen}, A.~I. and {Chen}, G.~M. and {Chen}, H.~S. and {Cheng}, L. and {Chou}, H.~Y. and {Choumilov}, E. and {Choutko}, V. and {Chung}, C.~H. and {Clark}, C. and {Clavero}, R. and {Coignet}, G. and {Consolandi}, C. and {Contin}, A. and {Corti}, C. and {Creus}, W. and {Crispoltoni}, M. and {Cui}, Z. and {Dadzie}, K. and {Dai}, Y.~M. and {Datta}, A. and {Delgado}, C. and {Della Torre}, S. and {Demirk{\"o}z}, M.~B. and {Derome}, L. and {Di Falco}, S. and {Dimiccoli}, F. and {D{\'\i}az}, C. and {von Doetinchem}, P. and {Dong}, F. and {Donnini}, F. and {Duranti}, M. and {D'Urso}, D. and {Egorov}, A. and {Eline}, A. and {Eronen}, T. and {Feng}, J. and {Fiandrini}, E. and {Fisher}, P. and {Formato}, V. and {Galaktionov}, Y. and {Gallucci}, G. and {Garc{\'\i}a-L{\'o}pez}, R.~J. and {Gargiulo}, C. and {Gast}, H. and {Gebauer}, I. and {Gervasi}, M. and {Ghelfi}, A. and {Giovacchini}, F. and {G{\'o}mez-Coral}, D.~M. and {Gong}, J. and {Goy}, C. and {Grabski}, V. and {Grandi}, D. and {Graziani}, M. and {Guo}, K.~H. and {Haino}, S. and {Han}, K.~C. and {He}, Z.~H. and {Heil}, M. and {Hsieh}, T.~H. and {Huang}, H. and {Huang}, Z.~C. and {Huh}, C. and {Incagli}, M. and {Ionica}, M. and {Jang}, W.~Y. and {Jia}, Yi and {Jinchi}, H. and {Kang}, S.~C. and {Kanishev}, K. and {Khiali}, B. and {Kim}, G.~N. and {Kim}, K.~S. and {Kirn}, Th. and {Konak}, C. and {Kounina}, O. and {Kounine}, A. and {Koutsenko}, V. and {Kulemzin}, A. and {La Vacca}, G. and {Laudi}, E. and {Laurenti}, G. and {Lazzizzera}, I. and {Lebedev}, A. and {Lee}, H.~T. and {Lee}, S.~C. and {Leluc}, C. and {Li}, H.~S. and {Li}, J.~Q. and {Li}, Q. and {Li}, T.~X. and {Li}, Y. and {Li}, Z.~H. and {Li}, Z.~Y. and {Lim}, S. and {Lin}, C.~H. and {Lipari}, P. and {Lippert}, T. and {Liu}, D. and {Liu}, Hu and {Lordello}, V.~D. and {Lu}, S.~Q. and {Lu}, Y.~S. and {Luebelsmeyer}, K. and {Luo}, F. and {Luo}, J.~Z. and {Lyu}, S.~S. and {Machate}, F. and {Ma{\~n}{\'a}}, C. and {Mar{\'\i}n}, J. and {Martin}, T. and {Mart{\'\i}nez}, G. and {Masi}, N. and {Maurin}, D. and {Menchaca-Rocha}, A. and {Meng}, Q. and {Mikuni}, V.~M. and {Mo}, D.~C. and {Mott}, P. and {Nelson}, T. and {Ni}, J.~Q. and {Nikonov}, N. and {Nozzoli}, F. and {Oliva}, A. and {Orcinha}, M. and {Palermo}, M. and {Palmonari}, F. and {Palomares}, C. and {Paniccia}, M. and {Pauluzzi}, M. and {Pensotti}, S. and {Perrina}, C. and {Phan}, H.~D. and {Picot-Clemente}, N. and {Pilo}, F. and {Pizzolotto}, C. and {Plyaskin}, V. and {Pohl}, M. and {Poireau}, V. and {Quadrani}, L. and {Qi}, X.~M. and {Qin}, X. and {Qu}, Z.~Y. and {R{\"a}ih{\"a}}, T. and {Rancoita}, P.~G. and {Rapin}, D. and {Ricol}, J.~S. and {Rosier-Lees}, S. and {Rozhkov}, A. and {Rozza}, D. and {Sagdeev}, R. and {Schael}, S. and {Schmidt}, S.~M. and {Schulz von Dratzig}, A. and {Schwering}, G. and {Seo}, E.~S. and {Shan}, B.~S. and {Shi}, J.~Y. and {Siedenburg}, T.},
        title = "{Observation of New Properties of Secondary Cosmic Rays Lithium, Beryllium, and Boron by the Alpha Magnetic Spectrometer on the International Space Station}",
      journal = {\prl},
         year = 2018,
        month = jan,
       volume = {120},
       number = {2},
          eid = {021101},
        pages = {021101},
          doi = {10.1103/PhysRevLett.120.021101},
       adsurl = {https://ui.adsabs.harvard.edu/abs/2018PhRvL.120b1101A}
}

@ARTICLE{2018PhRvL.121e1102A,
       author = {{Aguilar}, M. and {Cavasonza}, L. Ali and {Ambrosi}, G. and {Arruda}, L. and {Attig}, N. and {Aupetit}, S. and {Azzarello}, P. and {Bachlechner}, A. and {Barao}, F. and {Barrau}, A. and {Barrin}, L. and {Bartoloni}, A. and {Basara}, L. and {Ba{\c{s}}e{\v{g}}mez-du Pree}, S. and {Battarbee}, M. and {Battiston}, R. and {Becker}, U. and {Behlmann}, M. and {Beischer}, B. and {Berdugo}, J. and {Bertucci}, B. and {Bindel}, K.~F. and {Bindi}, V. and {de Boer}, W. and {Bollweg}, K. and {Bonnivard}, V. and {Borgia}, B. and {Boschini}, M.~J. and {Bourquin}, M. and {Bueno}, E.~F. and {Burger}, J. and {Cadoux}, F. and {Cai}, X.~D. and {Capell}, M. and {Caroff}, S. and {Casaus}, J. and {Castellini}, G. and {Cervelli}, F. and {Chae}, M.~J. and {Chang}, Y.~H. and {Chen}, A.~I. and {Chen}, G.~M. and {Chen}, H.~S. and {Chen}, Y. and {Cheng}, L. and {Chou}, H.~Y. and {Choumilov}, E. and {Choutko}, V. and {Chung}, C.~H. and {Clark}, C. and {Clavero}, R. and {Coignet}, G. and {Consolandi}, C. and {Contin}, A. and {Corti}, C. and {Creus}, W. and {Crispoltoni}, M. and {Cui}, Z. and {Dadzie}, K. and {Dai}, Y.~M. and {Datta}, A. and {Delgado}, C. and {Della Torre}, S. and {Demirk{\"o}z}, M.~B. and {Derome}, L. and {Di Falco}, S. and {Dimiccoli}, F. and {D{\'\i}az}, C. and {von Doetinchem}, P. and {Dong}, F. and {Donnini}, F. and {Duranti}, M. and {D'Urso}, D. and {Egorov}, A. and {Eline}, A. and {Eronen}, T. and {Feng}, J. and {Fiandrini}, E. and {Fisher}, P. and {Formato}, V. and {Galaktionov}, Y. and {Gallucci}, G. and {Garc{\'\i}a-L{\'o}pez}, R.~J. and {Gargiulo}, C. and {Gast}, H. and {Gebauer}, I. and {Gervasi}, M. and {Ghelfi}, A. and {Giovacchini}, F. and {G{\'o}mez-Coral}, D.~M. and {Gong}, J. and {Goy}, C. and {Grabski}, V. and {Grandi}, D. and {Graziani}, M. and {Guo}, K.~H. and {Haino}, S. and {Han}, K.~C. and {He}, Z.~H. and {Heil}, M. and {Hsieh}, T.~H. and {Huang}, H. and {Huang}, Z.~C. and {Huh}, C. and {Incagli}, M. and {Ionica}, M. and {Jang}, W.~Y. and {Jia}, Yi and {Jinchi}, H. and {Kang}, S.~C. and {Kanishev}, K. and {Khiali}, B. and {Kim}, G.~N. and {Kim}, K.~S. and {Kirn}, Th. and {Konak}, C. and {Kounina}, O. and {Kounine}, A. and {Koutsenko}, V. and {Kulemzin}, A. and {La Vacca}, G. and {Laudi}, E. and {Laurenti}, G. and {Lazzizzera}, I. and {Lebedev}, A. and {Lee}, H.~T. and {Lee}, S.~C. and {Leluc}, C. and {Li}, H.~S. and {Li}, J.~Q. and {Li}, Q. and {Li}, T.~X. and {Li}, Z.~H. and {Li}, Z.~Y. and {Lim}, S. and {Lin}, C.~H. and {Lipari}, P. and {Lippert}, T. and {Liu}, D. and {Liu}, Hu and {Lordello}, V.~D. and {Lu}, S.~Q. and {Lu}, Y.~S. and {Luebelsmeyer}, K. and {Luo}, F. and {Luo}, J.~Z. and {Lyu}, S.~S. and {Machate}, F. and {Ma{\~n}{\'a}}, C. and {Mar{\'\i}n}, J. and {Martin}, T. and {Mart{\'\i}nez}, G. and {Masi}, N. and {Maurin}, D. and {Menchaca-Rocha}, A. and {Meng}, Q. and {Mikuni}, V.~M. and {Mo}, D.~C. and {Mott}, P. and {Nelson}, T. and {Ni}, J.~Q. and {Nikonov}, N. and {Nozzoli}, F. and {Oliva}, A. and {Orcinha}, M. and {Palermo}, M. and {Palmonari}, F. and {Palomares}, C. and {Paniccia}, M. and {Pauluzzi}, M. and {Pensotti}, S. and {Perrina}, C. and {Phan}, H.~D. and {Picot-Clemente}, N. and {Pilo}, F. and {Pizzolotto}, C. and {Plyaskin}, V. and {Pohl}, M. and {Poireau}, V. and {Quadrani}, L. and {Qi}, X.~M. and {Qin}, X. and {Qu}, Z.~Y. and {R{\"a}ih{\"a}}, T. and {Rancoita}, P.~G. and {Rapin}, D. and {Ricol}, J.~S. and {Rosier-Lees}, S. and {Rozhkov}, A. and {Rozza}, D. and {Sagdeev}, R. and {Schael}, S. and {Schmidt}, S.~M. and {von Dratzig}, A. Schulz and {Schwering}, G. and {Seo}, E.~S. and {Shan}, B.~S. and {Shi}, J.~Y. and {Siedenburg}, T. and {Son}, D.},
        title = "{Observation of Complex Time Structures in the Cosmic-Ray Electron and Positron Fluxes with the Alpha Magnetic Spectrometer on the International Space Station}",
      journal = {\prl},
         year = 2018,
        month = aug,
       volume = {121},
       number = {5},
          eid = {051102},
        pages = {051102},
          doi = {10.1103/PhysRevLett.121.051102},
       adsurl = {https://ui.adsabs.harvard.edu/abs/2018PhRvL.121e1102A}
}

@ARTICLE{2023APh...14402776T,
       author = {{Thaler}, J. and {Kissmann}, R. and {Reimer}, O.},
        title = "{Cosmic-ray propagation under consideration of a spatially resolved source distribution}",
      journal = {Astroparticle Physics},
         year = 2023,
        month = jan,
       volume = {144},
          eid = {102776},
        pages = {102776},
          doi = {10.1016/j.astropartphys.2022.102776},
archivePrefix = {arXiv},
       eprint = {2209.02295},
 primaryClass = {astro-ph.HE},
       adsurl = {https://ui.adsabs.harvard.edu/abs/2023APh...14402776T}
}

@ARTICLE{2024MNRAS.52710897G,
       author = {{Girichidis}, Philipp and {Werhahn}, Maria and {Pfrommer}, Christoph and {Pakmor}, R{\"u}diger and {Springel}, Volker},
        title = "{Spectrally resolved cosmic rays - III. Dynamical impact and properties of the circumgalactic medium}",
      journal = {\mnras},
         year = 2024,
        month = feb,
       volume = {527},
       number = {4},
        pages = {10897-10920},
          doi = {10.1093/mnras/stad3628},
archivePrefix = {arXiv},
       eprint = {2303.03417},
 primaryClass = {astro-ph.GA},
       adsurl = {https://ui.adsabs.harvard.edu/abs/2024MNRAS.52710897G}
}

@ARTICLE{2023MNRAS.526..160J,
       author = {{Jacobs}, Hanno and {Mertsch}, Philipp and {Phan}, Vo Hong Minh},
        title = "{Unstable cosmic ray nuclei constrain low-diffusion zones in the Galactic disc}",
      journal = {\mnras},
         year = 2023,
        month = nov,
       volume = {526},
       number = {1},
        pages = {160-174},
          doi = {10.1093/mnras/stad2719},
archivePrefix = {arXiv},
       eprint = {2305.10337},
 primaryClass = {astro-ph.HE},
       adsurl = {https://ui.adsabs.harvard.edu/abs/2023MNRAS.526..160J}
}

@BOOK{2016crpp.book.....G,
   author = { {Gaisser}, Thomas K. and {Engel}, Ralph and {Resconi}, Elisa },
    title = "{ Cosmic Rays and Particle Physics }",
booktitle = { Cosmic Rays and Particle Physics },
  PUBLISHER = {Cambridge, UK: Cambridge University Press},
     year = 2016,
    month = jun,
   adsurl = {http://adsabs.harvard.edu/abs/2011hea..book.....L}
}

@ARTICLE{2019PhRvD..99j3023E,
       author = {{Evoli}, Carmelo and {Aloisio}, Roberto and {Blasi}, Pasquale},
        title = "{Galactic cosmic rays after the AMS-02 observations}",
      journal = {\prd},
         year = 2019,
        month = may,
       volume = {99},
       number = {10},
          eid = {103023},
        pages = {103023},
          doi = {10.1103/PhysRevD.99.103023},
archivePrefix = {arXiv},
       eprint = {1904.10220},
 primaryClass = {astro-ph.HE},
       adsurl = {https://ui.adsabs.harvard.edu/abs/2019PhRvD..99j3023E}
}

@ARTICLE{2020PhRvD.101b3013E,
       author = {{Evoli}, Carmelo and {Morlino}, Giovanni and {Blasi}, Pasquale and {Aloisio}, Roberto},
        title = "{AMS-02 beryllium data and its implication for cosmic ray transport}",
      journal = {\prd},
         year = 2020,
        month = jan,
       volume = {101},
       number = {2},
          eid = {023013},
        pages = {023013},
          doi = {10.1103/PhysRevD.101.023013},
archivePrefix = {arXiv},
       eprint = {1910.04113},
 primaryClass = {astro-ph.HE},
       adsurl = {https://ui.adsabs.harvard.edu/abs/2020PhRvD.101b3013E}
}

@ARTICLE{2018PhRvC..98c4611G,
       author = {{G{\'e}nolini}, Yoann and {Maurin}, David and {Moskalenko}, Igor V. and {Unger}, Michael},
        title = "{Current status and desired precision of the isotopic production cross sections relevant to astrophysics of cosmic rays: Li, Be, B, C, and N}",
      journal = {\prc},
         year = 2018,
        month = sep,
       volume = {98},
       number = {3},
          eid = {034611},
        pages = {034611},
          doi = {10.1103/PhysRevC.98.034611},
archivePrefix = {arXiv},
       eprint = {1803.04686},
 primaryClass = {astro-ph.HE},
       adsurl = {https://ui.adsabs.harvard.edu/abs/2018PhRvC..98c4611G}
}

@ARTICLE{2021ApJ...922...11A,
       author = {{Armillotta}, Lucia and {Ostriker}, Eve C. and {Jiang}, Yan-Fei},
        title = "{Cosmic-Ray Transport in Simulations of Star-forming Galactic Disks}",
      journal = {\apj},
         year = 2021,
        month = nov,
       volume = {922},
       number = {1},
          eid = {11},
        pages = {11},
          doi = {10.3847/1538-4357/ac1db2},
archivePrefix = {arXiv},
       eprint = {2108.09356},
 primaryClass = {astro-ph.HE},
       adsurl = {https://ui.adsabs.harvard.edu/abs/2021ApJ...922...11A}
}

@ARTICLE{2022ApJ...929..170A,
       author = {{Armillotta}, Lucia and {Ostriker}, Eve C. and {Jiang}, Yan-Fei},
        title = "{Cosmic-Ray Transport in Varying Galactic Environments}",
      journal = {\apj},
         year = 2022,
        month = apr,
       volume = {929},
       number = {2},
          eid = {170},
        pages = {170},
          doi = {10.3847/1538-4357/ac5fa9},
archivePrefix = {arXiv},
       eprint = {2203.11949},
 primaryClass = {astro-ph.GA},
       adsurl = {https://ui.adsabs.harvard.edu/abs/2022ApJ...929..170A}
}

@ARTICLE{2024ApJ...964...99A,
       author = {{Armillotta}, Lucia and {Ostriker}, Eve C. and {Kim}, Chang-Goo and {Jiang}, Yan-Fei},
        title = "{Cosmic-Ray Acceleration of Galactic Outflows in Multiphase Gas}",
      journal = {\apj},
         year = 2024,
        month = mar,
       volume = {964},
       number = {1},
          eid = {99},
        pages = {99},
          doi = {10.3847/1538-4357/ad1e5c},
archivePrefix = {arXiv},
       eprint = {2401.04169},
 primaryClass = {astro-ph.GA},
       adsurl = {https://ui.adsabs.harvard.edu/abs/2024ApJ...964...99A}
}

@ARTICLE{2021MNRAS.508.4269P,
       author = {{Peschken}, N. and {Hanasz}, M. and {Naab}, T. and {W{\'o}lta{\'n}ski}, D. and {Gawryszczak}, A.},
        title = "{The angular momentum structure of CR-driven galactic outflows triggered by stream accretion}",
      journal = {\mnras},
         year = 2021,
        month = dec,
       volume = {508},
       number = {3},
        pages = {4269-4281},
          doi = {10.1093/mnras/stab2784},
archivePrefix = {arXiv},
       eprint = {2109.12360},
 primaryClass = {astro-ph.GA},
       adsurl = {https://ui.adsabs.harvard.edu/abs/2021MNRAS.508.4269P}
}

@Article{	  2023mnras.522.5529p,
  author	= {{Peschken}, N. and {Hanasz}, M. and {Naab}, T. and
		  {W{\'o}lta{\'n}ski}, D. and {Gawryszczak}, A.},
  title		= "{The phase structure of cosmic ray driven outflows in
		  stream fed disc galaxies}",
  journal	= {\mnras},
  year		= 2023,
  month		= jul,
  volume	= {522},
  number	= {4},
  pages		= {5529-5545},
  doi		= {10.1093/mnras/stad1358},
  archiveprefix	= {arXiv},
  eprint	= {2210.17328},
  primaryclass	= {astro-ph.GA},
  adsurl	= {https://ui.adsabs.harvard.edu/abs/2023MNRAS.522.5529P}
}

@ARTICLE{1991RSPSA.434....9K,
       author = {{Kolmogorov}, A.~N.},
        title = "{The Local Structure of Turbulence in Incompressible Viscous Fluid for Very Large Reynolds Numbers}",
      journal = {Proceedings of the Royal Society of London Series A},
         year = 1991,
        month = jul,
       volume = {434},
       number = {1890},
        pages = {9-13},
          doi = {10.1098/rspa.1991.0075},
       adsurl = {https://ui.adsabs.harvard.edu/abs/1991RSPSA.434....9K}
}

@ARTICLE{2023arXiv230900298E,
   author = {   {Evoli}, Carmelo. and {Dupletsa}, U. },
    title = "{ Phenomenological models of Cosmic Ray transport in Galaxies  }",
archivePrefix = "arXiv",
 primaryClass = "astro-ph.IM",
     year = 2023,
    month = sep,
      doi = "10.48550/arXiv.2309.00298",
   adsurl = {https://ui.adsabs.harvard.edu/abs/2023arXiv230900298E/abstract},
}

@BOOK{2002cra..book.....S,
   author = { {Schlickeiser}, Reinhard  },
    title = "{ Cosmic Ray astrohysics }",
booktitle = { Cosmic Ray astrophysics },
  PUBLISHER = { Astronomy and Astrophysics Library},
     year = 2002,
   adsurl = {https://ui.adsabs.harvard.edu/abs/2002cra..book.....S/abstract}
}

@ARTICLE{2009ApJS..181..391T,
       author = {{Townsend}, R.~H.~D.},
        title = "{An Exact Integration Scheme for Radiative Cooling in Hydrodynamical Simulations}",
      journal = {\apjs},
         year = 2009,
        month = apr,
       volume = {181},
       number = {2},
        pages = {391-397},
          doi = {10.1088/0067-0049/181/2/391},
archivePrefix = {arXiv},
       eprint = {0901.3146},
 primaryClass = {astro-ph.SR},
       adsurl = {https://ui.adsabs.harvard.edu/abs/2009ApJS..181..391T}
}

@ARTICLE{2013ApJ...770...25A,
       author = {{Agertz}, Oscar and {Kravtsov}, Andrey V. and {Leitner}, Samuel N. and {Gnedin}, Nickolay Y.},
        title = "{Toward a Complete Accounting of Energy and Momentum from Stellar Feedback in Galaxy Formation Simulations}",
      journal = {\apj},
         year = 2013,
        month = jun,
       volume = {770},
       number = {1},
          eid = {25},
        pages = {25},
          doi = {10.1088/0004-637X/770/1/25},
archivePrefix = {arXiv},
       eprint = {1210.4957},
 primaryClass = {astro-ph.CO},
       adsurl = {https://ui.adsabs.harvard.edu/abs/2013ApJ...770...25A}
}

@ARTICLE{2023A&A...670A.158S,
       author = {{Stein}, M. and {Heesen}, V. and {Dettmar}, R. -J. and {Stein}, Y. and {Br{\"u}ggen}, M. and {Beck}, R. and {Adebahr}, B. and {Wiegert}, T. and {Vargas}, C.~J. and {Bomans}, D.~J. and {Li}, J. and {English}, J. and {Chy{\.z}y}, K.~T. and {Paladino}, R. and {Tabatabaei}, F.~S. and {Strong}, A.},
        title = "{CHANG-ES. XXVI. Insights into cosmic-ray transport from radio halos in edge-on galaxies}",
      journal = {\aap},
         year = 2023,
        month = feb,
       volume = {670},
          eid = {A158},
        pages = {A158},
          doi = {10.1051/0004-6361/202243906},
archivePrefix = {arXiv},
       eprint = {2210.07709},
 primaryClass = {astro-ph.GA},
       adsurl = {https://ui.adsabs.harvard.edu/abs/2023A&A...670A.158S}
}

@ARTICLE{1972Phy....58..379G,
       author = {{Gould}, R.~J.},
        title = "{Energy loss of a relativistic ion in a plasma}",
      journal = {Physica},
         year = 1972,
        month = apr,
       volume = {58},
       number = {3},
        pages = {379-383},
          doi = {10.1016/0031-8914(72)90159-0},
       adsurl = {https://ui.adsabs.harvard.edu/abs/1972Phy....58..379G}
}

@ARTICLE{2025arXiv251203385P,
       author = {{Porter}, Troy A. and {Moskalenko}, Igor V. and {Cummings}, Alan C. and {J{\'o}hannesson}, Gu{\dh}laugur},
        title = "{Voyager 1 Data Reveals Signatures of the Local Gas and Cosmic-Ray Source Distributions}",
      journal = {arXiv e-prints},
         year = 2025,
        month = dec,
          eid = {arXiv:2512.03385},
        pages = {arXiv:2512.03385},
          doi = {10.48550/arXiv.2512.03385},
archivePrefix = {arXiv},
       eprint = {2512.03385},
 primaryClass = {astro-ph.HE},
       adsurl = {https://ui.adsabs.harvard.edu/abs/2025arXiv251203385P}
}

@ARTICLE{2020LRCA....6....1M,
       author = {{Marcowith}, Alexandre and {Ferrand}, Gilles and {Grech}, Mickael and {Meliani}, Zakaria and {Plotnikov}, Illya and {Walder}, Rolf},
        title = "{Multi-scale simulations of particle acceleration in astrophysical systems}",
      journal = {Living Reviews in Computational Astrophysics},
         year = 2020,
        month = mar,
       volume = {6},
       number = {1},
          eid = {1},
        pages = {1},
          doi = {10.1007/s41115-020-0007-6},
archivePrefix = {arXiv},
       eprint = {2002.09411},
 primaryClass = {astro-ph.HE},
       adsurl = {https://ui.adsabs.harvard.edu/abs/2020LRCA....6....1M}
}

@ARTICLE{2017A&A...606A..22N,
       author = {{Neronov}, Andrii and {Malyshev}, Denys and {Semikoz}, Dmitri V.},
        title = "{Cosmic-ray spectrum in the local Galaxy}",
      journal = {\aap},
         year = 2017,
        month = sep,
       volume = {606},
          eid = {A22},
        pages = {A22},
          doi = {10.1051/0004-6361/201731149},
archivePrefix = {arXiv},
       eprint = {1705.02200},
 primaryClass = {astro-ph.HE},
       adsurl = {https://ui.adsabs.harvard.edu/abs/2017A&A...606A..22N}
}

@ARTICLE{2015arXiv150507601N,
       author = {{Neronov}, A. and {Malyshev}, D.},
        title = "{Hard spectrum of cosmic rays in the Disks of Milky Way and Large Magellanic Cloud}",
      journal = {arXiv e-prints},
         year = 2015,
        month = may,
          eid = {arXiv:1505.07601},
        pages = {arXiv:1505.07601},
          doi = {10.48550/arXiv.1505.07601},
archivePrefix = {arXiv},
       eprint = {1505.07601},
 primaryClass = {astro-ph.HE},
       adsurl = {https://ui.adsabs.harvard.edu/abs/2015arXiv150507601N}
}
\bibliographystyle{aasjournal}
%
%%%%%%%%%%%%%%%%%%%%%%%%%%%%%%%%%%%%%%%%%%%%%%%%%%%%%%%%%%%%%%%%%%%%%%
\appendix

\section{Detailed numerical method for multiple spectrally-resolved CR species} \label{App.A}
In this appendix, we detail the numerical method that was used to resolve equations (\ref{eq.4}) and (\ref{eq.5}), developed by \cite{2001CoPhC.141...17M} and adapted for PIERNIK in \cite{2021ApJS..253...18O}. The scheme and the algorithm have been adapted to model several spectrally-resolved CR hadronic species and their corresponding energy gain/loss processes in the appropriate momentum range.
\subsection{Numerical scheme.}
\label{App.A1}
We provide a numerical scheme modeling the evolution of CRs in the momentum space in the presence of energy-loss processes and spallation decay of primaries to secondaries. Equations (\ref{eq.5}) and (\ref{eq.7}) only with terms in the momentum space have the following form in a single bin $l$:

\begin{eqnarray}
\frac{\mathrm{d}n^\mathrm{prim}_{l}}{\mathrm{d}t}&=& \left[\left(\frac{1}{3}\nabla\cdot\vec{v} + b^{\mathrm{prim}}_l(p)  \right)4 \pi p^2 f^{\mathrm{prim}}_l(p) \right]^{p_R}_{p_L} - \sum_{\mathrm{sec}} \Gamma^{\mathrm{prim, sec}}_{\mathrm{l}}n^\mathrm{prim}_{l} + Q^\mathrm{prim}_\mathrm{SN} \label{eq.A1} \\  \frac{\mathrm{d}e^\mathrm{prim}_{l}}{\mathrm{d}t} &=& \left[\left(\frac{1}{3}\nabla\cdot\vec{v} + b^{\mathrm{prim}}_l(p)  \right)4 \pi p^2 f^{\mathrm{prim}}_l(p) T^{\mathrm{prim}}_l(p) \right]^{p_R}_{p_L}  - \int^{p_{\mathrm{R}}}_{p_{\mathrm{L}}}4\pi p^2\mathrm{d}p\;f^{\mathrm{prim}}_l(p)b^{\mathrm{prim}}_l(p) \beta_l^\mathrm{prim}(p)c + S^\mathrm{prim}_\mathrm{SN}
\label{eq.A2} \\ && - \sum_{\mathrm{sec}} \Gamma^{\mathrm{prim, sec}}_{\mathrm{l}}e^\mathrm{prim}_{l}\nonumber \\
\frac{\mathrm{d}n^\mathrm{sec}_{l}}{\mathrm{d}t}&=& \left[\left(\frac{1}{3}\nabla\cdot\vec{v} + b^{\mathrm{sec}}_l(p)  \right)4 \pi p^2 f^{\mathrm{sec}}_l(p) \right]^{p_R}_{p_L} - \left<\frac{1}{\gamma^\mathrm{sec}(p)\tau^\mathrm{sec}}\right>_n n^\mathrm{sec}_{l} + \sum_{\mathrm{prim}} (\tilde{Q}^{\mathrm{prim, sec}}_l \Gamma^{\mathrm{prim, sec}}_{l}n^\mathrm{prim}_{l} \label{eq.A3} \\  && + (1 - \tilde{Q}^{\mathrm{prim, sec}}_{l+1} )\Gamma^{\mathrm{prim, sec}}_{l+1}n^\mathrm{prim}_{l+1})\nonumber \\ \frac{\mathrm{d}e^\mathrm{sec}_{l}}{\mathrm{d}t} &=& \left[\left(\frac{1}{3}\nabla\cdot\vec{v} + b^{\mathrm{sec}}_l(p)  \right)4 \pi p^2 f^{\mathrm{sec}}_l(p) T^{\mathrm{sec}}_l(p) \right]^{p_R}_{p_L} - \int^{p_{\mathrm{R}}}_{p_{\mathrm{L}}}4\pi p^2\mathrm{d}p\;f^{\mathrm{sec}}_l(p)b^{\mathrm{sec}}_l(p)\beta_l^\mathrm{prim}(p)c
\label{eq.A4} \\ && - \left<\frac{1}{\gamma^\mathrm{sec}(p)\tau^\mathrm{sec}}\right> _e e^\mathrm{sec}_{l} + \sum_{\mathrm{prim}} (\tilde{S}^{\mathrm{prim, sec}}_l \Gamma^{\mathrm{prim, sec}}_{l}e^\mathrm{prim}_{l} + (1 - \tilde{S}^{\mathrm{prim, sec}}_{l+1} )\Gamma^{\mathrm{prim, sec}}_{l+1}e^\mathrm{prim}_{l+1})   \nonumber \\ \nonumber
\end{eqnarray}

where is included radioactive decay loss for secondaries, spallation loss for primaries and source for secondaries, with $\beta^a_l(p)=p/\sqrt{p^2+m_a^2c^2}$, $\Gamma_l^{\mathrm{prim},\mathrm{sec}}=n_{\mathrm{ISM}}\sigma^{\mathrm{prim}}_{\mathrm{sec}} \beta_l^\mathrm{prim}(p) c$ that are the reaction rates depending on the ISM gas density, the cross sections and the speed of primaries, with $\beta_l^\mathrm{prim} \approx 1$ but changes significantly for $p \leq m_{\mathrm{prim}}c$. $Q^\mathrm{prim}_\mathrm{SN}$ and $Q^\mathrm{prim}_\mathrm{SN}$ are the SN injection sources of primaries.
Since primary nuclei decay into lighter secondaries, the total parent particle energy cannot be transferred into the daughter one (see Equation (\ref{eq.11})). In our code, we, therefore, treat spallation decay as a non-conservative process. The consequence is that the momentum range in each bin for primaries and secondaries is different. The momentum space of secondaries is shifted towards lower energies due to the conservation of momentum per nucleon in spallation reactions compared to the momentum of primaries, following $p_{l^{'}} = (A_\mathrm{sec}/ A_\mathrm{prim})p_l$ (see Equation (\ref{eq.9})). To correctly represent secondaries in bins $l$, we correct the secondary source terms by adding $\tilde{Q}^{\mathrm{prim, sec}}_l$ and $\tilde{S}^{\mathrm{prim, sec}}_l$, in which the ratios represent the shift between bins in the spallation sources. To numerically integrate these equations, we use the following numerical scheme:
\begin{eqnarray}
n^\mathrm{prim}_{l}(t+ \Delta t) &=& n^\mathrm{prim}_{l}(t) - (\mathrm{d}n^\mathrm{prim}_{u,l+1/2}(\Delta t) - \mathrm{d}n^\mathrm{prim}_{u,l - 1/2}(\Delta t)) - \sum_{\mathrm{sec}} \Gamma^{\mathrm{prim, sec}}_{\mathrm{l}}n^\mathrm{prim}_{l} \Delta t + Q^\mathrm{prim}_\mathrm{SN}\Delta t \label{eq.A5} \\
e^\mathrm{prim}_{l}(t+ \Delta t) &=& e^\mathrm{prim}_{l}(t) - (\mathrm{d}e^\mathrm{prim}_{u,l+1/2}(\Delta t) - \mathrm{d}e^\mathrm{prim}_{u,l - 1/2}(\Delta t)) - R^\mathrm{prim}_l e^\mathrm{prim}_l \Delta t  \label{eq.A6} \\ &&- \sum_{\mathrm{sec}} \Gamma^{\mathrm{prim, sec}}_{\mathrm{l}}e^\mathrm{prim}_{l} \Delta t+ S^\mathrm{prim}_\mathrm{SN} \Delta t \nonumber \\  n^\mathrm{sec}_{l}(t+ \Delta t) &=& n^\mathrm{sec}_{l}(t) - (\mathrm{d}n^\mathrm{sec}_{u,l+1/2}(\Delta t) - \mathrm{d}n^\mathrm{sec}_{u,l - 1/2}(\Delta t)) + \sum_{\mathrm{prim}} (\tilde{Q}^{\mathrm{prim, sec}}_l \Gamma^{\mathrm{prim, sec}}_{l}n^\mathrm{prim}_{l} \label{eq.A7}  \\  && + (1 - \tilde{Q}^{\mathrm{prim, sec}}_{l+1} )\Gamma^{\mathrm{prim, sec}}_{l+1}n^\mathrm{prim}_{l+1}) \Delta t - \left<\frac{1}{\gamma^\mathrm{sec}(p)\tau^\mathrm{sec}}\right>_n n^\mathrm{sec}_{l}(t)\Delta t \nonumber \\
 e^\mathrm{sec}_{l}(t+ \Delta t) &=& e_{l}^{\mathrm{sec}}(t) - (\mathrm{d}e^\mathrm{sec}_{u,l+1/2}(\Delta t) - \mathrm{d}e^\mathrm{sec}_{u,l - 1/2}(\Delta t)) - R^\mathrm{sec}_l e^\mathrm{sec}_l \Delta t \label{eq.A8} + \sum_{\mathrm{prim}} (\tilde{Q}^{\mathrm{prim, sec}}_l \Gamma^{\mathrm{prim, sec}}_{l}e^\mathrm{prim}_{l} \\ && + (1 - \tilde{S}^{\mathrm{prim, sec}}_{l+1} )\Gamma^{\mathrm{prim, sec}}_{l+1}e^\mathrm{prim}_{l+1} ) \Delta t - \left<\frac{1}{\gamma^\mathrm{sec}(p)\tau^\mathrm{sec}}\right>_e e^\mathrm{sec}_{l}(t)\Delta t \nonumber
\end{eqnarray}
There, $\mathrm{d}n^\mathrm{prim}_{u,l\pm1/2}$ and $\mathrm{d}e^\mathrm{prim}_{u,l\pm1/2}$ are the bin interface fluxes of particle number and energy density integrated over $\Delta t$ in presence of energy losses: synchrotron, inverse Compton (IC), adiabatic expansion, and hadronic losses \citep[see][]{2011hea..book.....L}. $R_l$ represents the energy loss rate $\dot{e}_l/e_l$ resulting from (\ref{eq.A4}). The expression of $R^a_l$ for a specie labeled by $a$ is given by:
\begin{equation}
R^a_l = \frac{1}{e^a_l}\int^{p_{\mathrm{R}}}_{p_{\mathrm{L}}}4\pi p^2\mathrm{d}p\;f^a(p)b^{a}(p)\beta^a(p)c
\label{eq.A9}
\end{equation}
There, $b(p)=-\mathrm{d}p/\mathrm{d}t$ represents the energy loss terms for nuclei. With electrons and hadrons, it leads to the following general differential equation for momentum :
\begin{equation}
-\frac{\mathrm{d}p}{\mathrm{d}t} = \frac{1}{3}\nabla \cdot \vec{v}\;p + \frac{4 \sigma^a_T}{3(m_a c)^2}u_\mathrm{R}p^2 + \left(-\frac{\mathrm{d}p}{\mathrm{d}t}\right)_C
    \label{eq.A10}
\end{equation}
where the first term with the divergence of the velocity field represents the adiabatic process, while the second term represents radiative losses, $m_a=A_a m_p$ is the mass of nuclei, and $\sigma^a_T$ is the generalized Thomson scattering cross section. For protons and nuclei, $\sigma^a_T = (Z_a^4/A_a^2)(m_e/m_p)^2\sigma_T$ \citep[see][]{1994A&A...286..983M}, $m_e$ and $m_p$ are the mass of electrons and protons, and $u_R$ is the radiation energy density of magnetic fields, radiation fields, and CMB. For electrons, $\sigma^a_T=\sigma_T$. Although the generalization of the Thomson scattering is presented, it is only relevant for electrons and negligible for other particles in our context since $(m_e/m_p)^2 \ll 1$. The third term is the Coulomb energy losses. Its numerical implementation is described in Appendix~\ref{App.B}.

\subsection{Dimensionless variables}
\label{App.A2}

For numerical convenience, we resolve equations \ref{eq.4} and \ref{eq.5} for both trans-relativistic and ultra-relativistic ranges with the numerical scheme from the previous section using dimensionless variables for momentum and kinetic energy, re-scaled with the proton mass :
\begin{eqnarray}
    \tilde{p} = \frac{p}{m_pc},\quad g^a(\tilde{p})=\frac{T^a(p)}{m_pc^2}=\sqrt{\tilde{p}^2 + A_a^2} - A_a
    \label{eq.A11}
\end{eqnarray}
If $a$ labels electrons, we take the convention $A_e=m_e/m_p$. In such conditions, the dimensionless number and energy density in a bin l for specie $a$ read:
\begin{equation}
    \tilde{n}^a_l = \int^{\tilde{p}_{\mathrm{R}}}_{\tilde{p}_{\mathrm{L}}} 4\pi\tilde{p}^2 f^a(\tilde{p}) \mathrm{d}\tilde{p} = \frac{n^a}{(m_pc)^3}
    \label{eq.A12}
\end{equation}
\begin{equation}
    \tilde{e}^a_l= \int^{\tilde{p}_{\mathrm{R}}}_{\tilde{p}_{\mathrm{L}}} 4\pi\tilde{p}^2 f^a(\tilde{p})g^a(\tilde{p})\mathrm{d}\tilde{p}=\frac{e^a}{(m_pc)^3 m_p c^2}
    \label{eq.A13}
\end{equation}
The $R_l$ term reads :
\begin{eqnarray}
    R^a_l &=& \frac{1}{(m_pc)^3 m_p c^2 \tilde{e}^a_l}(m_pc)^3\int^{\tilde{p}_{\mathrm{R}}}_{\tilde{p}_{\mathrm{L}}}4\pi\tilde{p}^2\mathrm{d}\tilde{p}\;f^a(p)\left(-m_pc \frac{\mathrm{d}\tilde{p}}{\mathrm{d}t}\right)^{a}_l\frac{c\tilde{p}}{\sqrt{\tilde{p}^2+A_a^2}} \\ &=& \frac{1}{\tilde{e}^a_l}\int^{\tilde{p}_{\mathrm{R}}}_{\tilde{p}_{\mathrm{L}}}4\pi\tilde{p}^2\mathrm{d}\tilde{p}\;f^a(\tilde{p})\left(-\frac{\mathrm{d}\tilde{p}}{\mathrm{d}t}\right)^{a}_l\frac{\tilde{p}}{\sqrt{\tilde{p}^2+A_a^2}} \nonumber
    \label{eq.A14}
\end{eqnarray}
where the equation for energy loss is adapted in the following way:
\begin{equation}
    -\frac{\mathrm{d}\tilde{p}}{\mathrm{d}t} = \frac{1}{3}\nabla \cdot \vec{v} \;\tilde{p} + \frac{4 \sigma^a_T}{3A_a^2m_pc}u_\mathrm{R}\tilde{p}^2
    \label{eq.A15}
\end{equation}

The radioactive decay terms can be written in the following way:
\begin{eqnarray}
\left<\frac{1}{\gamma^a\tau^a}\right> _n = \frac{1}{\tilde{n}^a} \int^{\tilde{p}_{\mathrm{R}}}_{\tilde{p}_{\mathrm{L}}} 4\pi\tilde{p}^2\mathrm{d}\tilde{p} \;\frac{f^a(\tilde{p})}{\sqrt{1+\left(\tilde{p}/A_a\right)^2}\tau^a} = \frac{A_a}{\tilde{n}^a \tau^a} \int^{\tilde{p}_{\mathrm{R}}}_{\tilde{p}_{\mathrm{L}}} 4\pi\tilde{p}^2\mathrm{d}\tilde{p} \;\frac{f^a(\tilde{p})}{\tilde{p}}\frac{\mathrm{d}g^a}{\mathrm{d}\tilde{p}}
\label{eq.A16}
\end{eqnarray}
\begin{eqnarray}
\left<\frac{1}{\gamma^a\tau^a}\right>_e = \frac{1}{\tilde{e}^a} \int^{\tilde{p}_{\mathrm{R}}}_{\tilde{p}_{\mathrm{L}}} 4\pi\tilde{p}^2\mathrm{d}\tilde{p} \;\frac{f^a(\tilde{p})g^a(\tilde{p})}{\sqrt{1+\left(\tilde{p}/A_a\right)^2}\tau^a} = \frac{A_a}{\tilde{e}^a \tau^a} \int^{\tilde{p}_{\mathrm{R}}}_{\tilde{p}_{\mathrm{L}}} 4\pi\tilde{p}^2\mathrm{d}\tilde{p} \;\frac{f^a(\tilde{p})g^a(\tilde{p})}{\tilde{p}}\frac{\mathrm{d}g^a}{\mathrm{d}\tilde{p}}
\label{eq.A17}
\end{eqnarray}
All species share the same $\tilde{p}$ array in the code. However, they all have their mass number $A_a$, so each species has its kinetic energy, $\gamma^a$ factor, and momentum per nucleon that we distinguish.
\subsection{Solving with the piece-wise power-law approach}
\label{App.A3}
We implement the numerical scheme by applying the piece-wise power-law method to the distribution function and kinetic energy for each momentum bin $l$ in the range $\tilde{p} \in [\tilde{p}_{l-1/2},\tilde{p}_{l+1/2}]$:
\begin{equation}
f^a(\tilde{p}) = f^a_{l-1/2} \left(\frac{\tilde{p}}{\tilde{p}_{l-1/2}}\right)^{-q^a_l},
\quad
g^a(\tilde{p}) = g^a_{l-1/2} \left(\frac{\tilde{p}}{\tilde{p}_{l-1/2}}\right)^{s^a_l}
\label{eq.A18}
\end{equation}
where $f^a_{l-1/2}$ and $g^a_{l-1/2}$ are the values at the left edge of the bin, $q^a_l$ and $s^a_l$ are given by equations (\ref{eq.13}) and (\ref{eq.18}) respectively. We can calculate the number density in the bin $l$ as:
\begin{equation}
\tilde{n}^a_l = 4\pi \tilde{p}_{l-1/2}^3 f^a_{l-1/2} \times \left\{
    \begin{array}{ll}
        \left(\left(\frac{\tilde{p}_{l+1/2}}{\tilde{p}_{l-1/2}}\right)^{3-q^a_l} - 1 \right)/(3-q^a_l) & \mbox{if }  3 \neq q^a_l\\
        \mathrm{ln}\left(\frac{\tilde{p}_{l+1/2}}{\tilde{p}_{l-1/2}}\right) &  \mbox{if }  3 = q^a_l
    \end{array}
\right.
\label{eq.A19}
\end{equation}
We proceed similarly for the energy density:
\begin{equation}
\tilde{e}^a_l = 4\pi \tilde{p}_{l-1/2}^3 g^a_{l-1/2} f^a_{l-1/2} \times \left\{
    \begin{array}{ll}
        \left(\left(\frac{\tilde{p}_{l+1/2}}{\tilde{p}_{l-1/2}}\right)^{3 + s^a_l -q^a_l} - 1 \right)/ (3 + s^a_l -q^a_l) & \mbox{if }  3 + s^a_l \neq q^a_l \\
        \mathrm{ln}\left(\frac{\tilde{p}_{l+1/2}}{\tilde{p}_{l-1/2}}\right) &  \mbox{if } 3 + s^a_l = q^a_l
    \end{array}
\right.
\label{eq.A20}
\end{equation}
The factor $R^a_l$ can also be computed:
\begin{eqnarray}
R^a_l & = &\frac{1}{\tilde{e}^a_l}\int^{\tilde{p}_{\mathrm{R}}}_{\tilde{p}_{\mathrm{L}}}4\pi\tilde{p}^2\mathrm{d}\tilde{p}\;f^a(\tilde{p}) \left(-\frac{\mathrm{d}\tilde{p}}{\mathrm{d}t}\right)^{a}_l\frac{\mathrm{d}g^a}{\mathrm{\mathrm{d}\tilde{p}}} \\ \nonumber & = & \frac{1}{3}\nabla \cdot \vec{v}\; s^a_l  + \frac{1}{\tilde{e}^a_l} \frac{4 \sigma^a_T}{3A_a^2m_pc}u_\mathrm{R} 4 \pi \tilde{p}^4_{l-1/2} f^a_{l-1/2} s^a_l g^a_{l-1/2} \\ \nonumber & & \times \left\{
    \begin{array}{ll}
        \left(\left(\frac{\tilde{p}_{l+1/2}}{\tilde{p}_{l-1/2}}\right)^{4 + s^a_l -q^a_l} - 1 \right)/(4 + s^a_l -q^a_l) & \mbox{if }  4 + s^a_l \neq q^a_l \\
        \mathrm{ln}\left(\frac{\tilde{p}_{l+1/2}}{\tilde{p}_{l-1/2}}\right) &  \mbox{if } 4 + s^a_l = q^a_l
    \end{array}
\right.
\label{eq.A21}
\end{eqnarray}
The particle flux in the momentum bins coming from energy losses is modeled first by resolving Equation (\ref{eq.A11}). In the interval $(t,t+\Delta t)$, the solutions are, for each specie a:
\begin{eqnarray}
&& \tilde{p}^a(t+\Delta t)_{\mathrm{rad}} = \tilde{p}^a(t)\left(1 + \frac{4 \sigma^a_T}{3A_a^2m_pc}u_\mathrm{R} \tilde{p}(t)\Delta t\right)^{-1} \label{eq.A22} \\
&& \tilde{p}^a(t+\Delta t)_{\mathrm{adiab}} = \tilde{p}^a(t)\:\mathrm{exp}\left(-\frac{1}{3}\nabla \cdot \vec{v} \Delta t \right) \label{eq.A23} \\ \nonumber
\end{eqnarray}
Each particle labeled by $a$ will have a different time evolution for momentum: the intensity of Thomson scattering depends on the particle mass. It will be relevant only for electrons, while hadronic loss only applies to protons. Only adiabatic cooling is the same for all species.

We must model the number and energy density of particles crossing the bin edges in $(t,t+\Delta t)$. We then define the upstream momentum $\tilde{p}^a_{u,l-1/2}$, defined as follow: particles of momentum $\tilde{p}^a_{u,l-1/2}$ at time $t$ will reach the bin edge $\tilde{p}_{l-1/2}$ at $t + \Delta t$. In equations \ref{eq.A22}, and \ref{eq.A23}, $\tilde{p}^a_{u,l-1/2}$ is identified at $\tilde{p}^a(t)$, and $\tilde{p}_{l-1/2}$ as $\tilde{p}^a(t + \Delta t)$. For cooling, we obtain the flux by integrating the number density in the interval $[\tilde{p}_{l-1/2},\tilde{p}^a_{u,l-1/2}]$ :
\begin{equation}
    \mathrm{d}\tilde{n}^a_{u,l-1/2} = 4\pi \tilde{p}_{l-1/2}^3 f^a_{l-1/2} \times \left\{
    \begin{array}{ll}
        \left(\left(\frac{\tilde{p}^a_{u,l-1/2}}{\tilde{p}_{l-1/2}}\right)^{3-q^a_l} - 1 \right)/(3-q^a_l) & \mbox{if }  3 \neq q^a_l\\
        \mathrm{ln}\left(\frac{\tilde{p}^a_{u,l-1/2}}{\tilde{p}_{l-1/2}}\right) &  \mbox{if }  3 = q^a_l
    \end{array}
\right.
\label{eq.A24}
\end{equation}
We also have for the energy density flux:
\begin{equation}
\mathrm{d}\tilde{e}^a_{u,l-1/2} = 4\pi \tilde{p}_{l-1/2}^3  f^a_{l-1/2}g^a_{l-1/2} \times \left\{
    \begin{array}{ll}
        \left(\left(\frac{\tilde{p}^a_{u,l-1/2}}{\tilde{p}_{l-1/2}}\right)^{3 + s^a_l -q^a_l} - 1 \right)/(3 + s^a_l -q^a_l) & \mbox{if }  3 + s^a_l \neq q^a_l \\
        \mathrm{ln}\left(\frac{\tilde{p}^a_{u,l-1/2}}{\tilde{p}_{l-1/2}}\right) &  \mbox{if } 3 + s^a_l = q^a_l
    \end{array}
\right.
\label{eq.A25}
\end{equation}
For heating, we integrate number density in the range $[\tilde{p}^a_{u,l-1/2},\tilde{p}_{l-1/2}]$:
\begin{equation}
    \mathrm{d}\tilde{n}^a_{u,l-1/2} = 4\pi (\tilde{p}^a_{u,l-1/2})^3 f^a_{l-3/2}\left(\frac{\tilde{p}_{l-3/2}}{\tilde{p}^a_{u,l-1/2}}\right)^{q^a_l} \times \left\{
    \begin{array}{ll}
        \left(\left(\frac{\tilde{p}^a_{u,l-1/2}}{\tilde{p}_{l-1/2}}\right)^{3-q^a_l} - 1 \right)/(3-q^a_l) & \mbox{if }  3 \neq q^a_l\\
        \mathrm{ln}\left(\frac{\tilde{p}^a_{u,l-1/2}}{\tilde{p}_{l-1/2}}\right) &  \mbox{if }  3 = q^a_l
    \end{array}
\right.
\label{eq.A26}
\end{equation}
The energy density flux reads:
\begin{equation}
\mathrm{d}\tilde{e}^a_{u,l-1/2} = 4\pi (\tilde{p}^a_{u,l-1/2})^3 f^a_{l-3/2}g^a_{l-3/2}\left(\frac{\tilde{p}_{l-3/2}}{\tilde{p}^a_{u,l-1/2}}\right)^{q^a_l-s^a_l}\times \left\{
    \begin{array}{ll}
        \left(\left(\frac{\tilde{p}^a_{u,l-1/2}}{\tilde{p}_{l-1/2}}\right)^{3 + s^a_l -q^a_l} - 1 \right)/(3 + s^a_l -q^a_l) & \mbox{if }  3 + s^a_l \neq q^a_l \\
        \mathrm{ln}\left(\frac{\tilde{p}^a_{u,l-1/2}}{\tilde{p}_{l-1/2}}\right) &  \mbox{if } 3 + s^a_l = q^a_l
    \end{array}
\right.
\label{eq.A27}
\end{equation}
Radioactive decay losses follow the same method:
\begin{equation}
\left<\frac{1}{\gamma^a\tau^a}\right>_n = \frac{4\pi A_a}{\tilde{n}^a\tau^a}\tilde{p}^a_{l-1/2} S^a_l f^a_{l-1/2}g^a_{l-1/2} \times \left\{
    \begin{array}{ll}
        \left(\left(\frac{\tilde{p}^a_{l+1/2}}{\tilde{p}_{l-1/2}}\right)^{1 + s^a_l -q^a_l} - 1 \right)/(1 + s^a_l -q^a_l) & \mbox{if }  1 + s^a_l \neq q^a_l \\
        \mathrm{ln}\left(\frac{\tilde{p}^a_{l+1/2}}{\tilde{p}_{l-1/2}}\right) &  \mbox{if } 1 + s^a_l = q^a_l
    \end{array}
\right.
\label{eq.A28}
\end{equation}
\begin{equation}
\left<\frac{1}{\gamma^a\tau^a}\right>_e = \frac{4\pi A_a}{\tilde{e}^a\tau^a}\tilde{p}^a_{l-1/2} S^a_l f^a_{l-1/2}(g^a_{l-1/2})^2 \times \left\{
    \begin{array}{ll}
        \left(\left(\frac{\tilde{p}^a_{l+1/2}}{\tilde{p}_{l-1/2}}\right)^{1 + 2s^a_l -q^a_l} - 1 \right)/(1 + 2s^a_l -q^a_l) & \mbox{if }  1 + 2s^a_l \neq q^a_l \\
        \mathrm{ln}\left(\frac{\tilde{p}^a_{l+1/2}}{\tilde{p}_{l-1/2}}\right) &  \mbox{if } 1 + 2s^a_l = q^a_l
    \end{array}
\right.
\label{eq.A29}
\end{equation}
\subsection{Recovering the spectral index and distribution function}
\label{App.A4}
So far, we have reproduced the number and energy density $n^a_l$ and $e^a_l$ of particles using the distribution function $f^a_{l-1/2}$ and the slope $q^a_l$. We can do the opposite operation and recover $f^a_{l-1/2}$ and $q^a_l$ using $n^a_l$ and $e^a_l$: in \cite{2021ApJS..253...18O}, the quantity $e_l/n_lcp_{l-1/2}$ was used to find the slope $q_l$ of electrons using a Newton-Raphson algorithm. This approach can be generalized with the $e^a_l/n^a_lT^a_{l-1/2}$. With the dimensionless variables from Equation (\ref{eq.A11}):
\begin{equation}
    \frac{e^a_l}{n^a_lT^a_{l-1/2}}=\frac{\tilde{e}^a_l}{\tilde{n}^a_lg^a_{l-1/2}}
\label{eq.A30}
\end{equation}
Using expressions (\ref{eq.A17}) and (\ref{eq.A14}), we build a generalized equation for the $q^a_l$ variable :
\begin{equation}
\frac{\tilde{e}^a_l}{\tilde{n}^a_lg^a_{l-1/2}} =
\left\{
    \begin{array}{ll}
        \frac{\left(\frac{\tilde{p}_{l+1/2}}{\tilde{p}_{l-1/2}}\right)^{s^a_l} - 1}{s^a_l\mathrm{ln}\left(\frac{\tilde{p}_{l+1/2}}{\tilde{p}_{l-1/2}}\right)} & \mbox{if } 3  = q^a_l \\
        \frac{-s^a_l\mathrm{ln}\left(\frac{\tilde{p}_{l+1/2}}{\tilde{p}_{l-1/2}}\right)}{\left(\frac{\tilde{p}_{l+1/2}}{\tilde{p}_{l-1/2}}\right)^{-s^a_l} - 1} &  \mbox{if } 3 + s^a_l = q^a_l  \\  \frac{3 - q^a_l}{3 + s^a_l -q^a_l}\frac{\left(\frac{\tilde{p}_{l+1/2}}{\tilde{p}_{l-1/2}}\right)^{3 + s^a_l -q^a_l} - 1 }{\left(\frac{\tilde{p}_{l+1/2}}{\tilde{p}_{l-1/2}}\right)^{3 -q^a_l} - 1 } & \mbox{otherwise} \\
    \end{array}
\right.
\label{eq.A31}
\end{equation}
We solve this equation for $q^a_l$ in the code using a linear interpolation method (see section (\ref{subs:piece_wise_power_law}).

\section{Coulomb energy loss}\label{App.B}
\subsection{Coulomb losses for CR nuclei}

In \cite{2020MNRAS.491..993G}, an approximation of Equation (\ref{eq.25}) was used for protons and numerically solved using a free cooling method. This method justified treating Coulomb losses separately. We propose to use the following approximation for CR nuclei:

 \begin{equation}
 -\left(\frac{\mathrm{d}p}{\mathrm{d}t}\right)_\mathrm{C}\approx 10^{-18} Z^2\frac{n_e}{c}\left(1+\left(\frac{p}{A\,\mathrm{GeV}\,c^{-1}}\right)^{-1.9}\right)\:\mathrm{erg}\;\mathrm{cm^3}\;\mathrm{s}^{-1}
 \label{eq.B1}
 \end{equation}
 where $n_e$ is the electron gas density. The cooling explicitly depends on atomic mass $A$ and atomic number $Z$. Observing equation (\ref{eq.B1}), cooling is asymptotically constant at relativistic energy, where $p\gg 1\,\GeV $ per nucleon. On the contrary, momentum dependency increases and becomes non-negligible for $p\ll 1\,\GeV $ per nucleon, making the losses significantly higher at this energy range.
The Figure (\ref{fig.B1}) shows the cooling time $\tau_C$ dependency on momentum $p/(m_pc)$ of protons and \Ct, \Ben, and \Bel isotopes. The simulation time is also displayed, indicating the relevant energy range for treating cooling. $\tau_C$ significantly drops at non-relativistic energies, below $10\; \GeV$ for nuclei.
 \begin{figure}[!ht]
 \centering
 \includegraphics[scale=0.8]{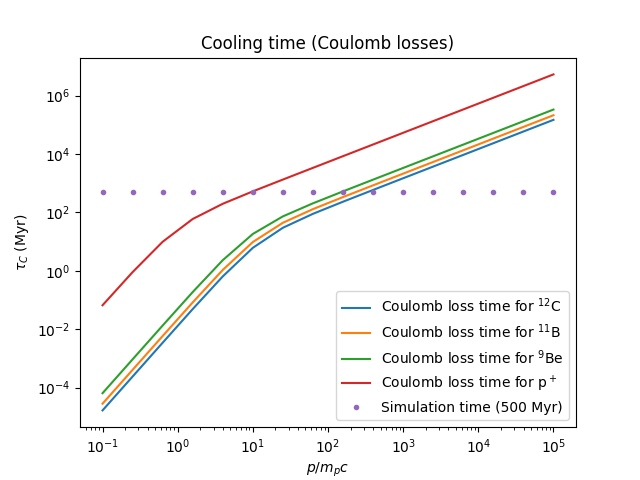}
 \caption{Coulomb loss time dependency on momentum for protons and three nuclei isotopes, estimated for $n_e=1\;\mathrm{cm}^{-3}$.}
 \label{fig.B1}
 \end{figure}
\subsection{Free cooling method}

We choose to compute Coulomb loss from formula (\ref{eq.B1}) using the free-cooling method in \cite{2020MNRAS.491..993G}. Consider a timestep $\Delta t$. Assuming number density conservation, $n^a(t)=n^a(t+\Delta t)$, and writing $p(t)=p_0$, $p(t+\Delta t)=p_1$, we can deduce:

\begin{equation}
4\pi \tilde{p}_0^2f^a(t, \tilde{p}_0)\mathrm{d}\tilde{p}_0 = 4\pi \tilde{p}_1^2f^a(t+\Delta t, \tilde{p}_1)\mathrm{d}\tilde{p}_1 \rightarrow f^a(t+\Delta t, \tilde{p}_1)=f^a(t, \tilde{p}_0)\left(\frac{\tilde{p}_0}{\tilde{p}_1}\right)^2\frac{\mathrm{d}\tilde{p}_0}{\mathrm{d}\tilde{p}_1}
\label{eq.B2}
\end{equation}
We know that Coulomb energy loss follows a power-law in momentum space:
\begin{equation}
-\left(\frac{\mathrm{d}\tilde{p}_0}{\mathrm{d}t}\right)_\mathrm{C} = Z^2 \Lambda_C \tilde{p}_0^h, \quad -\left(\frac{\mathrm{d}\tilde{p}_1}{\mathrm{d}t}\right)_\mathrm{C} = Z^2\Lambda_C \tilde{p}_1^h \rightarrow \frac{\mathrm{d}\tilde{p}_0}{\mathrm{d}\tilde{p}_1} = \left(\frac{\tilde{p_0}}{\tilde{p_1}}\right)^h
\label{eq.B3}
\end{equation}
where $h=-1.9$, $\Lambda_C = 10^{-18}n_e/c(A\,\GeV c^{-1}/m_pc)^{1-h}$. It follows:
\begin{equation}
f^a(t+\Delta t, \tilde{p}_1)=f^a(t, \tilde{p}_0)\left(\frac{\tilde{p}_0}{\tilde{p}_1}\right)^{2+h} = f^a(t, \tilde{p}_0)\left(\frac{\tilde{p}_0}{\tilde{p}_1}\right)^{0.1}
\label{eq.B4}
\end{equation}
We also need to find $\tilde{p}_1$. Solving the equation for $p$ by separating the variables, we do:
\begin{equation}
Z^2\Lambda_C \int_{\tilde{p}_0}^{\tilde{p}_1}\frac{\mathrm{d}\tilde{p}}{\tilde{p}^h} = \Delta t
\label{eq.B4}
\end{equation}
Which led to the following solution:
\begin{eqnarray}
\tilde{p}_1 &= & (\tilde{p}_0^{1-h} + (1-h)Z^2\Lambda_C\Delta t)^{1/(1-h)} \\
&  = & \left(\tilde{p}_0^{2.9} + 2.9Z^2\Lambda_C\Delta t\right)^{1/2.9} \\
\label{eq.B5}
\end{eqnarray}
So the final solution for $f^a$ is:
\begin{equation}
f^a(t+\Delta t, \tilde{p}_1) = f^a(t, \tilde{p}_0) \tilde{p}_0^{0.1}\left(\tilde{p}_0^{2.9} + 2.9Z^2\Lambda_C\Delta t\right)^{-0.1/2.9}
\label{eq.B6}
\end{equation}
The timestep $\Delta t$ must not exceed the cooling time of the left buffer bin, which has the minimal momentum. Therefore, computation of formula (\ref{eq.B6}) is executed with subcycling: if the cooling time in a given bin $\tau$ is smaller than $\Delta t$, the timestep is subdivided into $n=|\Delta t/\tau|$ substeps, and the bin update is applied iteratively at intervals of $\tau$ until the timestep $\Delta t$ is covered. %\note[MH]{Provide some details about the subcycling applied here.}
\end{document}